\documentclass[a4paper,11pt]{article}
\pdfoutput=1 

\usepackage{jheppub} 
\usepackage{Definitions}

\usepackage[T1]{fontenc} 

\usepackage{amsfonts,amsmath,amssymb}
\usepackage{graphicx}
\usepackage{subcaption}
\usepackage{xcolor}
\usepackage{amsmath,amsfonts,amssymb,amscd,mathrsfs,amsthm}
\usepackage{bbold}
\usepackage[]{graphicx}
\usepackage[percent]{overpic}
\usepackage{xcolor,dsfont}
\usepackage{mathtools}
\usepackage{simplewick}

\usepackage{tikz}
\usetikzlibrary{decorations.pathmorphing,snakes}
\usetikzlibrary{decorations.markings}
\usepgflibrary{shapes.geometric}

\def\qaq{\quad \text{and} \quad}

\def\wh{\widehat}
\def\lra{\leftrightarrow}

\usepackage[utf8]{inputenc}
\usepackage[T1]{fontenc}

\def\zb{\bar{z}}

\newcommand{\abs}[1]{{|#1|}}

\newcommand{\PSX}{|\Delta,x\rangle}
\newcommand{\PSK}{|\Delta,k\rangle}

\newcommand{\brakkket}[1]{\langle #1 \rangle}

\newcommand{\ket}[1]{|#1\rangle}

\newcommand{\expec}[1]{\langle #1 \rangle}

\def\thd{\tfrac{d}{2}}
\def\toolazy{Källén–Lehmann }
\newcommand{\Disc}[1]{\text{Disc}[#1]}

\newcommand{\FF}[4]{{}_2F_1\!\left[{{#1,#2}~\atop~{#3}} \Big| #4 \right]}
\newcommand{\FFG}[3]{{}_pF_q\!\left[{{#1}~\atop~{#2}} \Big| #3 \right]}
\newcommand{\FFF}[3]{{}_3F_2\!\left[{{#1}~\atop~{#2}} \Big| #3 \right]}

\newcommand*{\Scale}[2][4]{\scalebox{#1}{$#2$}}%

\newtheorem*{theorem*}{Theorem}

\def\ldef{\mathrel{\mathop:}=}
\def\rdef{=\mathrel{\mathop:}}
\newcommand{\limu}[1]{\mathrel{\mathop{\sim}\limits_{\scriptstyle{#1}}}}

\newcommand{\reef}[1]{(\ref{#1})}
\def\beq{\begin{equation}} 
\def\eeq{\end{equation}}
\def\be{\begin{equation}} 
\def\ee{\end{equation}}
\def\baa{\begin{align}}
\def\eaa{\end{align}}
\def\nn{\nonumber} 
\def\bsub{\begin{subequations}}
\def\esub{\end{subequations}}
\newcommand{\Od}{\mathcal{O}^\dagger}
\newcommand{\Dp}{\Delta_\mathcal{O}}
\def\bml{\begin{multline}} 
\def\eml{\end{multline}}

\def\mbb{\mathbb}

\def\mca{\mathcal}

\def\mrm{\mathrm}
\def\msc{\mathscr}

\def\msf{\mathsf}
\def\ads2{AdS${}_2$}

\def\th{\tfrac{1}{2}}
\def\half{\frac{1}{2}}
\def\pd{\partial}

\def\DD{\Delta}
\def\Oo{\mathcal{O}}

\def\DD{\Delta}
\def\D{\Delta}

\newcommand{\hd}{\frac{d}{2}}
\newcommand{\eref}[1]{(\ref{#1})}
\newcommand{\OO}{\mathcal{O}}
\newcommand{\Ot}{\tilde{\mathcal{O}}}
\newcommand{\Real}{\mathbb{R}}
\newcommand{\braketmh}[3]{\langle #1|#2|#3 \rangle}
\newcommand{\brakketmh}[2]{\langle #1|#2\rangle}
\newcommand{\ketmh}[1]{|#1\rangle}
\newcommand{\bramh}[1]{\langle #1|}
\renewcommand{\Im}{\operatorname{Im}}

 \usetikzlibrary{shapes.misc}
\tikzset{cross/.style={cross out, draw=black, minimum size=2*(#1-\pgflinewidth), inner sep=0pt, outer sep=0pt},
cross/.default={3pt}}

\usepackage{dsfont}
\usepackage{array}
\usepackage{braket}
\usepackage{color,soul}
\usepackage{bm}
\usepackage[force]{feynmp-auto}
\usepackage[shortlabels]{enumitem}

\usepackage{etoolbox}

\makeatletter
\patchcmd{\ttlh@hang}{\parindent\z@}{\parindent\z@\leavevmode}{}{}
\patchcmd{\ttlh@hang}{\noindent}{}{}{}
\makeatother

\title{\boldmath Towards the non-perturbative cosmological bootstrap}

\author{Matthijs Hogervorst,}
\author{Jo\~ao Penedones,}
\author{Kamran Salehi Vaziri}

\affiliation{Fields and Strings Laboratory, Institute of Physics\\ École Polytechnique Fédéral de Lausanne (EPFL)
\\ Route de la Sorge, CH-1015 Lausanne, Switzerland}

\emailAdd{matthijs.hogervorst@epfl.ch}
\emailAdd{joao.penedones@epfl.ch}
\emailAdd{kamran.salehivaziri@epfl.ch}

\abstract{We study quantum field theory on a de Sitter spacetime dS$_{d+1}$ background. Our main tool is the Hilbert space decomposition in irreducible unitary representations of its isometry group $SO(d+1,1)$.
As the first application of the Hilbert space formalism, we recover the Källen-Lehmann spectral decomposition of the scalar bulk two-point function. In the process, we exhibit a relation between poles in the corresponding spectral densities and the boundary CFT data. 
Moreover, we derive an inversion formula for the spectral density through analytical continuation from the sphere and use it to find the spectral decompisiton for a few examples. 
Next, we study the conformal partial wave decomposition of the four-point functions of boundary operators. These correlation functions are very similar to the ones of standard conformal field theory, but have different positivity properties that follow from unitarity in de Sitter.
We conclude by proposing a non-perturbative conformal bootstrap approach to the study of these late-time four-point functions, and we illustrate our proposal with a concrete example for QFT in dS$_2$.}

\begin{document} 
\maketitle
\flushbottom

\section{Introduction}

de Sitter (dS) spacetime is the simplest model of an expanding universe~\cite{Linde:2007fr,Baumann:2009ds}. 
It is interesting to understand the behaviour of quantum fields in such a background spacetime.
Most studies so far focus on a perturbative treatment of interactions \cite{Maldacena:2002vr,Arkani-Hamed:2015bza,Arkani-Hamed:2018kmz,Pajer:2020wnj,Goodhew:2020hob,Sleight:2019hfp,Baumann:2019oyu,Baumann:2020dch}.
In this paper, we take the first
 steps towards a non-perturbative formulation of 
Quantum Field Theory (QFT) on a dS background.
Our approach builds on the well-known fact that late-time correlation functions transform as conformal correlation functions under the isometry group $SO(d+1,1)$ of dS$_{d+1}$ \cite{Arkani-Hamed:2015bza}. 
This suggest that one can employ conformal bootstrap methods to study QFT in dS.
We support this idea by writing down the crossing equations and the partial wave decomposition for late-time four-point functions of scalar operators (see section \ref{sec:FourPointFunction}). The main difference with respect to the usual conformal bootstrap follows from requiring unitary representations of $SO(d+1,1)$ as opposed to $SO(d,2)$~\cite{Luscher:1974ez}.

Let us briefly recall the main ingredients of the  conformal bootstrap approach \cite{Rattazzi:2008pe,Poland:2018epd} applicable to Conformal Field Theories (CFTs) in $\mbb{R}^d$. 
The central observables are four-point functions of primary operators. For simplicity, consider four identical scalar operators in Euclidean space,
\beq
\label{4ptsimple}
G(x_1,x_2,x_3,x_4)=\langle \mathcal{O}(x_1)  \mathcal{O}(x_2)  \mathcal{O}(x_3)  \mathcal{O}(x_4)  \rangle
=G(x_{\pi(1)},x_{\pi(2)},x_{\pi(3)},x_{\pi(4)})\,,
\eeq
such that crossing symmetry is just invariance under permutations $\pi$ of the points $x_i \in \mathbb{R}^d$.
Using the convergent Operator Product Expansion (OPE), one can derive the conformal block decomposition 
\beq
\label{CBdecom}
G(x_1,x_2,x_3,x_4)=\sum_{\Delta,\ell} \,C^2_{\Delta,\ell}\, G_{\Delta,\ell}^{12,34} (x_1,x_2,x_3,x_4)\,,
\qquad \qquad C^2_{\Delta,\ell} \ge 0\,,
\eeq
where $C_{\Delta,\ell}$ are theory dependent OPE coefficients and $G_{\Delta,\ell}^{12,34}$ are kinematic functions called conformal blocks.
$SO(d,2)$ unitarity implies that $C^2_{\Delta,\ell} \ge 0$ and imposes lower bounds on the dimensions $\DD$ that can appear in~\reef{CBdecom}. 
Remarkably, the compatibility of  crossing symmetry, unitarity and the conformal block expansion \eqref{CBdecom} leads to non-trivial bounds in the space of CFTs. For example, it leads to a very precise determination of critical exponents in the Ising and $O(N)$ models in three dimensions \cite{Kos:2016ysd}. 

QFT in dS contains observables like \eqref{4ptsimple}. These are obtained by studying four-point correlations functions in the late-time limit (see section \ref{sec:SphereCorrelation} for more details).
In this context, crossing symmetry still holds. In fact, invariance under permutation of the points $x_i \in \mathbb{R}^d$ is an immediate consequence of operators commuting at spacelike separation.
In the dS context, there is no convergent OPE that leads to a conformal block decomposition. On the other hand, we can use the resolution of the identity decomposed into unitary irreducible representations of $SO(d+1,1)$ to obtain
\beq
\label{PWdecom}
G(x_1,x_2,x_3,x_4)=\sum_{\ell}  \int d\nu \,I_{\ell}(\nu) \, \Psi_{\frac{d}{2}+i\nu,\ell}^{12,34} (x_1,x_2,x_3,x_4)\,,\qquad \qquad
I_{\ell}(\nu)\ge 0\,,
\eeq
where $\Psi$ is a kinematic function often termed conformal partial wave. For simplicity, here we assumed that  only principal series representations contribute to this four-point function.
$SO(d+1,1)$ unitarity implies positivity of the expansion coefficients $I_{\ell}(\nu)\ge 0$.
Our main message is that the similarity between these two setups
\begin{align}
{\rm \bf Conformal\  Bootstrap:} \qquad  \eqref{4ptsimple} \quad+ \quad \eqref{CBdecom}  \nonumber \\
{\rm \bf QFT\ in \ dS\  Bootstrap:}  \qquad \eqref{4ptsimple} \quad+\quad \eqref{PWdecom}   \nonumber
\end{align}
 suggests that one may be able to develop (numerical) conformal bootstrap methods to obtain non-perturbative constraints on the space of QFTs in dS.  In this work, we give the first steps in this program.

\subsubsection*{Outline}
We start by reviewing some basic facts about  free field theory and Conformal Field Theory (CFT) in dS. This motivates the discussion of the main (non-perturbative) properties of QFT in dS presented in section~\ref{sec:QFT in dS Pre}. In particular, we define boundary operators via the late-time expansion, emphasise the absence of a state-operator map and spell out the resolution of identity in~\reef{completeXPre} which is heavily used later on for spectral decomposition of two and four-point functions.

In section~\ref{sec: Bulk two-point function}, we study two-point functions of \textit{bulk} scalar operators where we derive the~\toolazy representation.  We explain how to analytically continue the two-point function from the sphere to dS, transforming the decomposition in spherical harmonics into the Källen-Lehmann dS representation and then find an inversion formula for the spectral density in~\reef{eq:RhoFormulaNew}. Next, we clarify the concept of boundary operators by relating them to spectral density poles and conclude with concrete examples of massive free field and CFT in bulk. 

Sections~\ref{sec:FourPointFunction} and~\ref{sec:Bootstrpping boundary correlators} are concerned with four-point functions of boundary operators where we derive positivity conditions imposed by unitarity. In section~\ref{Examples4pt}, we then calculate the partial wave coefficient of massive free theory in bulk as well as its correction to the leading order of $\lambda \phi^4$ interaction. It turns out that unitarity requires appearance of local terms in free theory discussed in section~\ref{sec:Local}. Lastly, in section~\ref{sec:Bootstrpping boundary correlators}, we focus on dS$_2$ to set up a de Sitter bootstrap program.  After reviewing basics of the one-dimensional CFTs in section~\ref{sec:CFT1}, we discuss the convergence issue of partial wave expansion in dS. This issue has been overcome by proposal of a regularization scheme in section~\ref{sec:Regularized Crossing}. In the end, a concrete example of a non-trivial bound on partial wave coefficients is presented in section~\ref{sec: dS numerical bootstrap}. Our work leaves many opens  questions several of which we discuss in section~\ref{sec:Discussion}.
\\

{\bf Note added:} In the course of this project, we became aware that the authors of~\cite{DiPietro:2021sjt} were working on related questions. We are grateful to them for several useful discussions and  highly recommend their upcoming paper to the reader.

\section{Quantum field theory in de Sitter}\label{sec:QFT in dS Pre}
This section starts with a review of the basics of QFT in a fixed de Sitter  background. After defining dS as a hypersurface in embedding space and introducing some commonly used coordinate systems, we discuss the isometry group of dS in detail. After that, we review the quantization of a massive free scalar field in de Sitter.   
In \ref{sec:nonpert}, we state some  non-perturbative properties of QFT in dS. Namely, we discuss the structure of the Hilbert space and correlation functions of bulk and boundary operators.
 
\subsection{de Sitter spacetime}\label{sec:dS coordinates}
de Sitter space in $d+1$ dimensions (or dS${}_{d+1}$) can be realized as the embedding of the set of points that are a distance $R$ from the origin\footnote{Often the Hubble scale $H = 1/R$ is used instead of $R$.} in  Minkowski space $\mathbb{M}^{d+1,1}$ with the signature $(-,+,\ldots,+)$:
\begin{equation}
\label{eq:embeddingDef}
\text{dS}_{d+1}: \qquad -({X^0})^2  \,+\,  ({X^1})^2  \,+\,  \ldots \,+\, ({X^{d+1}})^2= R^2~.
\end{equation}

Let us present three different coordinate systems that cover all or part of dS. To start, we may introduce \emph{global coordinates} as follows
\begin{equation}
X^0 = R \sinh t~, \qquad X^i = R \cosh t\,y^i
\end{equation}
in which $i=1,\ldots,d+1$ and $y^i \in \mbb{R}^{d+1}$ are unit vectors ($y^i y_i = 1$), so they span the $d$-sphere $S^d$ . The induced metric in global coordinates is given by
\begin{align}
  \label{eq:globalMetric}
ds^2 = R^2 \left(-dt^2 + \cosh^2 t\, d\Omega_d^2\right)\,,
\end{align}
where $d\Omega^2_d$ denotes the standard metric of the unit $S^d$. After the change of variable $\tan(\tau /2)=\tanh(t/2)$, we find 
\be\label{eq:dS global prime}
X^0 = R \frac{\sin\tau}{\cos\tau}~,\qquad X^i = R \frac{y^i}{\cos\tau}~
\ee
from which the metric reads 
\begin{align}
  \label{eq:conformalMetric}
ds^{2}=R^{2}{\frac{-d\tau ^{2}+d\Omega _{d}^{2}}{\cos^{2}\tau}}~,
\end{align}
with $\tau \in(-\pi/2,\pi/2)$ . We conclude that in these coordinates dS is conformally equivalent to (part of) the Minkowski cylinder. This observation is important in the analysis of conformal field theories in dS (see section \ref{sec:CFTdS}). 

Finally, it will be useful to foliate dS using flat slices. This can be done by 
\begin{equation}
X^0 =R \frac{\eta^2 -1- x^2}{2\eta}~,\qquad
X^{d+1} = R\frac{x^2 - 1 - \eta^2}{2\eta}~, \qquad
X^{\mu} = -R\frac{x^\mu}{\eta}
\label{Xetax}
\end{equation}
for $\mu = 1,\ldots,d$. For definiteness, we will pick the Poincar\'{e} patch covering $X^0 + X^{d+1} \geq 0$. So, strictly speaking, such foliations only cover half of de Sitter space. However, this region is causally complete, in the sense that it is impossible to send a message to the other patch with $X^0 + X^{d+1} < 0$. This parametrization is called  \emph{conformal} or \emph{Poincar\'{e} coordinates} in which $\eta < 0$ and $x^\mu \in \mbb{R}^d$.

\begin{figure}
     \begin{subfigure}[c]{0.45\textwidth}
         \includegraphics[scale=0.7]{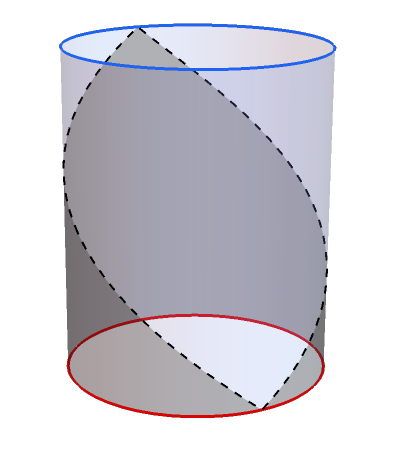}
     \end{subfigure}
\hspace{-5mm}
     \begin{subfigure}[c]{0.45\textwidth}
         \includegraphics [scale=0.5]{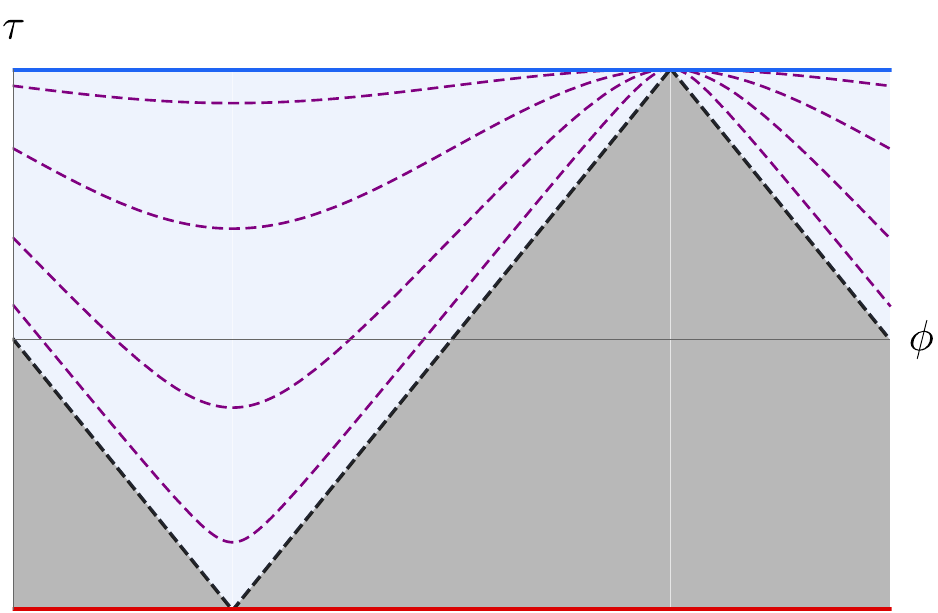}
    \end{subfigure}
              \caption{Left: de Sitter spacetime dS${}_{d+1}$ as a hollow Minkowski cylinder, cf.\@ equation~\reef{eq:conformalMetric}. Time $\tau$ runs upwards from $-\pi/2$ to $\pi/2$. Every horizonal timeslice corresponds to a copy of $S^d$. The infinite past (resp.\@ future) is shown as a solid red (blue) line. The light blue area is the Poincar\'{e} patch $X^0 + X^{d+1} \geq 0$; the boundary between the two patches is shown as a dashed line. Right: Penrose diagram of the same spacetime, specializing to  $d=1$. Spatial slices $S^1$ are parametrized by an angle $\phi \sim \phi + 2\pi$. Several timeslices of fixed $\eta < 0$ in the conformal coordinates~\reef{eq:flatMetric} are shown as thin purple lines. The left and right sides of the diagram are identified, owing to the periodicity of $\phi$.}
\label{fig:cylinder}
\end{figure}

 The coordinate $\eta$ plays the role of a conformal time, whereas the $x^\mu$ are spatial coordinates.  Poincare coordinates have the conformally flat metric of
 \begin{equation}
\label{eq:flatMetric}
ds^2 =R^2 \, \frac{-d\eta^2+d\vec{x}^2}{\eta^2}~.
 \end{equation}
This will be the main coordinate system we use throughout this paper, as it makes the conformal symmetry of the late-time boundary $\eta=0$ manifest. Global and conformal coordinates are related via the dictionary 
\begin{equation}
\eta = -\frac{1}{\sinh(t) + \cosh(t) y^{d+1}}~,\qquad
x^\mu = \frac{y^\mu}{\tanh(t) + y^{d+1}}~,
\end{equation}
which maps the late-time Poincar\'{e} patch to the subset of global coordinates satisfying $y^{d+1} +  \tanh(t) \geq 0$.  Figure~\ref{fig:cylinder} shows a picture of dS${}_{d+1}$ in the global coordinates of eq.~\reef{eq:conformalMetric}, along with a Penrose diagram which shows timeslices with $\eta = \text{constant}$. 

Let us draw your attention to the connection between dS and a sphere. At the level of embedding space, they are related by the Wick rotation $X^0 \to i X^0$. Moreover, the corresponding coordinates and metrics map to each other by the Wick rotation  $t \to i \theta $ in \eqref{eq:globalMetric} with $\theta \in [-\frac{\pi}{2}, \frac{\pi}{2})$.\footnote{This argument does not work for Poincare coordinates because these only cover half of the space, namely the points that satisfy $X^0 + X^{d+1} \geq 0$.} 
In other words, Euclidean dS is indeed the sphere.  

\subsubsection{Symmetries of dS} 
de Sitter space dS${}_{d+1}$ is manifestly invariant under $SO(d+1,1)$, as can be seen from its definition~\reef{eq:embeddingDef}. As such, it has $\th (d+2)(d+1)$ Killing vectors. The symmetry generators 
\begin{equation}\label{JAB}
  J_{AB} = X_A\frac{\partial}{\partial X^B} - X_B\frac{\partial}{\partial X^A},
  \quad
  A,B = 1,\ldots,d+2
\end{equation}
are rotations and boosts that preserve the dS hypersurface in the embedding space, and they obey commutation relations
\beq
[J_{AB},J_{CD}]=-\eta_{AC}J_{BD}-\eta_{BD}J_{AC}+\eta_{BC}J_{AD}+\eta_{AD}J_{BC} \label{Jcom}
\eeq
where $\eta_{AB} = \text{diag}(-1,1,\ldots,1)$. 
After relabeling the symmetry  generators as follows
\beq
\begin{gathered}
\label{eq:relabelGenerators}
D = J_{0,d+1}~, \qquad M_{\mu \nu} = J_{\mu\nu}~,\\
P_\mu = J_{0,\mu} + J_{d+1,\mu}~, \qquad K_\mu = J_{d+1,\mu} - J_{0,\mu}
\end{gathered}
\eeq
with $\mu,\nu = 1,\ldots,d$, we find that the new generators $D$, $P_\mu$, $K_\mu$, and $M_{\mu \nu}$  obey the familiar Euclidean conformal algebra:
\begin{equation}
\label{eq:CFTcommutation}
\begin{gathered}
  [D,P_\mu]= P_\mu~,
  \qquad
      [D,K_\mu]= -K_\mu~,
      \qquad
[K_\mu,P_\nu] = 2\delta_{\mu \nu}D - 2M_{\mu\nu}~,\\
[M_{\mu\nu},P_\rho] = \delta_{\nu\rho}P_\mu - \delta_{\mu\rho}P_\nu~,
\qquad
    [M_{\mu\nu},K_\rho] = \delta_{\nu\rho}K_\mu - \delta_{\mu\rho}K_\nu~,\\
[M_{\mu\nu},M_{\rho\sigma}]=\delta_{\nu\rho}M_{\mu\sigma}-\delta_{\mu\rho}M_{\nu\sigma}+\delta_{\nu\sigma}M_{\rho\mu}-\delta_{\mu\sigma}M_{\rho\nu}
\end{gathered}
\end{equation}
as well as  $[P_\mu,P_\nu] = 0~$,  $[K_\mu,K_\nu] = 0$ and $[D,M_{\mu\nu}]=0$. In our conventions, all these generators are  anti-hermitian. 

Expressed in flat coordinates $(\eta,x^\mu)$, the corresponding Killing vectors of dS${}_{d+1}$ can be expressed as follows:\footnote{Strictly speaking, the Killing vectors from Eq.~\reef{eq:Killing} need to be defined with an additional minus sign to be consistent with~\reef{eq:CFTcommutation}. The notation~\reef{eq:Killing} will prove to be convenient later on. Here we avoid equality sign and instead used ``$:$'' to emphasis these are not the isometry generators or Killing vectors $\hat{Q}$. Instead we use the notation $Q$ for the corresponding conserved charges.}
\beq
\label{eq:Killing}
\begin{aligned}
P_\mu &:  \frac{\partial}{\partial x^\mu}~,\\
M_{\mu \nu} &: x_\nu\frac{\partial}{\partial x^\mu}  -  x_\mu\frac{\partial}{\partial x^\nu}~,\\
D&:\eta \frac{\pd}{\pd \eta} +  x^\mu \frac{\partial}{\partial x^\mu}~,\\
K_\mu &: (\eta^2-x^2)\frac{\partial}{\partial x^\mu}   + 2  x_\mu \eta
\frac{\pd}{\pd \eta} + 2x_\mu x^\nu \frac{\partial}{\partial x^\nu} ~.
\qquad
\end{aligned}
\eeq
Note that at the late-time boundary $\eta=0$, the generators are the standard generators of the conformal algebra in flat space. We will exploit the conformal symmetry of late-time dS extensively throughout this paper.

Finally, local operators in de Sitter transform under the $SO(d+1,1)$ isometries according to~\reef{eq:Killing}. To be precise, a local scalar operator $\phi(\eta,x)$ transforms under the conformal generator $Q$ as
\beq
\label{eq:WardOperator}
[Q,\phi(\eta,x)] = \hat{Q} \cdot \phi(\eta,x)
\eeq
where $\hat{Q}$ is the Killing vector differential operator from Eq.~\reef{eq:Killing} --- for instance
\beq
    [P_\mu,\phi(\eta,x)] = \pd_\mu \phi(\eta,x)~,
    \quad
    [D,\phi(\eta,x)] = (\eta \partial_\eta + x \cdot \partial) \phi(\eta,x)
\eeq
and likewise for the other generators.

\subsection{Representation theory of $SO(d+1,1)$}\label{sec:reps}
Throughout this paper, we will need to deal with Hilbert spaces of QFTs in de Sitter. Such Hilbert spaces are organized into unitary irreducible representations of the dS isometry group, $SO(d+1,1)$. The representation theory of this group is rather complicated, owing to its non-compactness, but for our purposes we will only need to recall some basic facts about the most common representations. In general, we refer to~\cite{Dobrev:1976vr,Dobrev} for an in-depth discussion of $SO(d+1,1)$ group theory relevant to high-energy physics, or more recently~\cite{Basile:2016aen,Karateev:2018oml,Sengor:2019mbz}. For a more pedological review also see~\cite{Sun:2021rrs,Sun:2021thf}.  A technical and explicit discussion for general $d$ with a focus on special functions is presented in Ref.~\cite{Vilenkin}. Concerning the case of dS${}_2$, the representation theory of $SO(2,1)$ or its double cover $SL(2,\mbb{R})$ is discussed for example in~\cite{Knapp,HoweTan,Kitaev:2017hnr,Anous:2020nxu,Anninos:2019oka}.

As is well-known from $d$-dimensional CFT, one can construct infinite-dimensional representations of $SO(d+1,1)$ labeled by a dimension $\Delta$ and a representation $\varrho$ of $SO(d)$. In the present paper, only traceless symmetric tensor representations of $SO(d)$ will play a role, and these are labeled by an integer $\ell = 0,1,2,\ldots$, with $\ell = 0$ corresponding to the trivial representation. The dimension $\Delta$ can be any complex number, contrary to unitary CFTs where $\Delta$ is always real and positive. 
We define $\D$ so that the $SO(d+1,1)$ quadratic Casimir given by
\beq
\label{eq:Casimir}
C = D^2 - \th(K_\mu P^\mu + P_\mu K^\mu + M_{\mu \nu} M^{\mu \nu})
\eeq
has eigenvalue 
\beq
\label{eq:valofcas}
C(\Delta,\ell) = \Delta(\Delta-d) + \ell(\ell+d-2)~.
\eeq
So each  irreducible representation is defined by a pair of a dimension and a spin: $[\Delta,\ell]$. In any dimension $d$, there are two continuous families of unitary irreps:
\begin{itemize}
\item The \textbf{principal series} $\mathcal{P}$ has $\D=\frac{d}{2}+i\nu$ with $\nu\in\mathbb{R}$, and it exists for any spin $\ell$;
\item The \textbf{complementary series} $\mathcal{C}$ has $\Delta = \tfrac{d}{2} + c$ with $c \in \mbb{R}$, and the range of $c$ depends on $\ell$. To wit: 
\begin{description}
\item for spin $\ell =0$, $0< |c| \leq \hd$;
\item for spin $\ell\geq 1$, $0< |c| \leq \hd-1$.
\end{description}
The endpoints of the complementary series are known as {\bf exceptional series} of representations.
\item In odd $d$, there are in addition \textbf{discrete series} representations $\mathcal{D}$ with integer or half-integer values of $\Delta$.
\end{itemize}
Finally, we stress that the representation $[\Delta,\ell]$ and its so-called \emph{shadow} $[d-\Delta,\ell]$ are unitarily equivalent. This means that principal series irreps with $\Delta = \hd \pm i \nu$ can be identified, as well as complementary series irreps with $\Delta = \hd \pm c$. 

In $d=1$, the group $SL(2,\mbb{R}) \cong SU(1,1)$ has both ``even'' and ``odd'' principal series.\footnote{For an explicit definition of these irreps, see~\cite[Ch.\@ II \S 5]{Knapp}, where they are labeled as $\msc{P}^{\pm,iv}$. The irreps $\msc{P}^\pm$ are indistinguishable at the level of the Lie algebra, but they differ for finite group transformations.} The odd series of irreps does not factor down to an irrep of $SO(2,1) \cong PSL(2,\mbb{R}) \cong SL(2,\mbb{R})/\{\pm \mathds{1}\}$. 

\subsubsection*{Tensor products}
In this paper, we expand correlation functions using the Hilbert space decomposition into unitary irreducible representations. More precisely, we inject the resolution of identity~\reef{completeXPre} into correlation functions of scalar operators. We would like to know which irreps have a non-zero contribution to our correlation function expansion. In this work, we start from states that belong to scalar principal series representations and then analytically continue to complimentary series. We will come back to the case of the complimentary series shortly. Here we mention the list of tensor products of different representations that shed light on what we expect to appear in correlation functions. However, for the two-point and four-point functions, we present a direct reasoning of what representations to expect.

For dimensions $d\geq 2$, it is known that the tensor product of two \textit{scalar} principal series representations of the $SO(d+1,1)$ is decomposable into principal series representations~\cite{Dobrev:1976vr,Naimark,Martin} only. Schematically:
\bsub
\label{eq:Clebsch}
\begin{equation}
\mathcal{P}~\otimes~\mathcal{P} ~=~ \mathcal{P} \qquad \text{for}~  d\geq 2~.
\end{equation}
In the case of $d=1$ (that is to say dS${}_2$); however, the tensor product of two (even or odd) principal series irreps generally contains both principal series irreps, as well as discrete series irreps~\cite{Repka:1978}:
\begin{equation}
\mathcal{P}~\otimes~\mathcal{P} ~=~ \mathcal{P}~\oplus~\mathcal{D}~ \qquad \text{for}~  d=1~.
\end{equation}
For tensor products involving discrete series, we have schematically the tensor products\footnote{Strictly speaking, the discrete series are made up of two lowest weigh $\mathcal{D}^+$ and highest weight $\mathcal{D}^-$ representations. The decomposition of the  tensor products $\mathcal{D}^+\otimes \mathcal{D}^+$ and $\mathcal{D}^-\otimes \mathcal{D}^-$ only give $\mathcal{D}^+$ and $\mathcal{D}^-$ respectively. However, in addition to the discrete series, the principal series also show up in $\mathcal{D}^+\otimes \mathcal{D}^-$ decomposition.}
\begin{align}
\mathcal{P}~\otimes~\mathcal{D}~&= ~ \mathcal{P}~\oplus~\mathcal{D}~,\\
\mathcal{D}~\otimes~\mathcal{D}~ &= ~ \mathcal{P}~\oplus~\mathcal{D}.
\end{align}
\esub
In summary, in decomposition of scalar principal series for $d\geq2$ only principal series show up, while for $d=1$ one has to add discrete series too. 
Note that this discussion is about the \textit{bulk states}, not the boundary operators. In fact, because of the lack of bulk state-boundary operator map in dS, the boundary operators follow~\reef{eq:Clebsch} only in special cases like the free theory. In other words, in a generic interacting QFTs, late-time operators in dS do not necessarily fall into unitary irreps. 

There are indirect ways to see the results above. For example in the case of boundary four-point function, the injection of resolution of identity~\reef{completeXPre} leads to the conformal partial wave expansion (see~\ref{sec:FourPointFunction}). The completeness of partial waves that belong to principal series in $d\geq 2$ is another way to see the result above. For the case of $d=1$, one indeed needs to add discrete series to the expansion. See~\cite[A.3]{Simmons-Duffin:2017nub} for a more detailed discussion\footnote{The discussion there is about boundary operators with dimensions in the principal series or on the real line, but one can generalize the arguments to arbitrary dimensions by analytic continuation--taking pole crossing into account.}. Another indirect way to find the list above is to take a bulk CFT with symmetry group of $SO(d+1,2)$ and decompose its irreps to $SO(d+1,1)$ unitarity irreps. For the case of $d=1$, we explicitly construct the states and recover the results above in appendix~\ref{Sec:DiscreteSeries}.

\subsubsection*{Complementary series} 
The complementary series can be thought of as the analytic continuation of the principal series. Moreover, it is shown in~\cite{naimark1961decomposition,Dobrev:1976vr,Us:2022} that the complementary series would generally show up in tensor products as a discrete set of isolated points, not a continuous family:
\begin{align}
\mathcal{C}~\otimes~\mathcal{C}~&= ~ \mathcal{P}~\oplus~\mathcal{C}~\text{(isolated points)}~.
\end{align}  
This can also be seen in examples in~\cite[A.3]{Simmons-Duffin:2017nub} and section~\ref{sec:BulkCFTTwoPoint} for instance.  As we shall see, the free massive scalar field with $m^2 \geq 0$ has single-particle states that fall into principal or complementary series representations depending on the value of $m^2 R^2$. Since the Casimir eigenvalue is related to $m$ via $\D(d-\D)=m^2R^2$, ``light'' fields with mass $mR <d/2$ give rise to states in the complementary series  while ``heavy'' fields with mass $mR \geq d/2$ give rise to principal series states.

\subsection{Correlation functions and boundary operators}\label{sec:SphereCorrelation}
Correlation functions of local operators are one of the most basic observables in QFT.
In this paper, we are interested in the expectation values of local operators in the Bunch-Davies vacuum of de Sitter spacetime.
These can be conveniently defined by the analytic continuation of correlation functions of the same QFT on the Euclidean sphere $S^{d+1}$.
As discussed in~\ref{sec:dS coordinates}, in global coordinates this corresponds to writing $t= i \theta$, which transforms the metric   \eqref{eq:globalMetric} into the sphere metric
\beq
ds^2 = R^2 \left(d\theta^2 + \cos^2 \theta \,d\Omega_d^2 \right)= R^2 d\Omega_{d+1}^2~,
\eeq
where $\theta \in \left[-\frac{\pi}{2},\frac{\pi}{2} \right)$.
We can then write\footnote{For simplicity we restricted to scalar local operators.}
\begin{multline}
\langle \Omega | \phi_1(t_1,y_1) \dots \phi_n(t_n,y_n) | \Omega \rangle =\\
\lim_{0<\epsilon_n<\dots<\epsilon_1 \to 0} \langle \phi_1(\theta_1 = \epsilon_1-i t_1,y_1) \dots \phi_n(\theta_n = \epsilon_n-i t_n,y_n)\rangle_{S^{d+1}}~,
\label{FromStodS}
\end{multline}
recalling that $y_j \in S^d$.  We shall make heavy use of this approach in section~\ref{sec: Bulk two-point function}.

The space of local operators of a QFT is independent of the background geometry where it is placed. Moreover, for a UV-complete QFT defined as a relevant deformation of a UV CFT, the space of local operators is the one of the UV CFT.
In de Sitter, one can also define boundary operators by pushing bulk local operators to future (or past) infinity.
This is more conveniently stated in conformal coordinates as an expansion around $\eta=0$,
\begin{align}
\label{OPEdS}
\phi(\eta,x) &= \sum_\alpha b_{\phi \alpha} (-\eta)^{\Delta_\alpha} \left[ \mathcal{O}_\alpha(x) + c_1 \, \eta^2 \partial_x^2 \mathcal{O}_\alpha(x) 
+ c_2 \, \eta^4 (\partial_x^2)^2 \mathcal{O}_\alpha(x) +\dots \right]  \\
&= \sum_\alpha b_{\phi \alpha}\, (-\eta)^{\Delta_\alpha} {}_0F_1\!\left(\Delta_\alpha - \thd+1,\, \tfrac{1}{4} \eta^2 \pd_x^2\right)\! \Oo_\alpha(x)~. \nn
\end{align}
The operators $\Oo_\alpha$ are primary boundary operators, obeying $[K_\mu,\Oo_\alpha(0)] = 0$, whereas operators of the form $\Box^n \Oo_\alpha$ are $SO(d+1,1)$ descendants. In passing to the second line in~\reef{OPEdS} we used the fact that the coefficients $c_1$, $c_2$, \ldots are fixed by de Sitter isometries.\footnote{In practice, this can be done by using the expansion above to compute the two-point function 
\beq
\langle \Omega | \phi(\eta,x)\mathcal{O}_\alpha(y) | \Omega \rangle =  
b_{\phi \alpha} \frac{(-\eta)^{\Delta_\alpha}}{\left[ (x-y)^2-\eta^2 \right]^{\Delta_\alpha}}\,,
\eeq
which is fixed by symmetry. We normalize boundary operators to have unit two-point function. Also notice that the ${}_0F_1$ function in~\reef{OPEdS} can be recast as a Bessel function.}
 It is also easy to check that $\OO_\alpha$ satisfy the other commutation relation of primary operators of a conformal theory -- i.e. they define a conformal boundary theory. 
If the bulk operator $\phi$ is  hermitian, then the boundary operators $\mathcal{O}_\alpha$ can either be hermitian with real $\Delta_\alpha$ or appear in conjugate pairs $\mathcal{O}_\alpha$ and $\mathcal{O}_\alpha^\dagger$ with scaling dimensions  $\Delta_\alpha$ and  $\Delta_\alpha^*$. The dimensions $\Delta_\alpha$ of boundary operators should not be confused with the labels $\Delta$ of unitarity irreps in the Hilbert space. In particular, the values of $\Delta_\alpha$ are not restricted to be real or of the form $\frac{d}{2} + i \nu $ with $\nu\in \mathbb{R}$.
 
For QFT in Anti-de Sitter spacetime (or Boundary CFT), there is a similar expansion 
\be\label{eq:Bulk to Boundary EAdS}
\phi(z,x)=\sum_\alpha a_{\phi \alpha} z^{\D_\alpha} \left[\OO_\alpha(x)+ \text{descendants}\right]~.
\ee
The convergence of this type of Operator Product Expansion (OPE) can be established using a state-operator map~\cite{McAvity:1995zd, Lauria:2017wav, Paulos:2016fap}.  In dS, the convergence of the series~\reef{OPEdS} is more subtle. In particular, the OPE does not converge inside all matrix elements. For instance, using conformal symmetry we easily find that
\beq
\label{eq:parad0}
\braketmh{\Omega}{\phi(\eta,x)}{\tfrac{d}{2} + i \nu,y}  = c_\phi(i\nu) \left(\frac{-\eta}{|x-y|^2 - \eta^2} \right)^{\hd + i \nu},
\quad
c_\phi(i\nu) \in \mbb{C}.
\eeq
At the same time, 
\beq
\label{eq:parad}
\Delta_\alpha \neq \thd \pm i \nu
\quad
\Rightarrow
\quad
\braketmh{\Omega}{\Oo_\alpha(x)}{\tfrac{d}{2} + i \nu,y} = 0
\eeq
as also follows from a symmetry argument. If the OPE~\reef{OPEdS} converged, then~\reef{eq:parad} would imply that $c_\phi(i\nu)$ vanishes unless the late-time expansion~\reef{OPEdS} of $\phi$ contains an operator of dimension $\DD_\alpha = \hd \pm i \nu$. Yet we will see later that $c_\phi(i\nu)$ is in general a smooth, non-zero distribution for any non-trivial bulk operator $\phi$, even when its late-time expansion~\reef{OPEdS} does not contain any principal series operators. 

\subsection{Free scalar field in de Sitter}
\label{sec:scalar}
Let us start the discussion of QFT in dS by constructing an explicit example: the massive free scalar field. We will do so by canonically quantizing the theory in the flat slicing of Eq.~\reef{eq:flatMetric}. In the process, we will describe in detail the Hilbert space and its symmetry properties. 

In order to construct the free scalar in dS${}_{d+1}$, we start from the action 
\bsub
\label{eq:scalarAction}
\begin{align}
S&=- \int d^{d+1}\mathbf{x} \; \sqrt{-g} \left[\frac{1}{2}g^{\mu\nu}\partial_\mu\phi\partial_\nu\phi + \frac{1}{2} m^2\phi^2 \right]\\
&= R^{d-1} \int d^{d}x \int_{-\infty}^0 \frac{d\eta}{(-\eta)^{d+1}} \left[\eta^2 \left(\frac{1}{2}{\dot{\phi}}^2 -\frac{1}{2} \left(\nabla\phi\right)^2\right) - \half  R^2 m^2\phi^2\right]
\end{align}
\esub
where we define $\dot{\phi} \equiv \pd \phi/\pd \eta$.\footnote{Strictly speaking, in passing from the first to the second line in~\reef{eq:scalarAction}, we have discarded the early-time Poincar\'{e} patch covering ${X^0 + X^{d+1} < 0}$, but this will not influence the following discussion.} 
The Euler-Lagrange equation of motion for the field $\phi$ reads
\begin{equation}
  \label{eq:scalarEoM}
\eta^2 \ddot{\phi}(\vec{x},\eta) - \eta(d-1)\dot{\phi}(\vec{x},\eta) + \left( m^2 R^2 - \eta^2 \partial^2_{x} \right)\phi(\vec{x},\eta) = 0~.
\end{equation}
Introducing Fourier modes
\beq
\phi(x,\eta) = \frac{1}{R^{(d-1)/2}} \int\!\frac{d^dk}{(2\pi)^d} \, e^{i k \cdot x} \, \phi(k,\eta)
\eeq
the equation of motion reads 
\beq
\label{EoM Fourier}
\eta^2 \ddot{\phi}(\vec{k},\eta) - \eta(d-1)\dot{\phi}(\vec{k},\eta) + \left(\Delta(d-\Delta) + k^2 \eta^2 \right)\phi(\vec{k},\eta) = 0
\eeq
using the notation 
\be\label{eq:DeltadS}
\Delta(d-\Delta)= m^2R^2
\ee
for future convenience.

As we will see later, $\D$ can be interpreted as a scaling dimension once the limit $\eta \to 0$ is taken.  Depending on the value of $m^2 R^2$, the dimension $\Delta$ can either be real or complex. Let us discuss these cases separately. If $0 \leq m^2 R^2 < d^2/4$, then $\Delta$ takes values in the range $(0,d)$, which is the $\ell = 0$ complementary series. On the other hand, if $m^2 R^2 \geq d^2/4$ then $\Delta$ takes complex values: $\Delta = \hd + i \nu$ with $\nu \in \mbb{R}$. This is exactly the $\ell=0$ principal series. Remark that the label $\nu$ is only determined up to a sign. For a discussion of the $m^2 < 0$ case, we refer to~\cite{Epstein:2012zz}.

To proceed, we note that the solutions to the equation of motion can be written as Hankel functions. The exact mode decomposition reads
\beq
\label{eq:modeDecomposition}
\phi(\eta,k) = f_k(\eta) a_k^\dagger + \bar{f}_k(\eta) a_{-k} 
\eeq
where $a_k$ and $a_k^\dagger$ obey canonical commutation relations 
\beq
\label{eq:CCR}
[a_{\vec{k}},a^\dagger_{\vec{k}'}]=(2\pi)^d\delta^d(\vec{k}-\vec{k}')
\eeq
and $f_k$, $\bar{f}_k$ are solutions to \eqref{EoM Fourier}. As a second order differential equation, eq.~\reef{EoM Fourier} has two solutions: Hankel functions of the first and second kind. It turns out that by looking at early times $\eta\to-\infty$ and requiring the absence of the states with negative energy, one of the solutions is not allowed -- for instance, look at our detailed discussion of the wave function of a generic QFT in section~\ref{sec:KL} or~\cite[section 7.2]{Mukhanov:2007zz}.  One then finds that the solutions are
\bsub
\label{eq:explDecomp}
\beq
f_k(\eta)=(-\eta)^{d/2} \, h_{i\nu}(|k|\eta)~,\qquad \bar{f}_k(\eta)=(-\eta)^{d/2}\, \bar{h}_{i\nu} (|k|\eta)
\eeq
where
\beq
\label{eq:hhbardef}
h_{i\nu}(z) \ldef \frac{\sqrt{\pi}}{2} e^{\pi \nu/2} H^{(2)}_{i\nu}(-z)~, \qquad  \bar{h}_{i\nu}(z) \ldef \frac{\sqrt{\pi}}{2} e^{-\pi \nu/2} H^{(1)}_{i\nu}(-z)~.
\eeq
\esub
In particular, notice that $h_{i\nu}$ and $\bar{h}_{i\nu}$ are invariant under $\nu \mapsto -\nu$, which is expected since only the product $\Delta(d-\Delta) = d^2/4 + \nu^2$ is physical. Note that the normalization of the mode functions is chosen so that they obey
\bsub
\label{eq:fieldCCR}
\beq
f_k(\eta) \frac{d}{d\eta} \bar{f}_k(\eta) -  \bar{f}_k(\eta)\frac{d}{d\eta} {f}_k(\eta) = -i (-\eta)^{(d-1)}
\eeq
from which it follows that $\phi$ and its conjugate $\Pi$ satisfy canonical commutation relations:
\beq\label{eq:Canonical Pi and  phi}
[\phi(\eta,x),\Pi(\eta,x')] = i \delta^{(d)}(x-x'),
\quad
\Pi(\eta,x) = \frac{\delta S}{\delta \dot{\phi}} = (-R/\eta)^{d-1} \, \dot{\phi}(\eta,x)\,.
\eeq
\esub

At early times $\eta \to -\infty$, the field $\phi(\eta,x)$ behaves similarly to a massless scalar field in $(d+1)$-dimensional flat space:
\beq
\label{eq:fieldasymptotics}
\phi(\eta,x) \; \limu{\eta \to -\infty} \;  (-\eta/R)^{(d-1)/2}  \int\!\frac{d^dk}{(2\pi)^d \sqrt{2|k|}}  \left[e^{ik\cdot x + i \eta |k| + i \pi/4} \, a_k^\dagger + \text{h.c.}\right],
\eeq
The factor $-\eta/R$ is exactly the Weyl factor corresponding to the metric~\reef{eq:flatMetric}. This result can for instance be understood from the equation of motion~\reef{EoM Fourier}, since at early times both the damping term $\dot{\phi}$ and the mass term proportional to $\Delta(d-\Delta)$ become irrelevant. Finally, we define the Bunch-Davies vacuum $\ketmh{\Omega}$ to be the state annihilated by all $a_k$, so that correlators at $\eta \to -\infty$ are similar to ordinary Minkowski correlators.

\subsubsection{The Hilbert space of free theory}\label{sec:HilbertSpace}

Analogously to the quantization of a scalar field in flat space, the Hilbert state of the scalar theory in dS is a Fock space consisting of a zero-particle vacuum state $\ket{\Omega}$, single-particle states $a^\dagger_k \ket{\Omega}$ and multi-particle states $a_{k_1}^\dagger \dotsm a_{k_n}^\dagger \ket{\Omega}$. It will be instructive to study the properties of single-particle states, which we will denote by
\beq
\ket{\Delta,k} \ldef a_k^\dagger \ketmh{\Omega}~.
\eeq
These states inherit a normalization from~\reef{eq:CCR}, namely
\beq
\brakketmh{\Delta,k}{\Delta,k'} = (2\pi)^d \delta^{d}(k-k')~.
\eeq
We claim that the $\ket{\Delta,k}$ form an irreducible representation of the $SO(d+1,1)$ algebra. In order to obtain the transformation properties of the states in question, let us define wave functions  
\beq
\label{eq:wavefdef}
\Phi_k(\eta,x|\Delta) \ldef R^{(d-1)/2} \braketmh{\Omega}{\phi(\eta,x)}{\Delta,k}=    e^{-i\vec{k} \cdot \vec{x}} (-\eta)^{d/2} \, {h}_{i\nu}(|k| \eta)
\eeq
where the explicit expression on the RHS was obtained using~\reef{eq:modeDecomposition}, and we set $\D=\frac{d}{2}+i\nu$. We will use the expression~\reef{eq:wavefdef} to show how the states $\ket{\Delta,k}$ form a representation of $SO(d+1,1)$.

To start, notice that the vacuum state $\ket{\Omega}$ is annihilated by all generators. Moreover, since $\phi(\eta,x)$ is a local operator, it transforms under infinitesimal transformations as in~\reef{eq:WardOperator}. 
From the above facts, we deduce for example that
\bsub
\begin{align}
  -ik_\mu \Phi_{k}(\eta,x|\Delta)  &= \pd_\mu \Phi_{k}(\eta,x|\Delta)\\
  &= \braketmh{\Omega}{[P_\mu,\phi(\eta,x)]}{\Delta,k}\\
  &= 0- \braketmh{\Omega}{\phi(\eta,x) P_\mu}{\Delta,k}
\end{align}
\esub
hence it follows that
\bsub
\label{eq:genOnStates}
\beq
P_\mu \ketmh{\Delta,k} =  i k_\mu \ketmh{\Delta,k}~.
\eeq
For the other generators, we find that similarly 
\begin{align}
 D \ketmh{\Delta,k} &= -\left[k\cdot \pd + \thd \right] \ketmh{\Delta,k},\\
  K_\mu \ketmh{\Delta,k} &= i \left[k_\mu \pd^2 -2 (k\cdot \pd) \pd_\mu   -  d\, \pd_\mu +(\Delta- \thd)^2 \frac{k_\mu}{|k|^2} \right]\!\ketmh{\Delta,k}~,\\
M_{\mu \nu} \ketmh{\Delta,k} &= \left[k_\nu \pd_\mu - k_\mu \pd_\nu \right] \ketmh{\Delta,k}
\end{align}
\esub
where all derivatives act in $k$-space, that is to say $\pd_\mu = \pd/\pd k^\mu$. The derivation of the identities~\reef{eq:genOnStates} is tedious but straightforward. We perform these calculations in full detail in appendix~\ref{sec:Action of Generators}.

It is easy to check that the commutators of~\reef{eq:genOnStates} are consistent with the conformal algebra~\reef{eq:CFTcommutation}. Moreover, the Casimir~\reef{eq:Casimir} evaluates to
\beq\label{eq:CasimirEigenvalueNew}
C \ketmh{\Delta,k} = \Delta(\Delta-d) \ketmh{\Delta,k}~.
\eeq
The action~\reef{eq:genOnStates} is exactly the $\ell = 0$ representation of $SO(d+1,1)$ from section~\ref{sec:reps}. Multi-particle states can also be organized in representations of $SO(d+1,1)$. If $m^2$ is sufficiently large, then the single-particle state $\ketmh{\Delta,k}$ is in the principal series, because $\Delta = \hd + i \nu$ for some $\nu \in \mbb{R}$. In $d\ge 2$ dimensions, two-particle states are then a superposition of other principal series states $[\hd + i \nu',\ell]$ with $\nu' \in \mbb{R}$ and $\ell = 0,1,2,\ldots$~\cite{Dobrev}. For $d=1$, we expect that the Hilbert space of the theory also contains states in the discrete series, having integer $\Delta$. This observation will be important in section~\ref{sec:Bootstrpping boundary correlators} when we set up the bootstrap for QFT in dS$_2$.

\subsection{Non-perturbative QFT in de Sitter}
\label{sec:nonpert}

\subsubsection{Hilbert space}

In a general QFT, we expect that the Hilbert space falls into irreducible representations of the isometry group of its spacetime, plus any additional global symmetries of the theory in question. For a QFT on dS${}_{d+1}$, we therefore expect that all states form representations of $SO(d+1,1)$, like the single-particle states $\ket{\Delta,k}$ from section~\ref{sec:scalar}. In this section, will argue that after taking spin into account, the representation~\reef{eq:genOnStates} is essentially unique up to a choice of $\Delta$.
To prove this, let us write a generic state as $\ketmh{\Delta,k}_A$, where $A$ is an abstract $SO(d)$ index. Since the anti-hermitian momentum generators $P_\mu$ commute, we can diagonalize them
\beq
\label{eq:Pa}
P_\mu \ketmh{\Delta,k}_A = i k_\mu \ketmh{\Delta,k}_A
\eeq
as in~\reef{eq:genOnStates}. Next, let us briefly introduce some notation to describe spinning states $\ketmh{\Delta,k}_A$ where $A$ is an abstract $SO(d)$ index. Rotations act on such a state as
\beq
\label{eq:spinningAction}
M_{\mu \nu} \ketmh{\Delta,k}_A = (k_\nu \pd_\mu - k_\mu \pd_\nu + \Sigma_{\mu \nu})\ketmh{\Delta,k}_A
\eeq
where $\Sigma_{\mu \nu} = - \Sigma_{\nu\mu}$ acts on the $A$ indices and obeys the same commutation relations as $M_{\mu \nu}$. In the present paper, we will only deal with states that transform as traceless symmetric tensors of spin $\ell$. It will be convenient to use an index-free notation as follows: 
\beq
\label{eq:indexFree}
\ketmh{\DD,k,z} \ldef \ketmh{\DD,k}_{\mu_1 \dotsm \mu_\ell}\, z^{\mu_1} \dotsm z^{\mu_\ell}
\eeq
where the indices $\mu_1,\ldots,\mu_\ell$ run over $1,\ldots,d$. The tensor properties of the above state imply that
\beq
z^\mu  \frac{\pd}{\pd z^\mu} \ketmh{\DD,k,z} = \ell \, \ketmh{\DD,k,z}
\qaq
\frac{\pd}{\pd z^\mu} \frac{\pd}{\pd z_\mu}  \ketmh{\DD,k,z} = 0~.
\eeq
The spin operator $\Sigma_{\mu \nu}$ now acts as
\beq
\Sigma_{\mu \nu}\ketmh{\DD,k,z} = \left( z_\nu \frac{\pd}{\pd z^\mu} -  z_\mu \frac{\pd}{\pd z^\nu}\right)\ketmh{\DD,k,z}
\ee
such that
\be
-\th \Sigma_{\mu \nu}\Sigma^{\mu \nu}\ketmh{\DD,k,z} = \ell(\ell+d-2)\,\ketmh{\DD,k,z}
\eeq
which recovers the usual $SO(d)$ Casimir eigenvalue of a spin-$\ell$ representation. From equations~\reef{eq:Pa} and~\reef{eq:spinningAction}, the action of the other generators is fixed up to a single parameter. For instance, the generator $D$ should act as a scalar that assigns appropriate weights to $k^\mu$ and $\pd/\pd k^\mu$ because $[D,P_\mu] = P_\mu$. Hence $D$ should be of the form
\beq
\label{eq:Da} D \ketmh{\DD,k}_A = -\left(  k\cdot \pd + \beta \right) \ketmh{\DD,k}_A
\eeq
with some constant  $\beta$ to be determined.  Likewise, we can write down a completely general ansatz for $K_\mu$ which transforms as a vector and is built out of $k^\mu$, $\pd/\pd k^\mu$ and $\Sigma_{\mu \nu}$.
By imposing that $[D,K_\mu]$ and $[K_\mu,P_\nu]$ close as in~\reef{eq:CFTcommutation}, and that $[K_\mu,K_\nu] = 0$, we find that $K_\mu$ is fixed to
\begin{multline}
\label{eq:Ka}
K_\mu \ketmh{\DD,k}_A =\\ i \big[  k_\mu \pd^2  - 2 (k\cdot \pd) \pd_\mu -   d\, \pd_\mu + (\DD-\thd)^2 \frac{k_\mu}{|k|^2} - 2  \Sigma_{\mu \nu} \left(\pd^\nu \pm (\DD-\thd) \frac{k^\nu}{|k|^2} \right) \big]\! \ketmh{\DD,k}_A
\end{multline}
where $\Delta$  is now an arbitrary parameter. The requirement that $[K_\mu,P_\nu]$ reproduces the commutation relation~\reef{eq:CFTcommutation} fixes $\beta = d/2$ in~\reef{eq:Da}. The equations~\reef{eq:spinningAction}, \reef{eq:Pa}, \reef{eq:Da} and~\reef{eq:Ka} thus form the most general consistent representation of $SO(d+1,1)$ that diagonalize $P_\mu$. In addition, it is easy to see that the state $\ketmh{\Delta,k}_A$ will have conformal Casimir eigenvalue $\Delta(\Delta-d) - \th \Sigma_{\mu \nu}^2$, which for a spin-$\ell$ representation becomes  $\Delta(\Delta-d) + \ell(\ell+d-2)$.

Notice that in~\reef{eq:Ka} the action of $K_\mu$ is only determined up to a choice of sign, at least for spinning states where $\Sigma_{\mu \nu} \neq 0$: both sign choices respect the conformal algebra and lead to the same Casimir eigenvalue. Changing the sign is equivalent to redefining $\Delta \mapsto d - \Delta$. In what follows, we will choose the $+$ sign for definiteness.

Finally, the ground state $\ketmh{\Omega}$ of any QFT in dS must be annihilated by all of the symmetry generators, and as such it transforms as a trivial representation of dimension $\Delta = 0$, $\ell = 0$ and $k_\mu = 0$.

\subsubsection{Representations in position space}

Although the above representations look complicated, we can show that they take a more familiar form after introducing a specific Fourier-like transformation. To wit, define a new family of states as\footnote{These states can be found in the literature with different but equivalent definitions. see e.g.~\cite{Dobrev:1976vr,Sun:2021rrs,Sun:2021thf} }
\begin{equation}
\label{eq:primary state}
\ketmh{\Delta,x}_A \ldef \int\!\frac{d^dk}{(2\pi)^d} \; e^{i k\cdot x} \, |k|^{\Delta-d/2} \, \ketmh{\Delta,k}_A
\end{equation}
where a factor of $|k|^{\Delta-d/2}$ has been introduced for future convenience. 
We will argue that the state $\ketmh{\Delta,x}_A$ transforms just like a primary operator of dimension $\Delta$ in flat-space CFT. As a first hint, one readily computes that for a scalar state
\beq
\label{eq:tempint}
\brakketmh{\Delta,x}{\Delta,x'} = \int\!\frac{d^dk}{(2\pi)^d} \, e^{ik \cdot (x-x')} \, |k|^{\Delta + \bar{\Delta} - d}
\eeq 
provided that the $k$-space state $\ketmh{\Delta,k}$ is normalized such that 
\be
\brakketmh{\Delta,k}{\Delta,k'} = (2\pi)^d \delta^{d}(k-k')~. 
\ee
There are now two possibilities: if $\Delta$ is real (i.e.\@ when $\Delta$ is in the complementary series), then $\bar{\Delta} = \Delta$. On the other hand, if $\Delta$ is in the principal series then  $\bar{\Delta} = d- \Delta$. We conclude that
\beq
\label{eq:gramforx}
\brakketmh{\Delta,x}{\Delta,x'} = 
\begin{cases} \delta^{d}(x-x') & \Delta \in d/2 + i \mbb{R} \\ c_\Delta/|x-x'|^{2\Delta} & \Delta \in \mbb{R}
\end{cases}
\eeq
for some computable coefficient $c_\Delta$.\footnote{The integral~\reef{eq:tempint} diverges for $\Delta \in \mbb{R}$, so~\reef{eq:gramforx} is only true in the sense of distributions.}
For real $\Delta$ this is the form of a two-point function in flat-space CFT, but when $\Delta$ is in the principal series the states $\ketmh{\Delta,x}_A$ have a delta function normalization.

Let us make the above statement precise by computing the action of the $SO(d+1,1)$ generators. On a state of the form~\reef{eq:primary state}, $P_\mu$ acts as
\bsub
\beq\label{eq: PKD position}
P_\mu \ketmh{\Delta,x}_A =  \int\!d^dk \; e^{i k\cdot x} \, |k|^{\Delta-d/2} (ik_\mu) \ketmh{\Delta,k}_A = \frac{\pd}{\pd x^\mu} \, \ketmh{\Delta,x}_A
\eeq
and likewise
\begin{align}\label{eq: PKD positionP}
D \ketmh{\Delta,x}_A &= (x \cdot \pd + \Delta) \, \ketmh{\Delta,x}_A\\
M_{\mu \nu} \ketmh{\Delta,x}_A &= (x_\nu \pd_\mu - x_\mu \pd_\nu + \Sigma_{\mu \nu}) \ketmh{\Delta,x}_A\\
K_\mu \ketmh{\Delta,x}_A &= \left(2x_\mu (x \cdot \pd) - x^2 \pd_\mu + 2\Delta x_\mu - 2 \Sigma_{\mu \nu}\, x^\nu \right)\! \ketmh{\Delta,x}_A
\end{align}
\esub
where all derivatives act on $x$. We spell out the derivation for the case of the scalar states in appendix~\ref{sec:Action of Generators}. These formulas are exactly identical to those obtained by applying a fictitious CFT operator $\Oo_A^{(\Delta)}(x)$ of dimension $\Delta$ to the Bunch-Davies vacuum. From a practical point of view, this implies that any $n$-point amplitude
\bsub
\beq
\braketmh{\Omega}{\phi(\eta_1,x_1) \dotsm \phi(\eta_n,x_n)}{\Delta,x}_A
\eeq
has the exact same $SO(d+1,1)$ transformation properties as an $(n+1)$-point vacuum expectation value with an insertion of an operator $\Oo_A^{(\Delta)}(x)$:
\beq
\braketmh{\Omega}{\phi(\eta_1,x_1) \dotsm \phi(\eta_n,x_n) \Oo^{(\Delta)}_A(x)}{\Omega}~.
\eeq
\esub

For future reference, we remark that from~\reef{eq:gramforx} it follows that the resolution of the identity operator inside an irrep can then be written as
\beq\label{eq:identity x-space}
\Delta \in \hd + i \mbb{R}:
\quad
\int\!d^dx \; |\Delta,x \rangle_A  \,\,^A\langle \Delta,x |
\eeq
that after summing over all the irreps leads to:
\be
\label{completeXPre}
\boxed{\mathds{1} =  |\Omega \rangle \langle \Omega| +  \sum_{A} \int_{\hd-i\infty}^{\hd+i\infty} \frac{d\D}{2\pi i N_{x}(\D,\ell)} \int\!d^dx \; |\Delta,x \rangle_A  \,\,^A\langle \Delta,x |~ + ~\text{other irreps}}~
\ee
where the sum over $A$ formally stands for the sum over $SO(d)$ representations, which for our case of traceless symmetric representation will be simply a sum over spin $\ell$.  Here we did not explicitly spell out contributions from the irreps other than principal series as it will not be apparent in our discussion for scalar operators in the next sections.  $N_{x}(\D,\ell)$ is a positive normalization and we do not need its explicit expression. The subscript $x$ stands for the normalization in position space. There is a similar formula in momentum space with a positive normalization ${N_k(\Delta,\ell)}$ that we mention here for completeness and  future convenience:
\be
\label{completeKPre}
\mathds{1} = |\Omega \rangle \langle \Omega| + \sum_A \int\!\frac{d\Delta}{2\pi i} \frac{1}{N_k(\Delta,\ell)} \int\!\frac{d^dk}{(2\pi)^d} \; \ketmh{\Delta,k}_{A} \ ^{A} \bramh{\Delta,k} ~ + ~\text{other irreps}~.
\ee

\subsubsection{Conformal Field Theory in de Sitter}
\label{sec:CFTdS}

It is instructive to consider the case of a CFT on a de Sitter background. 
Given that the de Sitter metric \eqref{eq:flatMetric} is conformally flat, we can immediately write
\beq
\phi(\eta,x) = (-\eta/R)^{\Delta_\phi} \phi_{\rm flat} (\eta,x)\,,
\eeq
where we assumed that $\phi$ is a primary scalar operator of the bulk CFT and we denoted by $\phi_{\rm flat} (\eta,x)$ the same operator in flat Minkowski space with metric $ds^2=-d\eta^2+dx^2$.
The OPE \eqref{OPEdS} then follows from expanding $\phi_{\rm flat} (\eta,x)$ around the constant timeslice $\eta=0$. Clearly, in this case, the primary boundary operators $\mathcal{O}_\alpha$ are nothing but time derivatives of $\phi_{\rm flat}$. Thus a conformal primary of dimension $\Delta_\phi$ gives rise to a family of boundary operators with dimensions $\Delta_\alpha = \Delta_\phi + p$ with $p=0,1,2,\dots$.

This construction is useful because it gives us an infinite set of data to test any bootstrap approach to QFT in de Sitter.  In particular, any CFT correlation function with all operators inserted on a constant timeslice in Minkowski (or Euclidean) space can be interpreted as a correlation function of operators on the future boundary of de Sitter spacetime.

As mentioned above, the metric of de Sitter space \eqref{eq:conformalMetric} is a Weyl transformation of a part of the Minkowski cylinder. It is instructive to understand how a unitary conformal highest-weight representation on the Minkowski cylinder decomposes into irreps of the dS isometry group. For this purpose it is useful to think of the CFT living on the lightcone 
\beq
- (X^{-1} )^2 -  (X^{0} )^2+ (X^{1} )^2+\dots  (X^{d+1} )^2=0
\eeq
of the embedding space $\mathbb{R}^{d,2}$. Then dS${}_{d+1}$ is the section defined by $X^{-1}=R$ --- compare with \eqref{eq:embeddingDef} --- and the Minkowski cylinder is the universal cover of the section defined by $ (X^{-1} )^2 +  (X^{0} )^2 =R^2$.
The de Sitter isometry group $SO(d+1,1)$ can immediately be identified as the subgroup  of $SO(d+1,2)$ that leaves the coordinate $X^{-1}$ invariant. 

In appendix \ref{Sec:DiscreteSeries}, we focus on the $d=1$ case and build unitary irreps of $SO(2,1)$ inside the usual conformal family of $SO(2,2)$ labeled by the primary state $|\tilde{\Delta},\ell \rangle$ of dimension $\tilde{\Delta}$ and spin $\ell$.
We show that there are principal series irreps with $\Delta = \frac{1}{2}+i\nu$ for all $\nu \in \mathbb{R}$ and $|\ell|$ discrete series irreps as long as $|\ell| \ge 1$. We also found a complementary series irrep  if $\tilde{\Delta} < \frac{1}{2}$.
We leave for the future the instructive exercise of extending this analysis to  general spacetime dimension. 
.

\section{Bulk two-point function}\label{sec: Bulk two-point function}

In this section, we compute the two-point function~\reef{eq:2pt000} using the Hilbert space framework from section~\ref{sec:nonpert} and in particular the resolution of identity in~\reef{completeKPre}. This will lead to an expression for the two-point function in terms of a spectral integral with definite positivity properties, also known as the \toolazy decomposition.  After that, we relate the correlator~\reef{eq:2pt000} to its counterpart on sphere $S^{d+1}$ and its decomposition in terms of spherical harmonics.  After employing a Watson-Sommerfeld transformation, this leads to an explicit formula, the so-called \textit{inversion formula},  expressing the \toolazy spectral density in terms of an integral over the discontinuity of the two-point function. Finally, we analyze the \toolazy decomposition in several examples. 

Consider the two-point function of two scalar operators. Due to the translation and rotation invariance of dS, this correlation function should be of the form  
\beq
\label{eq:2pt000}
\expec{\Omega| \phi_1(\eta,x)\phi_2(\eta',x')|\Omega }  = G_{12}(\xi)
\eeq
where we define
\beq
\label{eq:xidef}
\xi = \frac{4R^2}{(X-X')^2 } = \frac{2}{1-\sigma} = \frac{4\eta \eta^\prime}{-(\eta-\eta^\prime)^2+|\vec{x}-\vec{x}^\prime|^2 }
\eeq
as another representation of the only $SO(d+1,1)$ invariant that can be built out of two bulk points: $\sigma=X.X'$. We used the conformal coordinates defined in~\reef{Xetax} for definiteness. Using $\xi$ instead of $\sigma$ is for later convenience. The invariant $\xi$ is positive ($\xi > 0$) when $X,X'$ are spacelike separated, negative ($\xi < 0$) when they are timelike separated and $\xi$ diverges when $X,X'$ are lightlike separated. As such an $i\epsilon$ prescription is required to define~\reef{eq:2pt000} properly when $\xi < 0$.

\subsection*{Free scalar two-point function}
To start, we can consider the two-point function of a free scalar field. The Wightman propagator is defined as the solution of the Klein-Gordon equation~\cite{Arkani-Hamed:2015bza,Spradlin:2001pw} 
\begin{equation}
\label{eq:laplace}
(\nabla^2-m^2)\brakkket{\phi(\eta,x)\phi(\eta',x^\prime)}_\text{f}=0 
\end{equation}
where $\nabla^2$ is the Laplace-Beltrami operator on dS$_{d+1}$.\footnote{The time-ordered propagator obeys instead~\reef{eq:laplace} with a  delta function source term.} The appropriately normalized solution to~\reef{eq:laplace} reads
\beq\label{eq:free two point function}
\brakkket{\phi(\eta,x)\phi(\eta',x')}_\text{f}= \frac{1}{R^{d-1}} \, G_{\text{f}}(\xi;\nu)~,
\ee
where
\be\label{eq:ThedSTwoPointFunction}
G_{\text{f}}(\xi;\nu) \ldef \frac{\Gamma(\frac{d}{2}+i\nu)\Gamma(\frac{d}{2}-i\nu)}{(4\pi)^{\frac{d+1}{2}}\Gamma(\frac{d+1}{2})} \;_2F_1\!\left(\frac{d}{2}+i\nu,\frac{d}{2}-i\nu;\, \frac{d+1}{2};\, 1-\frac{1}{\xi}\right)
\eeq
writing $m^2 R^2 = (d/2)^2 + \nu^2$ as before. The subscript "$\text{f}$" stands for free theory. This solution is regular as $\xi \to 1$, which corresponds to spacelike separated points in dS. 
The normalization is fixed by matching the singular behavior as $\xi\to\infty$ with the flat space propagator.
One may also derive this by Fourier transforming the two-point function to momentum space using \eref{eq:modeDecomposition}. For future reference, we note the Fourier decomposition in question:
\beq
\label{eq:WightFour}
G_\mrm{f}(\xi;\nu) = \frac{(\eta \eta')^{d/2}}{R^{d-1}} \int\!\frac{d^dk}{(2\pi)^d} \, e^{-i k \cdot (x-x')} \, \overline{h}_{i\nu}(|k|\eta){h}_{i\nu}(|k|\eta')~.
\eeq
From now on we will set $R=1$ unless otherwise noted. At this point, we notice that the Wightman correlator  gives rise to a specific $i\epsilon$ prescription: properly speaking
\beq
\label{eq:ieps}
\braket{\phi(\eta,x)\phi(\eta',x^\prime)}_\text{f} = G_{\text{f}}(\tilde{\xi};\nu)\,,
\quad
\tilde{\xi} = \frac{4\eta\eta'}{ - (\eta-\eta' - i \epsilon)^2 + |x-x'|^2}~.
\eeq
See for instance~\cite{Marolf:2010nz,Sleight:2019hfp}. In what follows we will not distinguish $\xi$ and $\tilde{\xi}$ unless mentioned otherwise.

\subsection{Källén–Lehmann decomposition}\label{sec:KL}
Let us now turn to the analysis of a generic two-point function of identical operators, $\braketmh{\Omega}{\phi(\eta,x)\phi(\eta',x')}{\Omega}$. We will assume that $\phi(\eta,x)$ is a Hermitian operator, although much of the argument holds as well for a generic two-point function $\expec{\phi_i(\eta,x) \phi_j(\eta',x')}$ of different scalar operators. We can analyze the $\braketmh{\Omega}{\phi(\eta,x)\phi(\eta',x')}{\Omega}$ correlator by inserting a resolution of the identity~\reef{completeKPre}:
\beq
\label{completeK}
\mathds{1} = |\Omega \rangle \langle \Omega| + \sum_\ell \int\!\frac{d\Delta}{2\pi i} \frac{1}{N(\Delta,\ell)} \int\!\frac{d^dk}{(2\pi)^d} \; \ketmh{\Delta,k}_{\mu_1 \dots \mu_\ell} \ ^{\mu_1 \dots \mu_\ell} \bramh{\Delta,k} + \ldots
\eeq
writing $\ldots$ for states with $SO(d)$ representations other than traceless symmetric tensors. In the above formula, we allow for an arbitrary normalization factor $N(\Delta,\ell) > 0$, depending on the normalization of the states $\ketmh{\Delta,k}_{\mu_1 \dots \mu_\ell}$ (which cannot depend on $k_\mu$). We also dropped the subscript $k$ compare to~\reef{completeKPre} to avoid clutter. Of course, there might be several irreps with the same quantum numbers $\{\Delta,\ell\}$, in which case an additional label $\alpha$ is needed to distinguish such states. We will not explicitly write such a label, but it is straightforward to adapt our analysis to this degenerate situation.
In~\reef{completeK}, we assume that only states in the principal series contribute, so the $\Delta$-integral runs from $d/2-i\infty$ to $d/2+i\infty$. This assumption seems to be correct in general; in specific examples we will briefly revisit this assumption. For a more complete discussion, we refer to section~\ref{sec:reps}. Moreover, when we find an inversion formula, based on completeness of Gegenbauer functions in the sphere, we basically prove the completeness of principal series as far as the inversion formula is convergent. 

After inserting the resolution of the identity~\reef{completeK} in the two-point function, one finds 
\begin{multline}
\label{2pt decom k}
\braketmh{\Omega}{\phi(\eta,x)\phi(\eta',x')}{\Omega} = \braketmh{\Omega}{\phi(\eta,x)}{\Omega}\braketmh{\Omega}{\phi(\eta',x')}{\Omega} \\
+\sum_\ell \int\!\frac{d\Delta}{2\pi i}\frac{1}{N(\Delta,\ell)} \int\!\frac{d^dk}{(2\pi)^d}  \; \langle \Omega | \phi(\eta,x) |\Delta,k  \rangle_{\mu_1 \dots \mu_\ell} \ ^{\mu_1 \dots \mu_\ell}  \langle \Delta,k | \phi(\eta',x') | \Omega  \rangle~.
\end{multline}
First of all, remark that the one-point functions $\braketmh{\Omega}{\phi(\eta',x')}{\Omega}$ do not depend on the coordinates $\eta$ and $x^\mu$ because $|\Omega\rangle$ is $SO(d+1,1)$ invariant.
Hence we can replace the second term with the constant $\expec{\phi}^2 \ldef \braketmh{\Omega}{\phi}{\Omega}^2$. From now on we consider theories with zero vacuum expectation value:
\be \expec{\phi}^2=0~.\ee

Moving to the second term of~\reef{2pt decom k}, one can show that matrix elements of the form $\braketmh{\Omega}{\phi(\eta,x)}{\Delta,k}_{\mu_1\cdots\mu_\ell}$ with $\ell \geq 1$ vanish. This can be proven either by using an explicit computation or by working in embedding space. Therefore, only states with $\ell = 0$ contribute, and the contribution of such a state is fixed by $SO(d+1,1)$ symmetry up to two constants. Using an $SO(d+1,1)$ symmetry argument, one can show that the most general form of the amplitude with the $\ell = 0$ state is given by
\beq
\label{eq:genMatElt}
\braketmh{\Omega}{\phi(\eta,x)}{\Delta,k} = e^{-i k \cdot x} \, (-\eta)^{d/2}  \left[ c_\phi(i\nu) \, \bar{h}_{i\nu}(\eta|k|) + c_\phi^\sharp(i\nu)\, h_{i\nu}(\eta|k|)  \right]~,
\eeq
for two undetermined coefficients $c_\phi(i\nu),\, c_\phi^\sharp(i\nu) \in \mbb{C}$ and $\Delta = \hd + i \nu$. We will now argue that $c_\phi^\sharp(i\nu)$ has to vanish in any unitary QFT. For this argument, consider the early-time limit $\eta \to -\infty$, where dS can be compared to flat space. Using the asymptotics of the Hankel functions, the matrix element behaves in this limit as
 \beq
 \label{eq:goToEarlyTime}
 \braketmh{\Omega}{\phi(\eta,x)}{\Delta,k} \; \limu{\eta \to -\infty} \; \frac{e^{-i k \cdot x} (-\eta)^{(d-1)/2} }{\sqrt{2\omega(k)}} \left[  c_\phi(i\nu)e^{ -i\eta \omega(k) - i\pi/4} + c^\sharp_\phi(i\nu)e^{ i\eta \omega(k)+ i\pi/4} \right]
 \eeq
with $\omega(k) \ldef |k|$ and two important phases $\pm i\eta \omega(k)$. The formula~\eqref{eq:goToEarlyTime} is reminiscent of flat-space QFT, where operators evolve in time as
 \beq
 \phi(t,x) = e^{iHt} \phi(0,x) e^{-i H t}~.
 \eeq
 Moreover, according to the Wightman axioms, the state $\phi(0,x) \ketmh{\Omega}$ can only have support inside the positive future lightcone. Consequently, if $\ketmh{E,k}$ is a state that diagonalizes $H$ and $P_\mu$, we must have
 \beq
 \hspace{-25mm}\text{flat space}:
 \quad
 \braketmh{\Omega}{\phi(t,x)}{E,k} \; \propto \; \Theta(E)e^{-i k \cdot x} e^{ -i E t}
 \eeq
 up to some constant that depends on the local operator $\phi$. We thus interpret the second term in~\reef{eq:goToEarlyTime} as originating from a state of negative energy, which would violate the Wightman axioms. Consequently, we have to require that $c^\sharp_\phi(i\nu) = 0$ for all $\nu$. 

We are now ready to compute the $k$-integral in~\reef{2pt decom k}. Since $\phi$ is a hermitian operator, it follows that
\beq\label{eq:wavefunction nonpert}
\braketmh{\Delta,k}{\phi(\eta',x')}{\Omega} = \braketmh{\Omega}{\phi(\eta',x')}{\Delta,k}^* = c_\phi(i\nu)^* \, e^{i k \cdot x'} (-\eta')^{d/2} \, h_{i\nu}(\eta'|k|)
\eeq
using the properties of the Hankel functions under complex conjugation. By performing the $k$-integral in~\reef{2pt decom k} that boils down to the Fourier transformation of the product of two Hankel functions in~\reef{eq:WightFour}, we conclude that
\beq
\label{eq:newKL}
\boxed{
\braketmh{\Omega}{\phi(\eta,x)\phi(\eta',x')}{\Omega} = \int_\mbb{R}\!\frac{d\nu}{2\pi} \; \rho_\phi(\thd + i\nu)  \, G_\mrm{f}({\xi};\nu)
}
\quad
\eeq
with
\be
\quad
\rho_\phi(\thd + i \nu) \ldef \frac{|c_\phi(i\nu)|^2 }{N(\hd + i\nu,0)} \geq 0~.
\eeq
This is the desired \toolazy decomposition which applies to any two-point function of bulk scalar operators. 
It is clear that similar \toolazy decompositions exist for all possible time-orderings.

In passing, let us comment on the apparent absence of states in the complementary series of $SO(d+1,1)$, having $0 \leq \Delta \leq d$, or even discrete series states. We did not explicitly include such states in the resolution of the identity~\reef{2pt decom k}. One can nevertheless accommodate for complementary series states in~\reef{eq:newKL}, by modifying the contour and integrating over small imaginary values of $\nu$. This is equivalent to looking for a pole crossing after analytical continuation off of the principal series.

Finally, we want to mention that~\reef{eq:newKL} is not a novel result. Versions of the \toolazy decomposition have already appeared in the literature, using different derivations and levels of mathematical rigor. An early reference to the \toolazy decomposition in dS appeared in~\cite{Bros:1990cu}, and later works using such a representation can be found in~\cite{Bros:1995js,Bros:1998ik,Bros:2010rku,Bros:2009bz,Marolf:2010zp,Marolf:2010nz,Hollands:2010pr,Hollands:2011we,Marolf:2012kh,Epstein:2012zz}.

\subsection{Late-time limit and boundary OPE}
\label{2ptLateTime}

Starting from the \toolazy representation~\reef{eq:newKL}, let us consider the late-time behavior of the correlator $\expec{\phi(\eta,x)\phi(\eta',x')}$ in the limit $\eta,\, \eta' \to 0^{-}$ at fixed $x,x'$. At the level of the invariant $\xi$ from~\reef{eq:xidef}, this corresponds to the limit $\xi \to 0^{+}$. Also notice that for sufficiently small $\eta$ and $\eta'$ the two insertions are spacelike separated, so there are no subtleties regarding $i\epsilon$ prescriptions.

Let us rewrite $G_\text{f}(\xi;\nu)$ in~\reef{eq:ThedSTwoPointFunction} in terms of a function $\psi_{\nu}(\xi)$ and its shadow ($\nu \mapsto -\nu$ or $\Delta \mapsto d-\Delta$):
\bsub
\label{eq:splitrep}
\beq
\label{eq:split1}
G_\text{f}(\xi;\nu) = \frac{\msf{g}(\thd + i\nu) \psi_{\hd + i\nu}(\xi) + (\nu \mapsto -\nu)}{2}
\eeq
with
\beq
\msf{g}(\Delta) = \frac{\Gamma(\hd - \Delta)\Gamma(\Delta)}{2^{2\Delta+1}\pi^{d/2+1}}
\qaq
\psi_{\Delta}(\xi) = \xi^{\Delta}\, {}_2F_1\!\left({{\Delta,\, \Delta-\th(d-1)}~\atop~{2\Delta-d+1}}\,\Bigg|\,\xi\right)~.
\eeq
\esub
The representation~\reef{eq:splitrep} is convenient to study the $\xi \to 0$ limit of the correlator because when $\xi$ is small the hypergeometric function simplifies and we can replace it with the leading term $\psi_{\Delta}(\xi) \approx \xi^{\Delta}$.

We would now like to perform the \toolazy integral~\reef{eq:newKL} by deforming the contour. As it stands, we can interpret the contour in~\reef{eq:newKL} as running upwards in the complex $\Delta$ plane, along the vertical line $\Re(\Delta) = d/2$. We would like to close the contour to the right by adding an arc at infinity and picking up any possible poles. As a first step, we therefore write
\beq
\label{eq:kl22}
\braketmh{\Omega}{\phi(\eta,x)\phi(\eta',x')}{\Omega}= \int_{\hd - i\infty}^{\hd + i \infty}\frac{d\Delta}{2\pi i } \; \rho_\phi(\Delta) \msf{g}(\Delta) \psi_\Delta(\xi)~,
\eeq
exploiting the shadow symmetry of the representation~\reef{eq:splitrep} to drop one of the terms. We claim that for sufficiently small $\xi$, the integration contour in~\reef{eq:kl22} can be deformed by closing the contour to the right. To prove this, we first notice that for sufficiently small $\xi > 0$, the function $\psi_\Delta(\xi)$ falls off for large real $\Delta$:
\beq
0 < \xi \ll 1:
\quad
\psi_\Delta(\xi) \limu{\Delta \to \infty} \msf{w}(\xi)^\Delta,
\quad
\msf{w}(\xi) = \frac{4\xi}{(1+\sqrt{1-\xi})^2}
\eeq
and for the limit in question $0 \leq \msf{w}(\xi) \approx \xi \ll 1$, so the special function $\psi_\Delta(\xi)$ indeed decays exponentially fast on the right half-plane. This statement does \emph{not} hold for the shadow function $\psi_{d-\Delta}(\xi)$. Next, let us investigate the function $\msf{g}(\Delta)$. On the real line, we have
\beq
\msf{g}(\Delta) \limu{\Delta \to \infty} \frac{1}{2 \pi^{d/2}}\cdot \frac{\Delta^{d/2-1}}{4^\Delta \sin\!\left(\pi(\hd-\Delta)\right)}
\eeq
up to a $\Delta$-independent coefficient. It follows that $\msf{g}(\Delta)$ has single poles at $\Delta = \hd + \mbb{N}$. Away from the real axis, the function $\msf{g}(\Delta)$ decays rapidly.

Finally, we need to make some assumptions about the behavior of the distribution $\rho_\phi(\Delta)$: (i) Originally $\rho_\phi(\Delta)$ is only defined on the axis $\Re(\Delta) = d/2$, but we assume that it can be analytically continued away from this axis. (ii)  $\rho_\phi(\Delta)$ does not grow too fast at infinity. This is not a strong condition as 
\be
 \msf{g}(\Delta) w(\Delta) \limu{\Delta \to \infty}  \frac{1}{2 \pi^{d/2}}\cdot \frac{\Delta^{d/2-1} \xi^\D}{4^\Delta \sin\!\left(\pi(\hd-\Delta)\right)}
\ee
for small $\xi$.  
(iii)we assume that
\beq
\label{eq:rhoZero}
\rho_\phi(d/2) =0
\eeq
in order to avoid picking up the pole at $\Delta = d/2$ coming from $\msf{g}(\Delta)$. The assumption~\reef{eq:rhoZero} seems to be satisfied in all known examples, cf.\@ later in this section. However, this assumption can be relaxed by explicitly adding the  residue of the possible pole at $\D=d/2$. (iv) we assume that $\rho_\phi$ is meromorphic, with single poles $\Delta_*$ on the right half-plane:
\beq
\rho_\phi(\Delta) \; \limu{\Delta \to \Delta_*} \frac{\text{Res}\, \rho_\phi(\Delta_*)}{\Delta - \Delta_*}~,
\quad
\Re(\Delta_*) > d/2~.
\eeq

Considering these assumptions now, we may deform the contour. By Cauchy's theorem, the $\expec{\phi \phi}$ correlator will pick up two series of poles: one family coming from the function $\msf{g}(\Delta)$ at $\Delta = \thd + \{1,2,3,\ldots\}$, and a second family of poles coming from the spectral density $\rho_\phi$. Bringing everything together, we have\footnote{The minus sign in the second term of~\reef{eq:repassum} arises from the fact that the contour is taken in the clockwise direction.}
\be
\begin{aligned}
\label{eq:repassum}
\braketmh{\Omega}{\phi(\eta,x)\phi(\eta',x')}{\Omega}  =  &- \sum_{\Delta_*} \text{Res}\, \rho_\phi(\Delta_*)\,  \msf{g}(\Delta_*)\,  \psi_{\Delta_*}(\xi) \\
&+ \sum_{n=1}^\infty \frac{(-1)^n \Gamma(\hd + n)}{ 2^{d+1+2n}\pi^{d/2}n!}\; \rho_\phi(\thd + n) \, \psi_{\hd + n}(\xi)~.
\end{aligned}
\ee
In particular in the late-time limit, setting $\eta = \eta'$ for convenience:
\beq
\label{eq:splitlimit}
\braketmh{\Omega}{\phi(\eta,x)\phi(\eta,x')}{\Omega} \; \limu{\eta\to 0^{-}} \;  - \sum_{\Delta_*} \text{Res}\,  \rho_\phi(\Delta_*) \, \msf{g}(\Delta_*) \,\left(\frac{-2\eta}{|x-x'|}\right)^{2\Delta_*} 
\eeq
omitting terms that are subleading as $\eta \to 0$.\footnote{There are two types of subleading terms. First, we have only kept the leading term in $\xi$ in the hypergeometric function $\psi_{i\nu}(\xi)$. Second, we have approximated the invariant $\xi$ by its leading piece in the limit $\eta \to 0$.} Here we did not write the contributions from residues of $ \msf{g}(\Delta_*)$. We will come back to that shortly at the end of this section. From~\reef{eq:splitlimit} it is clear that the leading late-time behavior of the $\expec{\phi \phi}$ correlator comes from poles in $\rho_\phi(\Delta)$ with the smallest real part, or to be precise the smallest $\Re(\Delta_* - \thd) > 0$. In addition, if $\rho_\phi(\hd + n) \neq 0$ then there are terms that scale as $(-\eta)^{d+2n}$ with $n \geq 1$. 

\subsubsection{Boundary OPE}
We may now derive the result above from the OPE~\reef{OPEdS}. This bulk-boundary OPE is not necessarily convergent, but we can still try to reproduce the late-time behavior of the $\expec{\phi(\eta,x)\phi(\eta,x')}$ correlator. The two-point function of conformal operators can only be non-vanishing if they have the same scaling dimension\footnote{Here we are assuming that $x$ and $y$ are separate points. So we do not deal with contact terms discussed in~\ref{sec:Local}}:
\beq
\expec{\Oo_\alpha(x) \Oo_{\alpha'}(y)} = \frac{\delta_{\alpha \alpha'}}{|x-y|^{2\Delta_\alpha}}
\eeq
which still holds when $\Delta_\alpha$ is a complex number. The double sum over boundary operators; therefore, collapses to a single sum, hence
\bsub
\label{eq:latetimeAlt}
\beq
\braketmh{\Omega}{\phi(\eta,x)\phi(\eta',x')}{\Omega} \; \sim \;   \sum_\alpha (b_{\phi \alpha})^2 (\eta \eta')^{\Delta_\alpha} \; \mca{D}_\alpha(\eta \pd_x) \mca{D}_\alpha(\eta' \pd_{x'})\, \frac{1}{|x-x'|^{2\Delta_\alpha}}
\eeq
where
\beq
\mca{D}_\alpha(\eta \pd_x) = {}_0F_1\!\left(\Delta_\alpha-d/2+1,\, \tfrac{1}{4}\eta^2 \pd_x^2\right) = 1 + \mrm{O}(\eta^2 \pd_x^2)~.
\eeq
\esub
As before, we are interested in the limit $\eta,\eta' \to 0^{-}$. Hence we can approximate the differential operator $\mca{D}_\alpha$ by its leading term, which leads to the asymptotic behavior
\beq
\label{eq:OPEexp}
\braketmh{\Omega}{\phi(\eta,x)\phi(\eta,x')}{\Omega} \; \limu{\eta \to 0^{-}} \;  \sum_\alpha (b_{\phi \alpha})^2 \left(\frac{-\eta}{|x-x'|}\right)^{2\Delta_\alpha} 
\eeq
For this expansion to match~\reef{eq:splitlimit}, we require first of all that the poles $\Delta_*$ equal the boundary operator spectrum $\{\Delta_\alpha\}$ exactly. Moreover, the residues of $\rho_\phi$ are related to the $b_{\phi \alpha}$ according to the dictionary
\beq
\boxed{
(b_{\phi \alpha})^2 = - 4^{\Delta_\alpha}  \msf{g}(\Delta_\alpha)\,\text{Res}\, \rho_\phi(\Delta_\alpha)
}~.
\eeq
This result should not be surprising. After all, the  special functions $\psi_\Delta(\xi)$ appearing in~\reef{eq:repassum} are nothing but boundary conformal blocks~\cite{McAvity:1995zd} in $d+1$ dimensions. To wit, the spectral integral and its counterpart~\reef{eq:repassum} appeared before in the BCFT context in a slightly different form~\cite{Hogervorst:2017kbj}.

Notice that neither $b_{\phi \alpha}$ nor $\text{Res}\, \rho_\phi(\Delta_\alpha)$ is required to be real-valued: in a generic QFT in dS they are complex-valued. Nevertheless, the hermiticity of $\phi$ implies that
\beq
\label{eq:causality}
\rho_\phi(\Delta)^* = \rho_\phi(\Delta^*)
\eeq
hence the residue of a pole at $\Delta_\alpha$ and its complex conjugate $\Delta_\alpha^*$ are necessarily related via complex conjugation.

Finally, the terms scaling as $\xi^{d/2+n}$ with $n=1,2,3,\ldots$ in~\reef{eq:repassum} and~\reef{eq:splitlimit} cannot be reproduced from the bulk-boundary OPE~\reef{OPEdS}. It is therefore natural to assume that
\bsub
\beq
\label{eq:ass}
\rho_\phi(\thd + n) = 0
\quad
\text{for all}
\quad
n=0,1,2,\ldots.
\eeq
We do not have a proof of this fact, beyond the fact that in all known examples
\beq
\rho_\phi(\thd + i \nu) \; \propto \; \nu \sinh(\pi \nu) = \frac{\pi}{\Gamma(\Delta - \thd)\Gamma(\thd - \Delta)}
\eeq
\esub
which indeed vanishes at $\Delta = d/2 + \mbb{N}$. Likely this phenomenon has a group-theoretical explanation. In the literature, it is common to write spectral integrals with a Plancherel measure, schematically \@ $1/(2\pi i)\sum_\ell \int\!d\Delta \, \mathfrak{P}(\Delta,\ell)$ --- see for instance~\cite[Eq.~(8.7)]{Dobrev} or~\cite[Eq.~(74)]{Mack:2009mi} and~\cite{Karateev:2018oml}. This measure is not physical: from our point of view, it amounts to a simple redefinition of $\rho_\phi(\Delta) \mapsto \rho_\phi(\Delta)/\mathfrak{P}(\Delta,0)$ which does not affect observables. Nevertheless, the analytic structure of $\rho_\phi(\Delta)$ is affected by this rescaling, and indeed the $\ell = 0$ Plancherel measure $\mathfrak{P}(\Delta,0)$ contains a factor $\nu \sinh(\pi \nu)$ which furnishes the desires zeroes~\reef{eq:ass}.

\subsection{Analytic continuation from $S^{d+1}$} \label{sec:two-point sphere}
As discussed in section \ref{sec:SphereCorrelation}, the dS correlation functions, and in particular two-point functions, can be  defined by the analytic continuation of correlation functions on the sphere $S^{d+1}$. In what follows, we use this fact to find a formula for the spectral density of a generic scalar field theory in dS as an integral over the discontinuity of the two-point function. 

Let us therefore consider the two-point function $\expec{\phi(X)\phi(X')}_{S^{d+1}}$  of a hermitian operator $\phi(X)$ on the sphere, where we parametrize $S^{d+1}$ by embedding space coordinates $X^A \in \mbb{R}^{d+2}$ obeying $X \cdot X = R^2$. Such a two-point function can only depend on the invariant
\beq
\label{eq:embeddingDistance}
x \ldef \frac{X \cdot X'}{R^2}~,
\quad
-1 \leq x \leq 1
\eeq
where $x=1$ (resp.\@ $x=-1$) corresponds to identical (resp.\@ antipodal) insertions. Consequently we write
\beq
\expec{\phi(X)\phi(X')}_{S^{d+1}} = \wh{G}(x)
\eeq
 for some function $\wh{G}(x)$ which is not determined by symmetries. This correlator maps to the dS two-point function from Eq.~\reef{eq:2pt000} via 
 \beq
 \label{eq:dstos}
 \xi = \frac{2}{1-x}~, \quad \text{or} \quad \wh{G}(x) = G\!\left(\xi = \frac{2}{1-x}\right).
 \eeq
From now on we will use this formula to identify both correlators, and write $G(x)$ instead of $\wh{G}(x)$ to avoid clutter.

It is well-known that any function of the invariant $x$ can be decomposed in terms of  $SO(d+2)$ Gegenbauer polynomials:
\begin{equation}
\label{eq:sphereExpansion}
G(x) = \sum_{J=0}^\infty a_{J} \, C_J^{\frac{d}{2}}(x)
\end{equation}
for some coefficients $a_J$ that depend on the $\expec{\phi \phi}$ correlator in question. The Gegenbauers form an orthogonal basis with respect to the norm
\beq
\label{eq:sphm}
|\!| f |\!|^2 \ldef \int_{-1}^1\!dx\, (1-x^2)^{(d-1)/2}\, |f(x)|^2~.
\eeq
Most physical correlators are not square-integrable with respect to the measure~\reef{eq:sphm} due to singularities near $x = 1$. To be precise, for a correlator $G$ to be square integrable, we need that\footnote{An equivalent condition for square integrability is that the coefficients $a_J$ decrease faster than $|a_J| \limu{J \to \infty} 1/J^{(d-1)/2}$, as follows from Parseval's theorem.}
\beq
\quad
G(\xi) \; \limu{\xi \to \infty} \; \xi^\gamma
\qaq
G(\xi) \; \limu{\xi \to 1^{+}} \; 1/(\xi-1)^{\gamma}
\quad
\text{with}
\quad
\gamma < (d+1)/4~.
\eeq
The Gegenbauer polynomials obey
\beq
\label{eq:kappaj}
|\!| C_J^{\hd} |\!|^2 = 1/\kappa_J~,
\quad
\kappa_J \ldef \frac{2^{d-1}J!(J + \hd)\Gamma(\hd)^2}{\pi \Gamma(d+J)}
\eeq
and in particular it follows that the coefficients $a_J$ can be recovered using orthogonality of Gegenbauer polynomials in~\reef{eq:ortho gegenbauer} as
\beq
\label{eq:EuclIF}
a_J = \kappa_J \int_{-1}^1\!dx\, (1-x^2)^{(d-1)/2} \, C_J^{\hd}(x) G(x)~.
\eeq
for $J = 0,1,2,\ldots$.

Let us print a formula for the $a_J$ in a specific case, taking $\phi$ to be a free massive scalar, so $G(x)$ is the function $G_\mrm{f}(\xi;\nu)$ from Eq.~\reef{eq:free two point function}. The correlator in question is not square-integrable in $d \geq 3$ dimensions: indeed the correlator grows as $G_\mrm{f}(\xi;\nu) \sim \xi^{(d-1)/2}$, so in $d \geq 3$ dimensions it does not represent a square-integrable function on $S^{d+1}$. Nevertheless one can compute the cofficients $a_J$ using the inversion formula~\reef{eq:EuclIF}, for instance by analytically continuing in $d$. This computation was carried out in~\cite{Marolf:2010zp}, yielding
\begin{equation}
\label{eq:marolfaj}
a_{J}= \frac{1}{R^{d-1}} \frac{\Gamma(\frac{d}{2})}{4\pi^{\frac{d}{2}+1}} \frac{2J+d}{J(J+d)+m^2R^2}~.
\end{equation}
We will shortly revisit the formula~\reef{eq:marolfaj} from a different point of view in~\reef{eq:phiSpec}.


In Eq.~\reef{eq:EuclIF}, we presented a formula to invert the expansion~\reef{eq:sphereExpansion}, expressing $a_J$ as an integral over the correlator $G(x)$. The inversion formula~\reef{eq:EuclIF} applies to integer $J$.  In appendix~\ref{inver form apx}, we obtain an alternative inversion formula that applies to \emph{complex} values of $J$. This inversion formula reads
\begin{equation}\label{The inversion formula}
a_J = \frac{1}{2\pi i} {\frac{\Gamma(\frac{d}{2}) \Gamma(J+1)}{\Gamma(J+\frac{d}{2}) 2^J}}\int_1^{\infty} dx \; \frac{(x+1)^{\frac{d}{2}-\frac{1}{2}}}{(x-1)^{J+\frac{d}{2}+\frac{1}{2}}}  \; \FF{J+d}{J+\frac{d}{2}+\frac{1}{2}}{2J+d+1}{\frac{2}{1-x}}  \Disc{G(x)}
\end{equation}
where the discontinuity $\Disc{G(x)}$ is defined as
\be\label{eq:DiscDef}
\Disc{f(x)} \ldef f(x+i\epsilon)-f(x-i\epsilon)~.
\ee
Since the RHS of~\reef{The inversion formula} is an analytic function of $J$,   the above identity extends $a_J$ to an analytic function of $J$ on the complex plane.

Let us briefly discuss the convergence of the integral in~\reef{The inversion formula}. Suppose that near $x=1$ and $x=\infty$ the discontinuity of $G(x)$ behaves as
\beq
\label{eq:discGas}
\mrm{Disc}\, G(x) \; \limu{x \to 1}\; \frac{1}{(x-1)^\delta}
\qaq
\mrm{Disc}\, G(x) \; \limu{x \to \infty}\; x^\varepsilon
\eeq
for some exponents $\delta$, $\varepsilon$. Then convergence requires that $\delta < 1$ and $\Re(J) > \varepsilon$, as follows from analyzing the $x \to 1,\infty$ asymptotics of the ${}_2F_1$ hypergeometric appearing in~\reef{The inversion formula}. Whenever $\Re(J) \leq \varepsilon$, the function $a_J$ can have singularities in the complex $J$-plane. Also notice that the integrand involves the correlator $G(x)$ analytically continued beyond the Euclidean region $-1 \leq x \leq 1$. In fact, $x \geq 1$ maps to $\xi < 0$, which describes timelike separated points in de Sitter. In practice; however, one can calculate $a_J$ for a two-point functions with $\delta \geq 1$ by means of analytical continuation from the convergent region, see e.g.~\reef{sec:ExampleInversion}. In what follows, we will rederive the \toolazy decomposition using the above inversion formula.

\subsubsection{Recovering the spectral density}\label{sec:Spectral density}
\begin{figure}[t]
\centering
\begin{tikzpicture}[scale=1]
\coordinate (origin) at (0,0);
\draw[line width=0.3mm,black,->] (origin) -- ++(4,0);
\draw[line width=0.3mm,black,->] (origin) -- ++(0,3);
\draw[line width=0.3mm,black,-] (origin) -- ++(-2,0) node (zminus) {};
\draw[line width=0.3mm,black,-] (origin) -- ++(0,-1) node (xminus) {};
\draw[line width=0.3mm,red!70,-,dashed] (-1.5,-1) node [red, below] {$-\frac{d}{2}$} --(-1.5,3);
\foreach \x in {0,...,3} 
\draw (\x,0) node[cross,black!70] {};
\foreach \x in {0,...,3} 
\draw[blue!70, thick,->] (\x,0) circle (0.25);
\node at (2.3,0.35) [blue!70,scale=1] {$ c_0 $} ;
\foreach \x in {0,...,3} 
\draw[blue!70, thick,-] (\x+0.25,0)--(\x+1-0.25,0);
\foreach \x in {0,...,3} 
\draw[blue!70, thick,->] (\x-0.05,0.25)--(\x-0.06,0.25);
\foreach \x in {0,...,3} 
\draw[blue!70, thick,->] (\x+0.05,-0.25)--(\x+0.06,-0.25);
\draw[line width=0.2mm,black,-] (3,2.5)--(3.5,2.5);
\draw[line width=0.2mm,black,-] (3,3)--(3,2.5)node [black, above right] {$J$};

\coordinate (origintwo) at (9,0);
\draw[line width=0.3mm,black,->] (origintwo) -- ++(4,0);
\draw[line width=0.3mm,black,->] (origintwo) -- ++(0,3);
\draw[line width=0.3mm,black,-] (origintwo) -- ++(-2,0) node (zminus) {};
\draw[line width=0.3mm,black,-] (origintwo) -- ++(0,-1) node (xminus) {};
\draw[line width=0.3mm,red!70,-,dashed] (9-1.5,-1) node [red, below] {$-\frac{d}{2}$} --(9-1.5,3);
\foreach \x in {0,...,3} 
\draw (9+\x,0) node[cross,black!70] {};

\draw [line width=0.3mm,black!50] (9,0.32) to[out=180,in=90] (8.65,0) ;
\draw [line width=0.3mm,black!50] (9,-0.32) to[out=180,in=-90] (8.65,0);
\draw [line width=0.3mm,black!50] (9,0.32) -- (11,0.32) node [black!50 ,above] {$c_1$} --(13,0.32) ;
\draw [line width=0.3mm,black!50] (9,-0.32) -- (13,-0.32) ;
\draw[black!50, thick,->] (11,0.32) --(10.95,0.32);
\draw[black!50, thick,->] (11,-0.32) --(11.05,-0.32);

\draw [line width=0.3mm,blue!70,dashed] (8.65,0.32) to[out=0,in=90] (9,0) ;
\draw [line width=0.3mm,blue!70,dashed] (8.65,-0.32) to[out=0,in=-90] (9,0);
\draw [line width=0.3mm,blue!70,dashed] (8.65,0.32) -- (9-1.2,0.32) ;
\draw [line width=0.3mm,blue!70,dashed] (8.65,-0.32) -- (9-1.2,-0.32) ;

\draw [line width=0.3mm,blue!70] (9-1.2,-1) -- (9-1.2,1.75) node [blue!70 ,right] {$c_2$} -- (9-1.2,2.9);
\draw[blue!70, thick,->] (9-1.2,1.75) --(9-1.2,1.7);
\draw[black!50, thick,<-] (9-1.5,2.9)--(9-1.35,2.9)node[black!50, above] {$\epsilon$};
\draw[black!50, thick,->] (9-1.35,2.9)--(9-1.2,2.9);

\draw[line width=0.2mm,black,-] (12,2.5)--(12.5,2.5);
\draw[line width=0.2mm,black,-] (12,3)--(12,2.5)node [black, above right] {$J$};
\end{tikzpicture}
\caption{Illustration of contour integrals of Watson-Sommerfeld transformation. Left: sum over non-negative integers as a set of contour integrals around the integers~\reef{eq:int over gtilde}. Right: Deforming the contour to a line integral with constant real part.}
\label{fig:ContourG}
\end{figure}
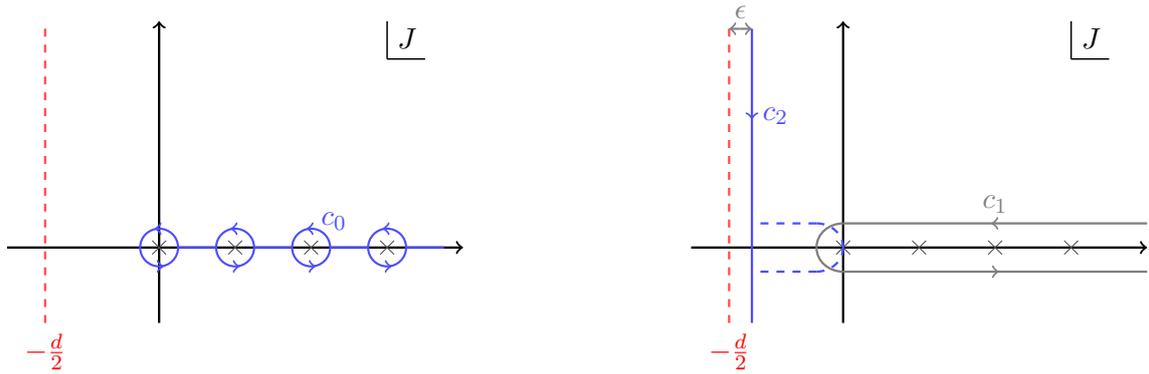

In order to derive the desired decomposition~\reef{eq:newKL}, let us turn our attention to the original expansion~\reef{eq:sphereExpansion}. The two-point function is a sum over non-negative integers: 
\begin{equation}
\label{eq:gs}
G(x) = \sum_{J=0}^\infty g_J(x)\,,
\qquad
g_J(x) \coloneqq a_J \, C_J^{\frac{d}{2}}(x)~.
\end{equation}
Suppose that we can extend $g_J(x)$ to a function $\tilde{g}_J(x)$ which is analytic in $J$ and coincides with $g_J(x)$ at integers: $\tilde{g}_J(x) = g_J(x)$ at $J=0,1,2,\ldots$. Moreover, we imagine that we are given a kernel $K(J)$ that is meromorphic, having poles at the non-negative integers with unit residue. We can then replace the sum~\reef{eq:gs} by the following integral:
\begin{equation}\label{eq:int over gtilde}
G(x)=\oint_{c_0} \frac{dJ}{2\pi i} \, K(J)\; \tilde{g}_J(x)
\end{equation}
where the contour $c_0$ consists of small circles around the non-negative integers, passed in the counterclockwise sense. Such a contour is illustrated in figure~\ref{fig:ContourG}. If we in addition assume that the product $K(J) \, \tilde{g}_J(x)$ decays sufficiently fast at large $|J|$, one can deform the contour to an integral over a line with fixed real part, e.g. $c_2$ in figure~\ref{fig:ContourG}. The act of expressing a discrete sum as contour integral in the complex plane is known as a \emph{Watson-Sommerfeld transformation}, see for instance~\cite{Marolf:2010zp}.

The discussion so far was general and did not involve details about the decomposition~\reef{eq:sphereExpansion} of the $\expec{\phi \phi}$ correlator. At this point, we will use some properties of the Gegenbauer polynomials, and we will propose an explicit kernel $K(J)$ as well as an analytic extension $\tilde{g}_J(x)$ of $g_J(x)$, to wit
\begin{equation}
\label{eq:KJdef}
K(J) \ldef \frac{\pi e^{i \pi J}}{\sin(\pi J)}
\qaq
\tilde{g}_J(x) \ldef e^{-i\pi J} a_J\,  C_J^{\frac{d}{2}}(-x)
\end{equation}
cf.~\cite[Eqs.~(20) and (21)]{Marolf:2010zp}.
For $J \notin \mbb{N}$, the functions $C_J^{\hd}(-x)$ are so-called Gegenbauer functions, which can be expressed as hypergeometric functions, cf.\@ equation~\reef{eq:def gegenbauer}. The choice of writing the function in terms of $-x$ will be apparent soon in~\reef{eq:FreeToGeg}. For integer $J$, the Gegenbauer functions reduce to the Gegenbauer polynomials that we have encountered so far, up to a sign:
\beq
J \in \mbb{N}:
\quad
C_J^{\hd}(-x) = (-1)^J C_J^{\hd}(x)
\quad
\Rightarrow
\quad
\tilde{g}_J(x) = g_J(x)
\eeq
as required. Moreover, it is easy to check that $K(J)$ from~\reef{eq:KJdef} has poles at integer $J$ with unit residue.

To see that we are able to deform the contour, let us comment on the large-$J$ behavior of the integrand in~\reef{eq:int over  gtilde}.  In appendix~\ref{LargeJ}, we show that the leading contribution at large $J$ of~\reef{The inversion formula} is dominated by the $x\to 1$ part of the integral. For a two-point function with a power-law singularity at $x=1$
\bsub
\label{eq:ajscaling}
\begin{equation}
G(x) \; \limu{x \to 1} \; \frac{1}{(1-x)^\delta}~,
\end{equation}
the large-$J$ behaviour of $a_J$ is given by\footnote{The proof in question assumes that $\delta < 1$. We expect~\reef{eq:ajscaling} to hold for larger values of $\delta$ as well. This can be seen, for example, if one calculates $a_J$ for a CFT in the bulk.}
\begin{equation}
\lim_{J\to\infty} a_J \; \sim \; \frac{1}{|J|^{d-2\delta}}
\end{equation}
\esub
up to a $J$-independent constant. We are now ready to analyze the product $K(J) \tilde{g}_J(x)$ at large $J$:
\beq
K(J) \tilde{g}_J(x) \approx \frac{e^{-\!\arccos(x)|\Im(J)|}}{|J|^{d/2-2\delta+1}} 
\eeq
so away from the real axis, the function decreases exponentially, provided that $x$ is in the Euclidean region $(-1,1)$. For sufficiently small $\delta$ the function decays as a power law along the real axis as well. It is therefore possible to deform the contour $c_0$ to $c_2$, as in Figure~\ref{fig:ContourG}.

At this point, let us go back to the expression of the free theory two-point function in eq.~\reef{eq:free two point function}. The formula in question is valid both for $S^{d+1}$ and dS${}_{d+1}$--as long as the insertions are spacelike separated, otherwise an $i\epsilon$ prescription is required. Given the definition of the Gegenbauer functions~\reef{eq:def gegenbauer} and~\reef{eq:embeddingDistance}, one can rewrite the propagator as
\begin{equation}\label{eq:FreeToGeg}
G_{\text{f}}(\xi;\nu) = \frac{\Gamma(\frac{d}{2})}{4\pi^{\frac{d}{2}} \sin(\pi \D)} \, C_{-\D}^{\frac{d}{2}}(-x)~,
\quad
\Delta = \hd + i \nu~.
\end{equation}
using the invariant $\xi$ instead of $x$ for convenience from~\reef{eq:dstos}. Identifying $-J$ with $\Delta$, we can therefore recast Eq.~\reef{eq:int over gtilde} as an integral of $a_{-\Delta}$ running over the principal series spectrum $\Re(\Delta) = d/2$, to wit
\begin{equation}\label{eq:integral over Delta and free}
G(\xi) = \int^{\frac{d}{2}+i\infty}_{\frac{d}{2}-i\infty} \frac{d\D}{2\pi i}\; \frac{4\pi^{\frac{d}{2}+1}}{\Gamma(\frac{d}{2})} \, a_{-\D}\,  G_{\text{f}}(\xi;\nu)~,
\end{equation}
Note that the minus sign $a_J \mapsto a_{-\Delta}$ has changed the orientation of the $c_2$ contour.  Of course, we recognize the above equation~\reef{eq:integral over Delta and free} as the \toolazy decomposition~\reef{eq:newKL}, after identifying
\beq
\label{eq:jac0}
\rho_\phi(\thd + i \nu) = \frac{2\pi^{\hd+1}}{\Gamma(\hd)} \lim_{\epsilon\to 0^+} \left( a_{i\nu - \hd+\epsilon} +a_{-i\nu - \hd+\epsilon} \right) ~
\eeq
where we used the symmetry of the free propagator $G_{\text{f}}(\xi;\nu)=G_{\text{f}}(\xi;-\nu)$ to replace $a_{-\D}=a_{-i\nu - \hd}$ in \eqref{eq:integral over Delta and free} by the shadow symmetric combination. We also kept the $\epsilon$ regulator that is important if $a_J$ has singularities on the line $\Re(J) = -\hd$ as depicted in figure \ref{fig:ContourG}. 
In case of no singularity on $\Re(\D)=d/2$, one can use the hypergeometric identities~\reef{eq:2f1inverseZ} and~\reef{eq:2f1identityW} to rewrite~\reef{eq:jac0} as
\be\label{eq:RhoFormulaNew}
\boxed{\rho_\phi(\D)=\beta(\D) \int_1^\infty dx \;  \FF{1-\D}{1-d+\D}{\frac{3-d}{2}}{\frac{1-x}{2}} \Disc{G(x)}}
\ee
where 
\be\label{eq:RhoFormulaNew2}
\beta(\D) = \frac{-i (4\pi)^{(d+1)/2}}{\Gamma(\frac{3-d}{2})}\frac{ \Gamma(1-\D)\Gamma(1-d+\D)}{\Gamma(\D-\hd)\Gamma(\hd-\D)}
\ee
writing $\Delta = d/2 + i \nu$. This is the promised inversion formula for spectral density. Let us remind you that the integration is over timelike separation $x=X.X^\prime>1$ which corresponds to the timelike separated points in dS. So we expected a discontinuity in the two-point function in this region. We will use this formula in section~\ref{sec:ExampleInversion} to find for instance the spectral density for a two-point function of a scalar operator of a bulk CFT. 

The derivation in Sec.~\ref{sec:KL} was based on symmetry properties of the dS Hilbert space alone which the positivity of $\rho_\phi(\hd + i \nu)$ at real values of $\nu$ is manifest there. The present derivation was based on the analytic continuation of correlators from $S^{d+1}$ to dS${}_{d+1}$ which leads to an explicit formula for  $\rho_\phi(\Delta)$ at complex values $\Delta$ but does not show a positivity manifestly.

Note that we assumed that the complimentary series are not apparent here; however, contour $c_2$ in figure~\ref{fig:ContourG} can be extended (to the dashed line) to take the complimentary series into account. As is discussed in section~\ref{sec:reps}, the complimentary series can be seen as an analytic continuation of the principal series that will appear as pole crossings over the solid line $c_2$ in figure~\ref{fig:ContourG}. In section~\reef{sec:ExampleInversion} we will see that this precisely happens for a range of dimensions of operators in a bulk CFT.
\subsection{Examples}\label{sec:ExampleInversion}

In the final part of this section, we will consider the \toolazy decomposition   in two different settings. First, we will consider the $\expec{\phi \phi}$ and $\expec{\phi^2 \phi^2}$ correlator in the theory of a free massive scalar $\phi$, followed by the analysis of a generic conformally invariant two-point function of scalar operators in dS bulk.

\subsubsection{Massive free boson: $\expec{\phi \phi} $}

As a first example, consider the correlator $\expec{\phi \phi} = G_\mrm{f}$ where $\phi$ is a free massive field in the bulk with mass $m$. Let us write $\Delta_\phi(d-\Delta_\phi) = m^2 R^2$ and set $\Delta_\phi = \hd + i \mu$ in order to avoid overloading the labels $\Delta$ and $\nu$. There are two possible ways to obtain the distribution $a_{-\Delta}$ for complex values of $\Delta$. On the one hand, in Eq.~\reef{eq:marolfaj}, a formula for $a_J$ at integer $J$ was presented, and the formula at hand can be analytically continued simply by replacing $J \mapsto -\Delta$. Alternatively, one can explicitly perform the integral~\reef{The inversion formula}, as is done in appendix~\ref{sec:free massive aJ appendix}. Regardless of the chosen method, the result reads
\beq
\label{eq:phiSpec}
 \frac{4\pi^{\hd+1}}{\Gamma(\hd)} a_{-\Delta} =   \frac{2\D-d}{\D(d-\D)-m^2R^2} =   - \frac{1}{\Delta - \Delta_\phi} - \frac{1}{\Delta -d + \Delta_\phi}~.
\eeq
This has poles at $\Delta = \Delta_\phi$ and $\Delta = d-\Delta_\phi$, which fall exactly on the axis of integration $\Re(\Delta) = \hd$.
Using the prescription \eqref{eq:jac0}, we find
\beq
\frac{1}{2\pi}\, \rho_\mrm{f}(\thd + i \nu) = \frac{\delta(\mu + \nu) + \delta(\mu - \nu)}{2}
\eeq
which reproduces the correct answer. In the case where $m^2 < d^2/4$ such that $\Delta_\phi$ is on the complementary series, it is straightforward to adapt the above argument to obtain a similar result. This can be done by two different means, one is take the contour $c_2$ to include dashed line around complimentary series in figure~\ref{fig:ContourG} or simply take the results from principal series and analytically continue $\D_\phi$ away from principal series to complimentary series. 

\subsubsection{Massive free boson: $\expec{\phi^2 \phi^2} $}
Next, we can consider the two-point function of the (normal-ordered) operator $\phi^2$ in the Gaussian theory. By Wick's theorem
\beq
\label{eq:miley7things}
\expec{\phi^2(\eta,x)\phi^2(\eta',x')} = 2 \, G_\mrm{f}(\xi;\mu)^2
\eeq
and as a matter of principle, the spectral density $\rho_{\phi^2}(\Delta)$ can be obtained by applying the inversion formula to the RHS of~\reef{eq:miley7things}. For the moment let us assume that $\D_\phi$ belongs to principal series. We will comment on the case of complementary series shortly. It turns out that $\rho_{\phi^2}(\Delta)$ has already been computed through other means in Ref.~\cite[Eq.~(3.25)]{Epstein:2012zz}\footnote{In that work, the \toolazy decomposition of the more general correlator $G_\mrm{f}(\xi;\mu_1)G_\mrm{f}(\xi;\mu_2)$ is presented, which reduces to~\reef{eq:phi2 spectral density} for $\mu_1 = \mu_2$.}. The resulting formula is given by
\begin{multline}\label{eq:phi2 spectral density}
\rho_{\phi^2}(\Delta)=\frac{\nu \sinh(\pi\nu)}{2^3 \pi^{\frac{d}{2}+2} \Gamma(\frac{d}{2})}\frac{\Gamma^2(\frac{\D}{2}) \Gamma^2(\frac{d-\D}{2})}{ \Gamma(\D)\Gamma(d-\D)}\\ \times
 \Gamma\!\left(\frac{2\D_\phi+\D-d}{2}\right)\Gamma\!\left(\frac{2\D_\phi-\D}{2}\right)\Gamma\!\left(\frac{d-2\D_\phi+\D}{2}\right)\Gamma\!\left(\frac{2d-2\D_\phi-\D}{2}\right)
\end{multline}
writing $\Delta = d/2 + i \nu$ as usual. It is easy to check that $\rho_{\phi^2}$ is invariant under $\Delta \mapsto d- \Delta$.  Moreover, the correlator is apparently completely representated by the principal series: the contour in~\reef{eq:newKL} does not need to be deformed to account for complementary series states.

At this point, we can analyze the spectrum of late-time operators appearing on the bulk-boundary OPE of $\phi^2 \sim \sum_\alpha \Oo_\alpha$. On the right half plane, the density has three families of single poles:
\bsub
\beq
\Delta = 2\Delta_\phi + 2\mbb{N}~,
\quad
\Delta = 2(d-\Delta_\phi) + 2\mbb{N}
\qaq
\Delta = d + 2\mbb{N}~.
\eeq
Because of their dimensions, the corresponding operators $\Oo_\alpha(x)$ can be interpreted as scalar ``double-trace'' operators of the late-time CFT, schematically
\beq
\Oo \Box^n\Oo~,
\quad
\Oo^\dagger \Box^n \Oo^\dagger
\qaq
\Oo^\dagger \Box^{n} \Oo + \Oo \Box^n \Oo^\dagger
\eeq
\esub
where $\Oo$ and $\Oo^\dagger$ have dimension $\Delta_\phi = d/2 + i \mu$ resp.\@ $d-\Delta_\phi = d/2 - i \mu$. Since the late-time CFT is a mean-field theory built out of the operators $\Oo$, $\Oo^\dagger$, this is exactly the expected result: there are no other $SO(d)$ scalar operators built out of two operators in the CFT in question that one can write down. Of course, the bulk-to-boundary OPE coefficients $b_{\phi^2 k}$ can be obtained from~\reef{eq:phi2 spectral density} by computing residues.

In the case where $\Delta_\phi$ is real and belongs to the complementary series, one can repeat the above analysis by analytic continuation.  Notice that \reef{eq:phi2 spectral density} has poles at
$$\D=2\D_\phi-d-2n$$
for non-negative integers $n$. When one analytically continues $\D_\phi$ to the real line a pole crossing in integral of~\reef{eq:newKL} can happen. More precisely, for $\frac{3d}{4}<\D_\phi<d$ one has to deform the contour to go around these poles. Similar to what was discussed above and in~\ref{sec:reps}, one might interpret these poles as the complementary series contribution to~\reef{eq:newKL}.


\subsubsection{Bulk CFT correlator}\label{sec:BulkCFTTwoPoint}
As the second applicatoon of the \toolazy representation, we consider the correlation function of the following form:
\begin{equation}
\label{eq:CRJ}
G_\delta(x)=\frac{1}{(1-x)^\delta}
\quad
\text{i.e.}
\quad
G_\delta(\xi) = \frac{1}{2^\delta}\,  \xi^\delta~.
\end{equation}
Such a correlator arises for instance when one constructs a bulk CFT in de Sitter: the correlator~\reef{eq:CRJ} corresponds to a scalar two-point function of an operator $\varphi$ of dimension $[\varphi] = \delta$.  Unitarity requires that
 \be
 \delta \geq \frac{d-1}{2}~, \quad (\text{unitarity}).
 \ee
Note that $\delta = (d-1)/2$ corresponds to a conformally coupled free boson. One can easily check that~\reef{eq:ThedSTwoPointFunction} boils down to~\reef{eq:CRJ} after requiring $\D_\phi=\delta = (d-1)/2$ up to some normalization.

The  spectral density $\rho_\delta(\Delta)$ for~\reef{eq:CRJ} can be computed in several ways, for instance using alpha space techniques~\cite{Hogervorst:2017kbj}. Alternatively, it can be computed starting from the inversion formula~\reef{eq:RhoFormulaNew}, making use of the fact that
\[
\text{Disc}\, G_\delta(x) = \frac{2i \sin(\pi \delta)}{(x-1)^\delta}~.
\]
The integral appearing in the inversion formula can be computed exactly using~\reef{eq:power 2f1 integral} yielding to
\beq
\label{eq:Spectral CFT2}
\rho_\delta(\Delta) = \frac{2^{d+2-\delta}\pi^{(d+1)/2}}{\Gamma(\delta)\Gamma(\delta-\hd+\th)} \, \nu \sinh(\pi \nu)\, \Gamma(\delta-\Delta)\Gamma(\delta-d+\Delta)~.
\eeq
As before, the spectral density has support on the axis $\Re(\Delta) = d/2$ and \textit{does not} require separate contributions from states in the complementary series -- except for $\delta<d/2$ that we will come back to it shortly. This appears to be specific to \emph{scalar} two-point functions. For the two-point functions of spinning bulk operators in dS${}_2$, it seems possible to have contributions of discrete series states, as is discussed in appendix~\ref{Sec:DiscreteSeries}.

The bulk-boundary OPE of the CFT operator $\varphi \sim \sum_\alpha \Oo_\alpha$ can be analyzed by closing the contour in~\reef{eq:newKL} and picking up poles on the right half-plane. For the density in question~\reef{eq:Spectral CFT2}, there is a single family of poles at
\beq
\label{eq:polesCFT}
\Delta = \delta + \mbb{N}~.
\eeq
An exception is given by the massless case $\delta = (d-1)/2$, where only the term with $\Delta = \delta$ arises. This set of boundary operators is precisely what we expect from the discussion in \ref{sec:CFTdS}.

Finally, some care must be taken when $(d-1)/2 < \delta < d/2$. In that case, the first pole in~\reef{eq:polesCFT} has $\Re(\Delta_1) = \Re(\delta) < d/2$, so it is located left of the axis $\Re(\Delta) = d/2$. To reproduce the full correlator $G(\xi)$, the contour in~\reef{eq:newKL} must be deformed to include this pole (and to exclude its shadow). 
This pole again can be interpreted as the contribution from complementary series states.
This is consistent with our analysis of the decomposition of an $SO(2,2)$ conformal family into irreps of $SO(2,1)$, in appendix~\ref{Sec:DiscreteSeries}.

\section{Boundary four-point function}
\label{sec:FourPointFunction}

The late-time expansion~\eqref{OPEdS} defines boundary operators $\mathcal{O}_\alpha$. The action of the conformal generators on these boundary operators is like that of Euclidean conformal generators on primary operators.
In particular, \eref{eq:Killing} shows that the late-time boundary operator $\mathcal{O}_\alpha (x)$ transforms as a primary operator with dimension $\D_\alpha$. The (infinite) set of correlation functions of the $\{\Oo_\alpha\}$ therefore defines a $d$-dimensional CFT on the $\eta = 0$ timeslice. This CFT lacks some useful features of flat-space CFT, e.g.\@ the state-operator correspondence and OPE convergence. Moreover, the late-time CFT does not have a stress-energy tensor $T_{\mu \nu}$.
Nevertheless, one still can use the conformal symmetry on the boundary to find non-trivial constraints.

In this section, by writing the complete set of states introduced previously, we expand the four-point function of boundary operators in conformal partial waves and using unitarity, we find positivity properties of their coefficients. We analyze the corresponding partial wave expansion extensively in the case of the free massive field, and furthermore we explore the $\lambda \phi^4$ theory in dS${}_{d+1}$ to leading order in $\lambda$. Along the way, we show that unitarity suggests the existence of local terms in two-point functions. 

\subsection{Partial wave expansion}\label{sec:FPF}
A four-point function of   boundary operators can be expressed in terms of conformal partial waves by adding a complete set of states \reef{eq:identity x-space}
\begin{equation}\label{eq:4point}
\begin{aligned}
\langle \mathcal{O}_1 \mathcal{O}_2 \mathcal{O}_3 \mathcal{O}_4 \rangle 
&= \braketmh{\Omega}{\Oo_1 \Oo_2}{\Omega}\braketmh{\Omega}{\Oo_3 \Oo_4}{\Omega} \\
&+\sum_\ell \int_{\hd}^{\hd+i\infty}  \frac{d\Delta}{2\pi i}  \frac{1}{N(\Delta,\ell)} \int  d^dx  \;  
\langle \mathcal{O}_1 \mathcal{O}_2 |\Delta,x\rangle_{\mu_1 \dots \mu_\ell} \ ^{\mu_1 \dots \mu_\ell} \langle \Delta,x  |\mathcal{O}_3 \mathcal{O}_4\rangle
\end{aligned}
\end{equation}
omitting the explicit $x_i$-dependence of the operators $\Oo_i(x_i)$ to avoid clutter. Once again we are assuming that the operators $\Oo_i$ are scalars, so only traceless symmetric tensor states are exchanged.  
We take only principal series states contribution to the decomposition of this four-point function; however, the complimentary series contribution can be recast by analytic continuation as discussed in~\ref{sec:reps}. In the next section, we revisit this decomposition by reintroducing the discrete series that appear only in $d=1$ for scalar four-point functions. We shall often omit the vacuum symbol $|\Omega \rangle$ to avoid cluttering.

We now establish explicitly the crucial fact that the matrix elements $\langle \mathcal{O}_1 \mathcal{O}_2 |\Delta,x,z\rangle$ and $\langle \Delta,x,z| \mathcal{O}_3 \mathcal{O}_4\rangle$ have the same structure as the three-point function $\langle \mathcal{O}_1 \mathcal{O}_2 \mathcal{O}(x)\rangle$ and $\langle \tilde{\mathcal{O}} (x) \mathcal{O}_3 \mathcal{O}_4\rangle$, where $\Oo$ is a fictional operator of dimension $\Delta$ and $\tilde{\mathcal{O}}$ its shadow~\cite{Simmons-Duffin:2012juh} of dimension $d-\Delta$.\footnote{Here we used that the three-point structure of $\brakkket{\OO^\dagger (x)\OO_3\OO_4}$ is proportional to $\brakkket{\tilde{\OO}(x)\OO_3\OO_4}$ when $\OO$ is living on principal series, having $\D \in \frac{d}{2}+i \mbb{R}$.} Here we used the index-free notation mentioned in~\reef{eq:indexFree}. We stress that $\Oo$ and $\tilde{\Oo}$ are not physical operators: they are only used to label certain conformally covariant objects. 
This follows from the fact that the action of isometries on  $|\Delta,x,z\rangle$ and $\OO(x)|\Omega\rangle$ are the same. The action of a general conformal charge on a correlator is
\begin{equation}
\begin{split} (\hat{Q}_1 + \hat{Q}_2 + \ldots + \hat{Q}_n )\langle \mathcal{O}_1 \cdots  \mathcal{O}_n \rangle &= \sum_i  \langle \mathcal{O}_1 \cdots [Q,\mathcal{O}_i] \cdots \mathcal{O}_n \rangle \\
&=\langle Q \, \mathcal{O}_1 \cdots  \mathcal{O}_n \rangle - \langle \mathcal{O}_1 \cdots  \mathcal{O}_n Q \rangle = 0
\end{split}
\end{equation}
in which $\hat{Q}_i$ is a differential operator acting on $x_i$, that is to say
\begin{equation}
[Q , \mathcal{O}_i(x_i)]  = \hat{Q}_i  \, \mathcal{O}_i(x_i)~.
\end{equation}
Similarly, we have
\begin{equation}\label{eq: diff of state rep}
\begin{split}
(\hat{Q}_{1}+\hat{Q}_{2}+\hat{Q}_{\D}) & \langle \mathcal{O}_1 \mathcal{O}_2 |\Delta,x,z\rangle  \\ 
= &\,\langle [Q,\mathcal{O}_1] \mathcal{O}_2 |\Delta,x,z\rangle + \langle \mathcal{O}_1 [Q,\mathcal{O}_2] |\Delta,x,z\rangle + \langle \mathcal{O}_1 \mathcal{O}_2 Q |\Delta,x,z\rangle \\
= &\,\langle Q \, \mathcal{O}_1 \mathcal{O}_2 |\Delta,x,z\rangle = 0
\end{split}
\end{equation}
in which we used the result of the previous section to substitute the action of differential operator with the Hilbert space operator $Q$ on state $|\Delta,x,z\rangle$. This is exactly the same differential equation one finds for a three-point function. Therefore, $\langle \mathcal{O}_1 \mathcal{O}_2 |\Delta,x,z\rangle $ is proportional to the conformal three-point structure \eref{three point structure}  which is fixed by the conformal symmetry:
\begin{equation}\label{eq:F of three point}
 \langle \mathcal{O}_1 \mathcal{O}_2 |\Delta,x,z\rangle = \mathcal{F}_{12}(\Delta,\ell) \; \langle \mathcal{O}_1 \mathcal{O}_2 \mathcal{O}(x,z)\rangle~,
\end{equation}
where $\mathcal{F}$ is   independent of position.
Using the shorthand notation $|x_{ij}| = |x_i - x_j|$, the three-point structure is given by
\bsub
\beq
 \langle \mathcal{O}_1(x_1) \mathcal{O}_2(x_2) \mathcal{O}_3(x_3,z)\rangle=
\langle \mathcal{O}_1(x_1) \mathcal{O}_2(x_2) \mathcal{O}^{\mu_1\ldots\mu_\ell}_3(x_3)\rangle
 z_{\mu_1} \dots z_{\mu_\ell} \,
\eeq
 with
\begin{gather}\label{three point structure}
\langle \mathcal{O}_1(x_1) \mathcal{O}_2(x_2) \mathcal{O}^{\mu_1\ldots\mu_\ell}_3(x_3)\rangle = \frac{Z^{\mu_1} \ldots Z^{\mu_\ell}-\text{traces}}{|x_{12}|^{\D_1+\D_2-\D_3}|x_{13}|^{\D_1+\D_3-\D_2}|x_{23}|^{\D_2+\D_3-\D_1}}~,\\
\quad Z^\mu\equiv \frac{|x_{13}||x_{23}|}{|x_{12}|} \left(\frac{x^\mu_{13}}{x^2_{13}}- \frac{x^\mu_{23}}{x^2_{23}}\right)~\,.
\end{gather}
\esub
Let us stress that the notation $\expec{\Oo_1 \Oo_2 \Oo(x,z)}$ in~\reef{eq:F of three point} does not refer to a physical correlation function: it is just a shorthand notation for the object~\reef{three point structure}.
Similarly, we can write 
\begin{align*}
\langle \mathcal{O}^\dagger_1 \mathcal{O}^\dagger_2|\Delta,x,z\rangle  &=  \mathcal{F}_{1^\dagger 2^\dagger}(\Delta,\ell)  \langle {\mathcal{O}}^\dagger_1 {\mathcal{O}}^\dagger_2 \mathcal{O}(x,z) \rangle ~,\\
\langle \Delta,x,z| \mathcal{O}_1 \mathcal{O}_2\rangle &= \mathcal{F}^*_{1^\dagger 2^\dagger}(\Delta,\ell) \langle \tilde{\mathcal{O}}(x,z) \mathcal{O}_1 \mathcal{O}_2\rangle~,
\end{align*}
where the second line is obtained from the first by complex conjugation. 
We also used $\langle {\mathcal{O}}^\dagger_1 {\mathcal{O}}^\dagger_2 \mathcal{O}(x) \rangle^*=\langle \tilde{\mathcal{O}}(x) \mathcal{O}_1 \mathcal{O}_2\rangle$ which can be explicitly checked from eq.~\eref{three point structure} when $\mathcal{O}$ is in the principal series.

Using the above facts, eq.~\eref{eq:4point} can be recast as
\begin{equation}\label{4p PWE}
\boxed{
\langle \mathcal{O}_1 \mathcal{O}_2 \mathcal{O}_3 \mathcal{O}_4 \rangle = \sum_\ell \int^{\frac{d}{2}+i\infty}_{\frac{d}{2}} \frac{d\Delta}{2\pi i} \; I_{\Delta,\ell} \;\Psi^{\D_i}_{\D,\ell}(x_i) + \brakkket{\OO_1\OO_2}\brakkket{\OO_3\OO_4}
}
\end{equation}
where we defined
\begin{align}\label{I def}
I_{\Delta,\ell}& \ldef \frac{\mathcal{F}_{1 2}(\Delta,\ell) \mathcal{F}^*_{3^\dagger 4^\dagger}({\Delta},\ell)}{N(\Delta,\ell)}~,\\
\Psi^{\D_i}_{\D,\ell} (x_i)& \ldef \int d^dx \; \langle \mathcal{O}_1(x_1) \mathcal{O}_2(x_2) \mathcal{O}_{\mu_1 \dots \mu_\ell}  (x)\rangle \langle {\tilde{\mathcal{O}}}^{\mu_1 \dots \mu_\ell}(x) \mathcal{O}_3(x_3) \mathcal{O}_4(x_4)  \rangle~. \label{partial wave}
\end{align}
We emphasize that unitarity leads to positivity properties of the partial wave coefficients $I_{\D,\ell}$.
In particular, we have 
\begin{equation}\label{eq: positivty}
I_{\Delta,\ell}\geq 0 \qquad \textbf{if}: \qquad
\OO_1 = \OO_3^\dagger \quad\text{and}\quad \OO_2 =\OO_4^\dagger~.
\end{equation}
Note that  $\langle \mathcal{O}_1 \mathcal{O}_2 |\Delta,x,z\rangle$ is symmetric under exchange of $\mathcal{O}_1$ and $\mathcal{O}_2$ because boundary operators commute, while the three-point structure \eref{three point structure} changes by the factor $(-1)^\ell$. This means $\mathcal{F}_{12}$ changes by the same factor under exchange of $\mathcal{O}_1$ and $\mathcal{O}_2$. This leads to
\begin{equation}\label{eq: positivty negativity}
\bar{I}_{\Delta,\ell}\equiv I_{\Delta,\ell} (-1)^\ell \geq 0 \qquad \textbf{if}: \qquad
\OO_1 = \OO_4^\dagger \quad\text{and}\quad \OO_2 =\OO_3^\dagger~.
\end{equation}
This positivity property is at the core of the bootstrap approach to dS late-time correlators  that will be presented in the next section.
The function $\Psi^{\D_i}_{\D,\ell}$ defined in \eref{partial wave} is a solution of the conformal Casimir equation, and is known as the Conformal Partial Wave.

The set of conformal partial waves with $\Delta$ running over the principal series forms a complete basis of the four-point correlation functions, in a way that can be made precise~\cite{Simmons-Duffin:2017nub}.\footnote{Whenever $\Re(\Delta_1 - \Delta_2)$ or $\Re(\Delta_3  - \Delta_4)$ are large, the question of completeness of the principal series of partial waves is subtle, see for instance~\cite[appendix A.3]{Simmons-Duffin:2017nub}.}
 In the case of $d=1$, we need to add discrete series states with $\D \in \mathbb{N}^+$ to have a complete set of states. Strictly speaking, eq~.\reef{4p PWE} will have an extra sum over positive integers. We will see this explicitly    in section \ref{sec:Bootstrpping boundary correlators}.

We would like to briefly mention some properties  of the conformal partial waves. The partial waves satisfy the orthogonality relation
\begin{equation}\label{PW Orthogonality}
\int \frac{d^dx_1 \dots d^dx_4}{\text{vol(SO($d+1,1$))}} \Psi^{\D_i}_{\D,\ell}(x_i) \Psi^{\tilde{\D}_i}_{\tilde{\D}^\prime,\ell^\prime}(x_i) = 2\pi n_{\D,\ell} \; \delta_{\ell,\ell^\prime} \delta(\nu-\nu^\prime)~,
\end{equation}
where $\D=\hd+i\nu$, $\D^\prime=\hd+i\nu^\prime$ and the normalization factor reads
\begin{equation}\label{nDelta}
n_{\D,\ell} = \frac{\pi^{d+1} \text{vol($S^{d-2}$)}}{\text{vol(SO($d-1$))}}\frac{(2\ell+d-2)\Gamma(\ell+d-2)\Gamma(\ell+1)}{2^{2\ell+d-2} \Gamma(\ell+\frac{d}{2})^2 (\D+\ell-1)(\tilde{\D}+\ell-1)} \frac{\Gamma(\D-\frac{d}{2})\Gamma(\tilde{\D}-\frac{d}{2})}{\Gamma(\D-1)\Gamma(\tilde{\D}-1)}~.
\end{equation}
Here we use the shorthand notation $\tilde{\Delta} = d-\Delta$ and~\cite{compactmanifolds}
\beq
\label{eq:volumes}
\text{vol}(S^{d-1}) = \frac{2\pi^{d/2}}{\Gamma(\thd)},
\quad
\text{vol}(SO(d-1)) = \frac{2^{d-2}\pi^{(d-2)(d+1)/4}}{\prod_{j=2}^{d-1} \Gamma(\tfrac{j}{2})}.
\eeq
The partial waves  can also be written in terms of conformal blocks,
\begin{gather}\label{PWBlock}
\Psi^{\D_i}_{\D,\ell}(x_i) =  K^{\Delta_3,\Delta_4}_{\tilde{\Delta},\ell} G^{\Delta_i}_{\Delta,\ell} (x_i) + K^{\Delta_1,\Delta_2}_{\Delta,\ell} G^{\Delta_i}_{\tilde{\Delta},\ell} (x_i)~,
\\
K^{\Delta_1,\Delta_2}_{\Delta,\ell} = \frac{\pi^{\frac{d}{2}} \Gamma(\D-\frac{d}{2}) \Gamma(\D+\ell-1) \Gamma(\frac{\tilde{\D}+\D_1-\D_2+\ell}{2})\Gamma(\frac{\tilde{\D}+\D_2-\D_1+\ell}{2})}{\Gamma(\D-1) \Gamma(d-\D+\ell) \Gamma(\frac{{\D}+\D_1-\D_2+\ell}{2})\Gamma(\frac{{\Delta}+\D_2-\D_1+\ell}{2})}~\label{Kind}
\end{gather}
where $G^{\Delta_i}_{\Delta,\ell} (x_i)$ is proportional to the usual conformal block $G_{\Delta,\ell}^{\Delta_i}(z,\zb)$, to be precise:
\beq\label{eq:x to z psi translation}
G^{\Delta_i}_{\Delta,\ell} (x_i) = \frac{1}{|x_{12}|^{\Delta_1 + \Delta_2} |x_{34}|^{\Delta_3 + \Delta_4}} \left(\frac{|x_{14}|}{|x_{24}|} \right)^{\Delta_2 - \Delta_1} \left(\frac{|x_{14}|}{|x_{13}|}\right)^{\Delta_3 - \Delta_4} \, G_{\Delta,\ell}^{\Delta_i}(z,\zb)~,
\eeq
and we have introduced cross ratios $z,\zb$ as
\beq
\frac{|x_{12}|^2|x_{34}|^2}{|x_{13}|^2|x_{24}|^2} = z\zb,
\quad
\frac{|x_{14}|^2|x_{23}|^2}{|x_{13}|^2|x_{24}|^2} = (1-z)(1-\zb)~.
\eeq
For small $z,\zb$, the above definition of the conformal blocks fixes their short-distance behavior to be
\bsub
\label{Gnormalization}
\begin{align}
&G^{\D_i}_{\D,\ell}(z,\bar{z}) \to (-1)^\ell \; \frac{\Gamma(\ell+1)\Gamma(\frac{d-2}{2})}{2^\ell \Gamma(\ell+\frac{d-2}{2})} \;  (z\bar{z})^{\frac{\D}{2}} \; C^{\frac{d-2}{2}}_\ell\!\left(\frac{z+\bar{z}}{2\sqrt{z\bar{z}}}\right)   & z\sim\bar{z} \ll 1~,\\
&G^{\D_i}_{\D,\ell}(z,\bar{z}) \to \left(-\half\right)^\ell \; z^{\frac{\D-\ell}{2}}  \bar{z}^{\frac{\D+\ell}{2}}   & z \ll \bar{z} \ll 1~.
\end{align}
\esub

\subsubsection*{OPE for boundary operators}

Combining \eqref{4p PWE} with \eqref{PWBlock} and using the fact that $I_{\D,\ell}$ is shadow symmetric, one can write
\begin{equation}
\label{eq:JP11}
\langle \mathcal{O}_1 \mathcal{O}_2 \mathcal{O}_3 \mathcal{O}_4 \rangle = \sum_\ell \int^{\frac{d}{2}+i\infty}_{\frac{d}{2}-i\infty} \frac{d\Delta}{2\pi i} \; I_{\Delta,\ell} \;K^{\Delta_3 \Delta_4}_{\tilde{\Delta},\ell} G^{\Delta_i}_{\Delta,\ell} (x_i) + \brakkket{\OO_1\OO_2}\brakkket{\OO_3\OO_4}~.
\end{equation}
Since the conformal block $G^{\Delta_i}_{\Delta,\ell} (x_i)$ decays exponentially when ${\rm Re}\,\D \to \infty$ (whilst keeping the $x_i$ fixed) we can deform the contour to the right and pick up poles along the way. This gives
\begin{equation} 
\langle \mathcal{O}_1 \mathcal{O}_2 \mathcal{O}_3 \mathcal{O}_4 \rangle = -\sum_\ell \sum_{\Delta_\alpha} \; {\rm Res}_{\Delta=\Delta_\alpha} I_{\Delta,\ell} \;K^{\Delta_3 \Delta_4}_{\tilde{\Delta}_\alpha,\ell} G^{\Delta_i}_{\Delta_\alpha,\ell} (x_i) + \brakkket{\OO_1\OO_2}\brakkket{\OO_3\OO_4}~.
\end{equation}
As discussed in~\cite{Costa:2012cb,Dobrev,Simmons-Duffin:2017nub}, there are non-trivial cancellations among poles of the conformal blocks and poles of the  partial wave coefficients. When the dust settles, what is left is the contribution from the dynamical (not spurious) poles of $I_{\Delta,\ell}$;  these therefore control the expansion in powers of $|x_1-x_2|^2$.

This gives rise to an OPE between boundary operators, and we can read off the dimension of the exchanged boundary operators from the position of the poles in the partial wave coefficients $I_{\Delta,\ell}$. This is similar to what we saw in section \ref{2ptLateTime} for the late time expansion of the bulk two-point function from the Källén-Lehnmann spectral decomposition.

\subsection{Examples of partial wave coefficients}
\label{Examples4pt}
Before using the partial wave expansion in crossing equations to find non-trivial bounds, we would like to present some simple examples to gain more intuition about the partial wave coefficients $I_{\D,\ell}$. In what follows, we first consider a free massive field in dS which leads to Mean Field Theory (MFT) type conformal correlators for late-time boundary operators.
We shall see that the positivity condition \eref{eq: positivty} requires a careful treatment of contact terms in late-time correlators.
Then, we consider a $\lambda\phi^4$ bulk interaction and calculate the partial wave coefficients to the leading order in $\lambda$. 
\subsubsection{Mean Field Theory}
Consider the following four-point function of late-time boundary operators  of a free massive scalar field in dS,
\begin{equation}\label{eq:oototo}
\brakkket{\OO_1 \Od_2 \Od_3\OO_4}
\end{equation}
where we used $\OO_i$ as a short notation for  $\OO(x_i)$. 
Since the bulk field is free, the four-point function is given by three Wick contractions. 
Of course, this has the same structure as MFT,
\begin{align}\label{eq:oototoMFT1}
\langle \mathcal{O}_1\mathcal{O}^\dagger_2\mathcal{O}^\dagger_3\mathcal{O}_4 \rangle_{\text{MFT}} = \langle \mathcal{O}_1\mathcal{O}^\dagger_2 \rangle \langle \mathcal{O}^\dagger_3\mathcal{O}_4\rangle 
+ \langle \mathcal{O}_1\mathcal{O}_4 \rangle \langle \mathcal{O}^\dagger_2\mathcal{O}^\dagger_3\rangle 
+ \langle \mathcal{O}_1\mathcal{O}^\dagger_3 \rangle \langle \mathcal{O}^\dagger_2\mathcal{O}_4\rangle 
\end{align}
which can also be expanded as
\be
\begin{aligned}\label{eq:oototoMFT2}
\langle \mathcal{O}_1\mathcal{O}^\dagger_2\mathcal{O}^\dagger_3\mathcal{O}_4 \rangle_{\text{MFT}}  = & \, \langle \mathcal{O}_1\mathcal{O}^\dagger_2 \rangle \langle \mathcal{O}^\dagger_3\mathcal{O}_4\rangle  \\
+& \sum_\ell \int_{\frac{d}{2}}^{\frac{d}{2}+i\infty} \frac{d\D}{2\pi i} I^{\text{MFT}}_{\D,\ell} \Psi^{\D_i}_{\D,\ell}(x_i)+ \sum_\ell \int_{\frac{d}{2}}^{\frac{d}{2}+i\infty} \frac{d\D}{2\pi i} I^\delta_{\D,\ell} \Psi^{\D_i}_{\D,\ell}(x_i)
\end{aligned}
\ee
Here we wrote the partial wave decomposition in the (12)(34) channel, identifying the expansion of each of the 3 terms in~\reef{eq:oototoMFT1}.
Namely, the first corresponds to the vacuum contribution, the second we call $I^{\text{MFT}}_{\D,\ell}$ and the third we denote as $I^\delta_{\D,\ell}$ because it is a pure contact term that will be discussed later in section~\ref{sec:Local}. Remark that the operators $\OO$ and $\OO^\dagger$ do not commute as we will see in~\reef{eq:commutation of o snd ot}. For the four-point function~\reef{eq:oototo} we implictly use the radial ordering and therefore MFT expansion of it has the unique form of \reef{eq:oototoMFT1}.

Let us first calculate $I^{\text{MFT}}_{\D,\ell}$. We have
\begin{align*} 
\langle \mathcal{O}_1\mathcal{O}_4 \rangle \langle \mathcal{O}^\dagger_2\mathcal{O}^\dagger_3\rangle =
\frac{1}{|x_1-x_4|^{d+2i\mu}}\frac{1}{|x_2-x_3|^{d-2i\mu}}
=  \sum_\ell \int_{\frac{d}{2}}^{\frac{d}{2}+i\infty} \frac{d\D}{2\pi i} I^{\text{MFT}}_{\D,\ell} \Psi^{\D_i}_{\D,\ell}(x_i) \,.
\end{align*}
Using the orthogonality relation \eref{PW Orthogonality}, one finds
\begin{align}\label{eq:IMFT}
I^{\text{MFT}}_{\D,\ell}&=\frac{1}{n_{\D,\ell}}\int\frac{d^dx_1 \ldots d^dx_5}{\text{vol(SO($d$+1,1))}} \brakkket{\OO_1\OO_4} \brakkket{\Ot_2\Ot_3} \brakkket{\Ot_1\OO_2 \tilde{O}_{\mu_1 \dots \mu_\ell}(x_5)}\brakkket{O^{\mu_1 \dots \mu_\ell}(x_5)\OO_3\Ot_4}\nonumber\\
&=\frac{S([\Ot]\OO O)S([\OO]\OO O) }{n_{\D,\ell}}\int\frac{d^dx_1 d^dx_2 d^dx_5}{\text{vol(SO($d$+1,1))}} \brakkket{\Ot_1\OO_2\tilde{O}_{\mu_1 \dots \mu_\ell}(x_5)}\brakkket{O^{\mu_1 \dots \mu_\ell}(x_5)\Ot_2 \OO_1}\nonumber\\
&= (-1)^\ell \frac{2^{\ell-1}\Gamma(\ell+\hd)}{\pi^{\hd}\Gamma(\ell+1)} \frac{\Gamma(i\mu)\Gamma(-i\mu)}{\Gamma(\hd+i\mu)\Gamma(\hd-i\mu)}\frac{(\D+\ell-1) (\tilde{\D}+\ell-1) \Gamma(\D-1)\Gamma(\tilde{\D}-1)}{\Gamma(\D-\hd)\Gamma(\tilde{\D}-\hd)}
\end{align}
where we used shorthand notation $\tilde{\Delta} = d-\Delta$ as well as $O$ to denote the (hypothetical) exchanged operator with spin $\ell$ in the integral representation of the conformal partial wave to contrast with external operator $\mathcal{O}$.
In addition, we used the identity~\cite{Karateev:2018oml} 
\beq
\begin{aligned}\label{eq:identity xxi}
\zeta_{d,\ell}\equiv& \int\frac{d^dx_1 d^dx_2 d^dx_5}{\text{vol(SO(d+1,1))}}
\brakkket{\OO_1(x_1)\OO_2(x_2)O_{5,\ell}(x_5)}\brakkket{\tilde{O}_{5,\ell}(x_5) \Ot_1(x_1) \Ot_2(x_2)}\\
=&\frac{\text{vol}(S^{d-2})}{\pi^{\hd-1}\text{vol}(SO(d-1))}\frac{\Gamma(\ell+d-2)}{2^{\ell+d-2}\Gamma(\ell+\hd-1)}~,
\end{aligned}
\eeq
and the notion of shadow transform $\textbf{S}[\OO(x)]$ that creates a linear map on the space of three-point functions as~\cite{Karateev:2018oml}\footnote{The shadow transformation is defined as $$\textbf{S}[\OO(x)] = \int d^dy \brakkket{\Ot(x) \Ot(y)} \OO(y)$$ 
where  $\brakkket{\Ot(x) \Ot(y)} = \frac{1}{\left|x-y\right|^{2d-2\D}}$ is two-point structure of operators $\Ot$ with dimension $\tilde{\D} = d- \D$.}
\begin{gather}
\brakkket{\textbf{S}[\OO_1](x_1)\OO_2(x_2)\OO_3(x_3)}= S([\OO_1]\OO_2\OO_3) \brakkket{\Ot_1(x_1)\OO_2(x_2)\OO_3(x_3)}~.
\end{gather}
In particular, for scalar operators $\OO_1$ and $\OO_2$ we have the explicit formula
\begin{gather}
S([\OO_1]\OO_2\OO_{3,\ell}) = \frac{\pi^{\frac{d}{2}}\Gamma(\D_1-\frac{d}{2})\Gamma(\frac{d-\D_1+\D_2-\D_3+\ell}{2})\Gamma(\frac{d-\D_1+\D_3-\D_2+\ell}{2})}{\Gamma(d-\D_1)\Gamma(\frac{\D_1+\D_2-\D_3+\ell}{2})\Gamma(\frac{\D_1+\D_3-\D_2+\ell}{2})}~.
\end{gather}
Note that in \reef{eq:IMFT}, we used the fact that by swapping $\OO_1$ and $\OO_2$ in the three-point structure defined in~\eref{three point structure}, one picks a factor of $(-1)^\ell$.

$I^{\text{MFT}}_{\D,\ell}$ is negative for odd spins. On the other hand, according to~\reef{eq: positivty} the partial wave coefficients of the correlator~\reef{eq:oototo} have to be non-negative for all spins and values of $\D=\hd+i\nu$ with $\nu\geq0$. 
In the next section, we shall see that the third term in~\eqref{eq:oototoMFT2} solves this problem.

Lastly, let us remark that we could start from a more general four point function of the type
\be
\brakkket{\OO_1(x_1)\OO_2(x_2)\OO_2(x_3)\OO_1(x_4)}~.
\ee
Then we would find~\cite{Karateev:2018oml}
\be\label{eq:generalIMFT}
\begin{split}
I^{\text{MFT}}_{\D,\ell} = & (-1)^\ell \frac{2^{l-1}\Gamma(\ell+\hd)}{\pi^\hd \Gamma(\ell+1)} \frac{\Gamma(\hd-\D_1)\Gamma(\hd-\D_2)}{\Gamma(\D_1)\Gamma(\D_2)}  \\
&\times \frac{(\D+\ell-1)(\tilde{\D}+\ell-1) \Gamma(\D-1)\Gamma(\tilde{\D}-1)}{\Gamma(\D-\hd) \Gamma(\tilde{\D}-\hd)} \frac{\Gamma(\frac{\ell-\D+\D_1+\D_2}{2}))\Gamma(\frac{\ell-\tilde{\D}+\D_1+\D_2}{2}))}{\Gamma(\frac{d+\ell+\D-\D_1-\D_2}{2})\Gamma(\frac{d+\ell+\tilde{\D}-\D_1-\D_2}{2})}~.
\end{split}
\ee

\subsubsection{Local terms in the Gaussian theory}\label{sec:Local}

At late times, the propagator of a massive field in dS${}_{d+1}$ contains a delta function term~\cite{Arkani-Hamed:2015bza}. In this section, we calculate this local term explicitly, starting from the momentum-space expression~\reef{eq:WightFour} of the propagator. We will also make contact with the boundary OPE~\reef{OPEdS}.

As before, we encode the mass $m^2$ of the scalar by the dimension $\Delta = d/2 + i \mu$. Expanding the Hankel functions in~\reef{eq:WightFour} around $\eta = 0$, we obtain
\beq
\expec{\phi(\eta,x)\phi(\eta,y)} \; \limu{\eta \to 0} \; (-\eta)^d \int\!\frac{d^d k}{(2\pi)^d}\; e^{-ik \cdot(x-y)} \left[ \frac{\Gamma(-i\mu)^2}{4 \pi} \left(\frac{-|k|\eta}{2}\right)^{2i\mu} + \text{c.c} + \frac{\coth(\pi \mu)}{2\mu}\right].
\eeq
Performing the $k$-integral using
\[
\int d^dx \; e^{ik \cdot k} |x|^{-2a} = \pi^{\frac{d}{2}} \frac{\Gamma(\frac{d}{2}-a)}{\Gamma(a)} \left({\frac{|k|}{2}}\right)^{2a-d}
\]
we find that
\begin{equation}\label{eq:pos two point}
\begin{aligned}
\langle \phi(\eta,x,)\phi(\eta,y) \rangle \limu{\eta \to 0^{-}} &(-\eta)^{d+2i\mu}  \, \frac{\Gamma(-i\mu)\Gamma(\frac{d}{2}+i\mu)}{4\pi^{\frac{d}{2}+1}} \frac{1}{|x-y|^{d+2i\mu}}~+~\text{c.c.} \\
+& (-\eta)^d \, \frac{\coth(\pi \mu)}{2\mu } \delta^{(d)}(x-y)~.
\end{aligned}
\end{equation}
In what follows, we will refer to the third term as a \emph{local} term.\footnote{The local term in \eqref{eq:pos two point} can be derived in an alternative way.
Recall that the two-point function can be written as $\langle \phi(x,\eta)\phi(y,\eta) \rangle=F(\xi)$ with $\xi=4\eta^2/|x-y|^2$. 
In the limit $\eta\to 0$, we can then write 
$\langle \phi(x,\eta)\phi(y,\eta) \rangle \sim (-\eta)^d \delta^{(d)}(x-y) \int d^dw F(4/|w|^2) +\dots $ where the remaining terms vanish when integrated over $\int d^dx$.
Using the explicit expression  $F=G_{\text{f}}(\xi;\mu)$ given in \eqref{eq:ThedSTwoPointFunction} one recovers the coefficient of the local term in  \eqref{eq:pos two point}.
}

This expression should be compared with the expectation from the late-time OPE \eqref{OPEdS}, which in the case of a free massive bulk field simplifies to
\begin{equation}
\phi(x,\eta) \limu{\eta \to 0^{-}}  b (-\eta)^{\D} \mathcal{O}(x) + b^*  (-\eta)^{\D^*} \mathcal{O}^\dagger(x)~,
\qquad
\D=\frac{d}{2}+i\mu~.
\end{equation} 
 The late time limit of the bulk two-point function is then given by
\begin{equation}
\begin{aligned}
\langle \phi(x,\eta)\phi(y,\eta) \rangle \limu{\eta \to 0^{-}}  &(-\eta)^{2\D} b^2 \langle \mathcal{O}(x) \mathcal{O}(y) \rangle ~+ ~(-\eta)^{2\D^*} {b^*}^2 \langle \mathcal{O}^\dagger(x) \mathcal{O}^\dagger(y) \rangle \\
 \quad~+~ &(-\eta)^d |b|^2  \left(\langle \mathcal{O}(x) \mathcal{O}^\dagger(y) \rangle +  \langle  \mathcal{O}^\dagger(y) \mathcal{O}(x) \rangle \right)~.
\end{aligned}
\end{equation}
Comparing with \eqref{eq:pos two point}, one finds the anti-commutator to be
\be\label{eq:anticommutator of o od}
\left\{  \OO(x) , \OO^\dagger(y)  \right\} = \frac{\coth(\pi\mu)}{2\mu |b|^2}\delta^{(d)}(x-y)~, \qquad b=\sqrt{\frac{\Gamma(-i\mu)\Gamma(\frac{d}{2}+i\mu)}{4\pi^{\frac{d}{2}+1}}}~
\ee
Let us draw your attention to the fact that $\OO(x)$ and $\OO(y)^\dagger$ do not commute at coinciding points $x\to y$\footnote{We thank Matteo Delladio and Victor Gorbenko for pointing out this fact\cite{VictorMatteo}.}. One way to see this is the  canonical commutation relations of $\phi$ and its conjugate $\Pi$ in~\reef{eq:Canonical Pi and  phi} which gives
\be\label{eq:commutation of o snd ot}
\left [ \OO(x) , \OO^\dagger(y) \right] = \frac{1}{2\mu |b|^2}\delta^{(d)}(x-y)~.
\ee
This together with~\reef{eq:anticommutator of o od} leads to
\bsub
\begin{align}
\langle \mathcal{O}(x) \mathcal{O}^\dagger(y) \rangle &= \frac{\coth(\pi\mu)+1}{4\mu |b|^2} \delta^{(d)}(x-y) \equiv  C_\delta \delta^{(d)}(x-y)  ~, \\
\langle \mathcal{O}^\dagger(x) \mathcal{O}(y) \rangle &= \frac{\coth(\pi\mu)-1}{4\mu |b|^2} \delta^{(d)}(x-y) \equiv  \bar{C}_\delta \delta^{(d)}(x-y)~.
\label{eq:OpOm}
\end{align}
\esub

Now, let us go back to~\eref{eq:oototoMFT2} and find $I^\delta_{\D,\ell}$. The calculation is very similar to the one of $I^{\text{MFT}}_{\D,\ell}$ in \reef{eq:IMFT} except that we have the delta function of~\reef{eq:OpOm} instead of the conformal two-point functions that can easily be calculated by~\reef{eq:identity xxi}:
\begin{equation}\label{eq:deltaIOOtOOt}
\begin{aligned}
I^\delta_{\D,\ell} &=\frac{C_\delta \bar{C}_\delta}{n_{\D,\ell}} \int\frac{d^dx_1 \ldots d^dx_5}{\text{vol(SO($d$+1,1))}}  \delta^{(d)}(x_1-x_3)\delta^{(d)}(x_2-x_4) \brakkket{\Ot_1\OO_2\tilde{O}_5}\brakkket{O_5\OO_3\Ot_4} \\
&= \zeta_{d,\ell} \frac{C_\delta \bar{C}_\delta}{n_{\D,\ell}} = (-1)^\ell I^{\text{MFT}}_{\D,\ell}>0~.
\end{aligned}
\end{equation}
This leads to the total partial wave coefficient
\begin{gather}
I_{\D,\ell}=I^{\text{MFT}}_{\D,\ell}+I^\delta_{\D,\ell} = \left(1+(-1)^\ell\right) I^{\delta}_{\D,\ell}~.
\end{gather}
This means that the total partial wave expansion vanishes for the odd spin and is positive for even spins which agrees with the positivity condition~\reef{eq: positivty}. Let us emphasise again that without the contribution of the local terms, the partial wave coefficient is negative for odd spins which disagrees with the unitary condition~\reef{eq: positivty}. In other words, the existence of the local terms is essential in unitary Mean Field Theories in dS.

Let us remark that one could consider the correlator   $\brakkket{\OO_1(x_1)\OO_2(x_2)\OO_2(x_3)\OO_1(x_4)}$ with two different bulk fields. In this case, we would find a similar expression for $I^{\text{MFT}}_{\D,\ell}$ as in \eref{eq:IMFT} but there would be no local contribution.
This is not in contradiction with unitarity as this correlator no longer fulfills the positivity condition \eref{eq: positivty}.

\subsubsection{Adding interactions; $\phi^4$ theory at leading order}
\label{contact partial wave}

\begin{figure}[htb]
\begin{center}
\includegraphics[scale=1]{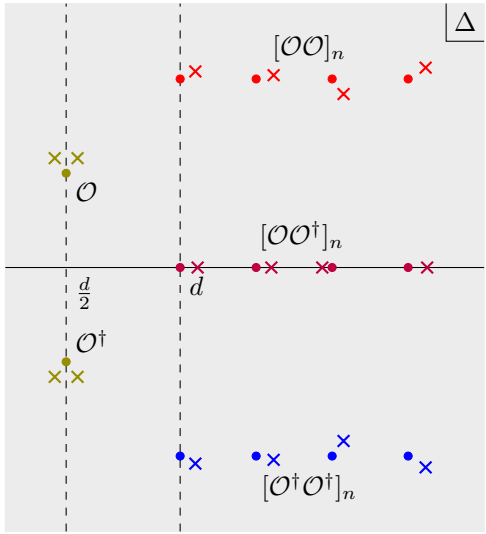}
\end{center}
\vspace{0mm}
\caption{\label{fig:poles} Analytic structure of the spectral density $I_{\Delta,\ell=0}$ in the case of a free and a weakly-coupled theory in dS. The solid circles are the locations of the poles for ``single-trace'' and ``double-trace'' operators of the dS mean field theory. The single-trace poles appear for instance in the two-point function of the bulk field. The three families of double-trace poles are visible in different correlators, namely $\expec{\Oo \Oo \Oo \Oo}$, $\expec{\Oo \Oo^\dagger \Oo \Oo^\dagger}$ and $\expec{\Oo^\dagger \Oo^\dagger \Oo^\dagger \Oo^\dagger}$. After turning on interactions, the locations of the poles shifts, indicating that boundary operators pick up anomalous dimensions. These shifted poles are shown as crosses in the figure. Of course, new poles may appear too.
}
\end{figure}

So far, we have considered the spectral decomposition of the correlator $\expec{\Oo \Oo^\dagger \Oo^\dagger \Oo}$, where $\Oo$ and $\Oo^\dagger$ were boundary operators with scaling dimensions $d/2 \pm i \mu$. This led to the spectral density $I_{\Delta,\ell}^\mrm{MFT}$ from Eq.~\reef{eq:IMFT}.\footnote{This process is a bit delicate since there are spurious poles at $\D=d+\ell+\mathbb{N}$. To distinguish these poles from physical poles, it is better to start from~\reef{eq:generalIMFT} where there exists poles at $\D=\D_1+\D_2+\ell+2\mathbb{N}$ and then take the limit of $\D_2\to\D_1^*=\hd-i\mu$. }  Closing the contour and picking up poles in the $\Delta$-plane, we find that the $x_{12} \to 0$ OPE limit of $\expec{\Oo \Oo^\dagger \Oo^\dagger \Oo}$ is governed by boundary operators with dimension
\bsub
\label{eq:MFTspec}
\beq
\Delta = d + \ell + 2\mbb{N},
\quad
\ell = 0,1,2,\ldots.
\eeq
Had we instead considered the correlators $\expec{\Oo \Oo \Oo \Oo}$ or $\expec{\Oo^\dagger \Oo^\dagger \Oo^\dagger \Oo^\dagger}$, using~\reef{eq:generalIMFT} we would have instead found double-trace operators with dimensions
\beq
d + 2i \mu + \ell + 2\mbb{N}
\quad
\text{resp.}
\quad
d - 2i \mu + \ell + 2\mbb{N}~.
\eeq
\esub
The locations of these three families of poles are depicted in figure~\ref{fig:poles}.

The above picture must be modified in interacting theories. If one can construct a QFT in dS${}_{d+1}$ that is controlled by a small coupling $\lambda \ll 1$, we expect that its spectrum is close to~\reef{eq:MFTspec}, up to corrections of order $\lambda$ (or $\lambda^2$, depending on the operator and interaction in question). Let us denote the dimensions of some boundary operator $\Oo_\alpha$ as $\Delta_\alpha(\lambda)$, such that $\Delta_\alpha(0) = \Delta_\alpha^\mrm{MFT}$. The shifting of poles is shown in figure~\ref{fig:poles}. 
We can ask how this behavior can be reproduced from perturbation theory. Including interactions, a general four-point function is modified according to
\bsub
\beq
\expec{\Oo_1\Oo_2 \Oo_3 \Oo_4}_\lambda = \expec{\Oo_1 \dotsm \Oo_4}_\mrm{MFT} + \lambda \mca{A}(x_1,\ldots,x_4) + \mrm{O}(\lambda^2)
\eeq
for some diagram $\mca{A}(x_1,\ldots,x_4)$, or by passing to the spectral representation
\beq
I_{\Delta,\ell}(\lambda) = I_{\Delta,\ell}^\mrm{MFT} + \lambda I_{\Delta,\ell}^\mca{A} + \mrm{O}(\lambda^2)~.
\eeq
\esub
Now suppose that the full spectral density $I_{\Delta,\ell}(\lambda)$ has a simple pole at $\Delta = \Delta_\alpha(\lambda)$ with residue $s(\lambda)$. Expanding around $\lambda = 0$, we then must have
\beq
\frac{s(\lambda)}{\Delta - \Delta_\alpha(\lambda)} = \frac{s(0)}{\Delta - \Delta_\alpha^\mrm{MFT}} + \lambda \left[ \frac{s'(0)}{\Delta - \Delta_\alpha^\mrm{MFT}}  +  \frac{s(0) \Delta'_\alpha(0)}{(\Delta - \Delta_\alpha^\mrm{MFT})^2}  \right]  + \mrm{O}(\lambda)^2.
\eeq
In particular, a double pole in the spectral density $I_{\Delta,\ell}^\mca{A}$ signifies the fact that $\Oo_\alpha$ has an anomalous dimension already at order $\lambda$.

To give an example of this phenomenon, let us consider $\phi^4$ theory in dS${}_{d+1}$. Despite the extensive literature calculating Witten diagrams in AdS (starting with~\cite{Liu:1998ty,Freedman:1998bj}), the knowledge of late-time correlators in dS has been primitive until recent years. A recent series of papers~\cite{Sleight:2019hfp,Sleight:2020obc,DiPietro:2021sjt} has shed light on the relation between tree-level diagrams in AdS and dS. For the case at hand, let us rewrite~\cite[(3.21)]{Sleight:2019mgd}, which states that for a general dS contact diagram
\bsub
\beq
\expec{\Oo_1(x_1) \Oo_2(x_2) \Oo_3(x_3)\Oo_4(x_4)}_\text{contact} =  -2\sin(\frac{\pi}{2} \zeta) \;  \prod_{i=1}^4 b_{i} \; D_{\Delta_1 \Delta_2 \Delta_3 \Delta_4}(x_i)
\eeq
where
\beq
\Delta_i = \hd + i \mu_i
\qaq
\zeta = d + i (\mu_1 + \ldots + \mu_4)~.
\eeq
and $b_i$ is given by~\reef{eq:anticommutator of o od} with $\mu=\mu_i$.
The special function that appears here,
\beq
D_{\Delta_1 \Delta_2 \Delta_3 \Delta_4}(x_1,\ldots,x_4) = \int_0^\infty\!\frac{dz}{z^{d+1}}\int_{\mbb{R}^d}d^dy\, \prod_{i=1}^4 \left(\frac{z}{z^2 + |y-x_i|^2} \right)^{\Delta_i}~,
\eeq
\esub
represents a contact diagram in Euclidean AdS known as the $D$-function.  For definiteness, let us compute the leading correction to the four-point function
\begin{equation}\label{eq:oootot}
\langle \mathcal{O}(x_1)\mathcal{O}(x_2)\mathcal{O}^\dagger(x_3)\mathcal{O}^\dagger(x_4)\rangle
\end{equation}
which according to~\reef{eq: positivty} has a positive spectral density. This is an example of a correlator of the above type, with $\mu_1 = \mu_2 = \mu$ and $\mu_3 = \mu_4 = - \mu$, such that $\zeta = d$ in the phase factor $\sin(\tfrac{\pi}{2}\zeta)$. Moreover, the $D$-function has a known spectral representation~\cite{Zhou:2018sfz}. Using these facts, we conclude that
\begin{multline}
\label{eq:contactFull}
I_{\Delta,\ell}^\text{contact} =  - \sin(\pi \thd)  \frac{\Gamma(\hd\pm i\mu)\Gamma(\pm i\mu)}{8\pi^{d+2}} \times \\
 \Gamma\!\left(\frac{\Delta}{2} \pm i \mu \right)\Gamma\!\left(\frac{d-\Delta}{2} \pm i \mu \right) \frac{\Gamma\!\left(\frac{\Delta}{2}\right)^2\Gamma\!\left(\frac{d-\Delta}{2}\right)^2}{\Gamma(\thd-\Delta)\Gamma(\Delta-\thd)}\, \delta_{\ell,0}~.
\end{multline}
Here we use the shorthand notation of $\Gamma(a\pm b):= \Gamma(a+b)\Gamma(a-b)$.
Interestingly, the above analysis seems to indicate that the diagram in question vanishes identically when $d$ is even.

In order to read off the physical content of the partial wave coefficient $I_{\Delta,0}^\mrm{contact}$, one has to multiply $I_{\Delta,\ell}^\text{contact}$ by the coefficient $K_{\tilde{\Delta},0}^{\Delta_3,\Delta_4}$, see for instance Eq.~\reef{eq:JP11}. In the $s$-channel, corresponding to $\Oo \times \Oo \to \Oo^\dagger \times \Oo^\dagger$, we find that the physical poles are at $\Delta = d \pm 2 i\mu + 2\mbb{N}$:
\beq
K_{\tilde{\Delta},0}^{\hd - i\mu,\hd-i\mu} \, I_{\Delta,0}^\text{contact} \; \limu{\Delta \to d \pm 2i\mu + 2n} \; \frac{\rho^\pm_n}{\Delta-d\mp 2i\mu -2n}.
\eeq
Since these are single poles, they do not have an interpretation of giving rise to anomalous dimensions: instead, they mean that the boundary OPE coefficients $c_{\Oo \Oo [\Oo^\dagger \Oo^\dagger]_{n,0}}$ and their counterparts with $\Oo \lra \Oo^\dagger$ are generated at order $\lambda$. In the cross-channel, corresponding to the exchange $\Oo \times \Oo^\dagger \to \Oo \times \Oo^\dagger$, we find both double and single poles at $\Delta = d + 2\mbb{N}$:
\beq
\label{eq:ipp}
K_{\tilde{\Delta},0}^{\hd + i\mu,\hd-i\mu} I_{\Delta,0}^\text{contact} \; \limu{\Delta \to d+2n} \; \frac{\sigma_n}{\Delta - d-2n} + \frac{\tau_n}{(\Delta -d-2n)^2},
\quad
n=0,1,2,\ldots
\eeq
but there are no other physical poles present. This indicates that the double-trace operators $[\Oo \Oo^\dagger]_{n,0}$ with spin $\ell = 0$ and dimension $\Delta = d+2\mbb{N}$ have their scaling dimension corrected at tree level. The presence of a single pole in~\reef{eq:ipp} indicates that their residues, i.e.\@ the OPE coefficients $c_{\Oo \Oo^\dagger [\Oo \Oo^\dagger]}^2$, also get renormalized.

\section{Setting up the QFT in dS Bootstrap}\label{sec:Bootstrpping boundary correlators}

The   CFT four-point functions enjoy crossing symmetry. In other words, the four-point function is invariant under permutations of the external operators. The conformal block in each channel do not transform trivially under these permutations. This results in a non-trivial set of equations called crossing equations. This is the basic idea behind the conformal bootstrap program \cite{Rattazzi:2008pe,Poland:2016chs}. In this section we, consider the  four-point functions with the identical operators and set up a set of bootstrap equations for their conformal partial wave expansions.
Let us see how the same philosophy works for QFTs in de Sitter

Consider the four-point function of late-time boundary  real operators
\beq
\langle \mathcal{O}(x_1) \mathcal{O}(x_2)  \mathcal{O}(x_3) \mathcal{O}(x_4) \rangle \,.
\eeq
Note that they do not necessarily need to belong to complementary series as they are boundary operators and are not subject to unitary irreps of the dS isometry group. Out of 24 permutations of partial wave expansions for scalar operators, there are 3 equivalence classes. This can be checked from explicit expression of partial waves in \eref{PWBlock}. We choose the channels $s,t$ and $u$ as representatives of these equivalence classes. Hence, we end up with two sets of non-trivial crossing equations
\begin{equation}\label{crossing}
\begin{split}
\sum_\ell \int^{\frac{d}{2}+i\infty}_{\frac{d}{2}}  \frac{d\Delta}{2\pi i} \; I_{\Delta,\ell} \;\Psi^s_{\D,\ell}(x_i) + D^s(x_i) = \sum_\ell \int^{\frac{d}{2}+i\infty}_{\frac{d}{2}}  \frac{d\Delta}{2\pi i} \; I_{\Delta,\ell} \;\Psi^t_{\D,\ell}(x_i) + D^t(x_i)~, \\
\sum_\ell \int^{\frac{d}{2}+i\infty}_{\frac{d}{2}}  \frac{d\Delta}{2\pi i} \; I_{\Delta,\ell} \;\Psi^s_{\D,\ell}(x_i) + D^s(x_i) = \sum_\ell \int^{\frac{d}{2}+i\infty}_{\frac{d}{2}}  \frac{d\Delta}{2\pi i} \; I_{\Delta,\ell} \;\Psi^u_{\D,\ell}(x_i) + D^u(x_i)~,
\end{split}
\end{equation}
where $D^{j}(x_i)$ is the contribution from the vacuum state in channel $j$:
\begin{align}
&D^{s}(x_i) = \frac{1}{x_{12}^{2\Dp}x_{34}^{2\Dp}}~, \qquad
D^{t}(x_i) = \frac{1}{x_{23}^{2\Dp}x_{14}^{2\Dp}}~, \qquad 
D^{u}(x_i) = \frac{1}{x_{13}^{2\Dp}x_{24}^{2\Dp}}~,
\end{align}
and we defined the $s,t$ and $u$ channel partial waves as follows
\begin{equation}\label{eq:channel labels}
\begin{split}
&\Psi^s_{\D,\ell}(x_i) = \Psi^{\Dp}_{\D,\ell}(x_1,x_2,x_3,x_4) \\
&\Psi^t_{\D,\ell}(x_i) = \Psi^{\Dp}_{\D,\ell}(x_3,x_2,x_1,x_4) 
\\ &\Psi^u_{\D,\ell}(x_i) = \Psi^{\Dp}_{\D,\ell}(x_1,x_3,x_2,x_4)~.
\end{split}
\end{equation}
As discussed in section~\ref{sec:FPF}, the partial wave expansion is derived by inserting a complete set of states in the four-point function and the unitarity of the bulk theory puts positivity constraints on  partial wave coefficients.\footnote{The constraints are more general for mixed correlators. The conformal bootstrap approach to mixed correlators has been studied in great detail, see for example   eq.~(2.10) of~\cite{Kos:2014bka}. A similar approach can be taken here by considering the analogy between $\mathcal{F}_{12}(\Delta,\ell)$ in~\eref{eq:F of three point} and the OPE coefficients $\lambda_{12\mathcal{O}}$ in the usual conformal bootstrap. However, one should be also careful about the fact that the operators are not necessarily commuting when they are different operators. This has been seen  in the simple case of a free theory in the previous section. We avoid such issue by concentrating on identical boundary operators in this section.} 
For simplicity, from now on we focus on
QFT on dS$_2$, \emph{i.e.} we take $d=1$. 
This has the important advantage of removing the infinite sums over spin $\ell$.
However, it forces us to take into account discrete series irreps of $SO(2,1)\cong SL(2,\Real)$~\cite{Simmons-Duffin:2017nub,Maldacena:2016hyu,Mazac:2018qmi,Repka:1978}. This is what we explain next.
We plan to extend the analysis to higher dimensions in the future.

\subsection{Review of CFT$_{1}$}\label{sec:CFT1}
We shall proceed with reviewing some basics of $d=1$ conformal partial waves similar to what we did in section \ref{sec:FPF}. In this section, we present the results for a generic scalar four-point function. In section~\ref{sec:Regularized Crossing}, we consider identical operators by setting $\D_1=\D_2=\D_3=\D_4=\D_{\mathcal{O}}$.
The four-point function, after stripping out the appropriate scaling factors, is
\begin{gather}\label{eq:CFT1FourPoint}
\langle \mathcal{O}_1(x_1) \mathcal{O}_2(x_2) \mathcal{O}_3(x_3) \mathcal{O}_4(x_4)\rangle = \frac{1}{|x_{12}|^{\D_{12}}|x_{34}|^{\D_{34}}} \left|\frac{x_{14}}{x_{24}}\right|^{\delta_{21}}\left|\frac{x_{14}}{x_{13}}\right|^{\delta_{34}}\mathcal{G}(z)
\end{gather}
with a single cross ratio  
\begin{gather}
 z=\frac{x_{12}x_{34}}{x_{13}x_{24}} \in \Real~,
\end{gather}
where we used $x_{ij}=x_i-x_j$, $\D_{ij}=\D_i+\D_j$ and $\delta_{ij}=\D_i-\D_j$. 
$\mathcal{G}(z)$ is singular at $z=0,1,\infty$ corresponding to coincident points. 

Let us expand the correlator $\mathcal{G}(z)$ in a complete set of eigenfunctions of the Casimir operator, orthogonal with respect to inner product~\cite{Mazac:2018qmi,Maldacena:2016hyu}
\begin{equation}\label{eq:innerprod} 
(f,g)= \int^\infty_{-\infty}dz z^{-2}f(z)g(z)~.
\end{equation}
These are the conformal partial waves introduced in the previous section. 
However, for $d=1$ the complete basis includes both principal and discrete series  ($\Delta \in \mathbb{N}$) with both parities, which we denote by spin $\ell\in \{0,1\}$~\cite{Simmons-Duffin:2017nub}.
These obey  the orthogonality relations
\begin{align}
(\Psi_{\half+ i\alpha,\ell} , \Psi_{\half+ i\beta,\ell^\prime}) &= 2\pi n_{\D,\ell} \; \delta_{\ell \ell^\prime} \delta(\alpha-\beta) &&\alpha,\beta \in \Real_{+}~,\\
(\Psi_{m,\ell},\Psi_{n,\ell^\prime})&= \frac{4\pi^2}{2m-1} \delta_{\ell \ell^\prime} \delta_{mn}   &&m,n \in \mathbb{N}~,
\end{align}
with vanishing inner product between partial waves in the discrete and principal series. Notice that in this equation $\delta$ is the Kronecker delta. The normalization factor $n_{\D,\ell}$ will be given below.
Using this basis, we can write the $s$-channel decomposition
\begin{equation}
\label{eq:PWE}
\mathcal{G}(z) = \sum_{\ell=0,1}\int_0^\infty \frac{d\nu}{2\pi} I^s_{\frac{1}{2}+i\nu,\ell} \Psi_{\half+i \nu,\ell}(z) 
+  \sum_{\substack{n\in\mathbb{N} \\ \ell=0,1}} \tilde{I}^s_{n,\ell} \Psi_{n,\ell}(z)\,,
\end{equation}
that replaces \eqref{4p PWE} in $d=1$. Here we use the superscript $s$ for partial wave coefficients to emphasis that this is an expansion in $s$-channel. 

The partial waves are given by integrals of the product of three-point structures as in \eqref{partial wave} and the stripped version can be found using~\reef{eq:x to z psi translation}.
More precisely, for $\ell=0$ we have 
\begin{align}
\nonumber
\Psi_{\D,0}(z) &= \left|\frac{x_{14}}{x_{24}}\right|^{\delta_{12}}\left|\frac{x_{14}}{x_{13}}\right|^{\delta_{43}} \int^\infty_{-\infty} dx_5 \frac{\abs{x_{12}}^\D}{\abs{x_{15}}^{\D+\delta_{12}}\abs{x_{25}}^{\D-\delta_{12}}}  \frac{\abs{x_{34}}^{1-\D}}{\abs{x_{35}}^{1-\D+\delta_{34}}\abs{x_{45}}^{1-\D-\delta_{34}}}
\\&= |z|^\D \int^\infty_{-\infty} dx \; \frac{|x-1|^{\D-1-\delta_{34}}}
{|x-z|^{\D-\delta_{12}} |x|^{\D+\delta_{12}}}~,
\label{eq:1dPartialEven}
\end{align}
where in the second line, we fixed the conformal gauge by setting $x_1=0$, $x_2=z$, $x_3=1$, $x_4=\infty$ and $x_5=x$. In the case $\ell=1$, the three-point structure has an extra numerator $Z$ that can be derived from the higher dimensional scalar-scalar-spin-$\ell$ correlator in \eref{three point structure}
\begin{equation}\label{eq:3ptOdd1D}
\langle \mathcal{O}_1(x_1) \mathcal{O}_2(x_2) \mathcal{O}_3(x_3)\rangle = \frac{Z}{x_{12}^{\D_1+\D_2-\D_3}x_{23}^{\D_2+\D_3-\D_1}x_{13}^{\D_1+\D_3-\D_2}}~,
\end{equation}
with  $Z=\frac{\abs{x_{13}}\abs{x_{23}}}{\abs{x_{12}}}(\frac{1}{x_{13}}-\frac{1}{x_{23}})= -\operatorname{sgn}(x_{13})\operatorname{sgn}(x_{23})\operatorname{sgn}(x_{12})$. Then one finds 
\begin{align}
\Psi_{\D,1}(z) &= \left(\frac{x_{14}}{x_{24}}\right)^{\delta_{12}}\left(\frac{x_{14}}{x_{13}}\right)^{\delta_{43}}  
\int^\infty_{-\infty} dx_5 \frac{\abs{x_{12}}^\D}{\abs{x_{15}}^{\D+\delta_{12}}\abs{x_{25}}^{\D-\delta_{12}}}  \frac{\abs{x_{34}}^{1-\D}}{\abs{x_{35}}^{1-\D+\delta_{34}}\abs{x_{45}}^{1-\D-\delta_{34}}} \Theta \nonumber
\\
\label{eq:1dPartialOdd}
&= |z|^\D \int^\infty_{-\infty} dx \; \frac{|x-1|^{\D-1-\delta_{34}}}
{|x-z|^{\D-\delta_{12}} |x|^{\D+\delta_{12}}} \operatorname{sgn} (x z (x-1) (z-x))~.
\end{align}
where $\Theta \equiv \operatorname{sgn} (x_{12}x_{15}x_{25} x_{34}x_{35}x_{45})$.

For $z\in(0,1)$, the partial waves with $\Delta$ on the principal series can be written as a linear combination of a conformal block and its shadow
\begin{equation}\label{eq:PsiKG}
\Psi_{\D,\ell}(z)=K^{\D_3,\D_4}_{1-\Delta,\ell} \; G_{\Delta,\ell}(z) + K^{\D_1,\D_2}_{\Delta,\ell} \; G_{1-\Delta,\ell}(z)~,
\end{equation}
where  
\begin{align}\label{eq:CB1d}
G_{\D,\ell}(z) &= (-1)^\ell z^{\Delta } \, _2F_1(\Delta-\delta_{12} ,\Delta+ \delta_{34}; 2\Delta ;z)~,\\
K^{\D_1,\D_2}_{\Delta,\ell} &= \frac{\sqrt{\pi}\Gamma(\D-\frac{1}{2})\Gamma(\D+\ell-1)}{\Gamma(\D-1)\Gamma(1-\D+\ell)} \frac{\Gamma(\frac{1-\D+\D_1-\D_2+\ell}{2})\Gamma(\frac{1-\D+\D_2-\D_1+\ell}{2})}{\Gamma(\frac{\D+\D_1-\D_2+\ell}{2})\Gamma(\frac{\D+\D_2-\D_1+\ell}{2})}~.
\end{align}
One way to find these expressions is to perform integrals~\eref{eq:1dPartialEven} and~\eref{eq:1dPartialOdd} explicitly. 
Alternatively, one can  set $d=1$ in the general formula \eqref{PWBlock}. For integer $\Delta$, corresponding to the discrete series, we have instead
\beq
n \in \mbb{N}:
\quad
\Psi_{n,\ell}(z) =  K^{\Delta_3,\Delta_4}_{1-n,\ell}\,  G_{n,\ell}(z)~.
\eeq

Finally, we would like to show that 
\begin{equation}
n_{\D,\ell} = \frac{4\pi \tan (\pi\D)}{2\D-1}~.
\end{equation}
 As it is stated in~\cite{Simmons-Duffin:2017nub}\footnote{There is a slight difference in notations: $I^{\text{here}}=I^{\text{there}} n^{\text{there}}$, $K^{\text{here}}=S^{\text{there}}=(-2)^J K^{\text{there}}$ but  $n^{\text{here}}=n^{\text{there}}$.}, $n_{\D,\ell}$ in general dimension $d$ can be written as 
 \begin{equation}
n_{\D,\ell} = \frac{\text{vol}(S^{d-2})(2\ell+d-2) \Gamma(\ell+d-2)}{\text{vol}(SO(d-1))} \frac{\pi\Gamma(\ell+1)}{2^{2\ell+d-2}\Gamma(\ell+\hd)^2} \; K^{\D_3,\D_4}_{\tilde{\D},\ell} K^{\tilde{\D}_3,\tilde{\D}_4}_{\D,\ell}~.
\end{equation}
In order to take the limit $d \to 1$ of this expression, we shall analytically continue in $d$ using the recursion relation
\begin{equation}
\text{vol}(SO(d))=\text{vol}(S^{d-1})\,\text{vol}(SO(d-1))~,
\end{equation}
and the fact that $\text{vol}(SO(2))=\text{vol}(S^1)=2\pi$.
This leads to the formal results $\text{vol}(SO(1))=1$ and  $\text{vol}(SO(0))=\half$.
Therefore, we find $\lim_{d\to 1} n_{\D,\ell}  =0 $ for all $\ell\ge 2$.
On the other hand,  we find
\begin{equation}
\lim_{d\to 1} n_{\D,0} = \lim_{d\to 1} n_{\D,1} =  \frac{4\pi \tan (\pi\D)}{2\D-1}~.
\end{equation}

\subsection{A toy example: almost MFT}
We would like to understand the convergence properties of the partial wave decomposition \eqref{eq:PWE}. This is very important for the goal of developing a numerical bootstrap approach to QFT in dS. 
With this in mind, let us consider the example of a weakly coupled massive scalar  field in dS$_2$. In this case, we expect boundary operators almost on the principal series, \emph{i.e.} $\D_\mathcal{O} = \D_{re}+i\D_{im}$ with $\D_{re} -\half \ll 1$. On the other hand, the imaginary part $\D_{im}$ can be large because it is related to the mass of the bulk field via 
$m^2 R^2 =\frac{1}{4} + \D_{im}^2 $, if we turn off interactions.

The disconnected part of  four-point function  $\langle \mathcal{O}_1 \mathcal{O}_2^\dagger \mathcal{O}_3 \mathcal{O}_4^\dagger \rangle _{\rm disc} = 
\langle \mathcal{O}_1  \mathcal{O}_3 \rangle \langle \mathcal{O}_2^\dagger \mathcal{O}_4^\dagger \rangle$ gives:\footnote{In the case of a single real operator $\mathcal{O}=\mathcal{O}^\dagger$, there are two more contributions from other channels. The (stripped) four-point function for identical external operators reads
\[
\mathcal{G}_{\rm disc}(z)= 1 + \left|z\right|^{2\D_\OO} + \left|\frac{z}{z-1}\right|^{2\D_\OO}~.
\]
The first term ($=1$, from the s-channel) is non-normalizable with respect to the inner product~\eref{eq:innerprod}. The spectral density $I^{(3)}_{\Delta,\ell}$ corresponding to the third term is equal to the density $I^{(2)}_{\Delta,\ell}$ up to a factor $(-1)^\ell$. This is a consequence of  the behavior of the partial waves under $z \mapsto z/(z-1)$.}
\begin{equation}
\mathcal{G}_{\rm disc}(z)= \left|z\right|^{\D_\OO+\D_\OO^*}= \left|z\right|^{2\D_{re}} ~.
\label{Gdisc}
\end{equation}
Notice that if $\D_{re} \neq \frac{d}{2}$ the  local terms discussed in section~\ref{sec:Local} are not allowed in the two-point function $\langle \OO \OO^\dagger \rangle$.
Using the orthogonality relations of $\Psi_\D$, one is able to calculate the partial wave coefficients. The basic integral to compute is the following
\bsub
\begin{align}
W_{\D,\ell}&=  \int^\infty_{-\infty} \frac{dz}{z^2} \; \mathcal{G}_\text{disc}(z) \Psi_{\D,\ell}(z) 
\\
&=  \int^\infty_{-\infty} dx \; \frac{|x-1|^{\D-1-2i\D_{im}}}{|x|^{\D+2i\D_{im}}} 
\int^\infty_{-\infty} \frac{dz}{z^2} \; \frac{|z|^{\D+2\D_{re}}}{|x-z|^{\D-2i\D_{im}}} \Scale[0.9]{\left(\delta_{\ell,0}+\delta_{\ell,1} \operatorname{sgn} (x z (x-1) (z-x))\right)}.
\end{align}
\esub
This integral can be done explicitly:\footnote{In practice, we divide the integration domain in  9 regions according to the position of $x$ with respect to $0$ and $1$ and the position of $z$ with respect to $0$ and $x$.}  
\begin{multline}
W_{\D,\ell}= \frac{2^{\ell} \sqrt{\pi} \Gamma(\ell+\half)}{\Gamma(\ell+1)}
\frac{\Gamma(\half + i\D_{im} - \D_{re})\Gamma(\half- i\D_{im} - \D_{re})}{\Gamma(\D_{re}+i\D_{im}) \Gamma(\D_{re}-i\D_{im})} \\
\times \frac{\Gamma(\frac{\ell-\D+2\D_{re}}{2})\Gamma(\frac{\ell-1+\D+2\D_{re}}{2})}{\Gamma(\frac{1+\ell+\D-2\D_{re}}{2})\Gamma(\frac{2+\ell-\D-2\D_{re}}{2})}~.
\end{multline}
Then, the principal series partial wave coefficients are given by 
\begin{align}
I^{\text{disc}}_{\D,\ell}= \frac{1}{n_{\D,\ell}} W_{\D,\ell}&=\frac{2^{\ell-2}\Gamma(\ell+\half)}{\sqrt{\pi}\Gamma(\ell+1)}
\frac{\Gamma(\half + i\D_{im} - \D_{re})\Gamma(\half- i\D_{im} - \D_{re})}{\Gamma(\D_{re}+i\D_{im}) \Gamma(\D_{re}-i\D_{im})} 
\frac{ (2\D-1)}{\tan (\pi \D)  }
\nonumber\\&\times 
\frac{\Gamma(\frac{\ell-\D+2\D_{re}}{2})\Gamma(\frac{\ell-1+\D+2\D_{re}}{2})}{\Gamma(\frac{1+\ell+\D-2\D_{re}}{2})\Gamma(\frac{2+\ell-\D-2\D_{re}}{2})}~,
\end{align}
for $\Delta=\half+i\nu$ and $\nu>0$,  and the discrete series by 
\begin{align}
\tilde{I}^\text{disc}_{n,\ell} &=\frac{2n-1}{4\pi^2} W_{n,\ell} = \frac{2^{\ell-2}  \Gamma(\ell+\half)}{\pi^\frac{3}{2} \Gamma(\ell+1)}
\frac{\Gamma(\half + i\D_{im} - \D_{re})\Gamma(\half- i\D_{im} - \D_{re})}{\Gamma(\D_{re}+i\D_{im}) \Gamma(\D_{re}-i\D_{im})} 
\nonumber\\&\times 
\frac{(2n-1)\Gamma(\frac{\ell-n+2\D_{re}}{2})\Gamma(\frac{\ell-1+n+2\D_{re}}{2})}{\Gamma(\frac{1+\ell+n-2\D_{re}}{2})\Gamma(\frac{2+\ell-n-2\D_{re}}{2})}~.
\end{align}
Notice that $I^{\text{disc}}_{\D,\ell}$
 is shadow symmetric (\emph{i.e} invariant under $\Delta \to 1-\Delta$) and has poles on the real line at $\D \in \mathbb{Z}$ and $\D=2\D_{re}+2k+\ell$ for $k\in\mathbb{N}$ and their shadow. 
The attentive reader may worry that these partial wave coefficients do not satisfy the unitarity condition $I_{\half+i\nu,\ell}(-1)^\ell \ge 0$ for $\nu \in \mathbb{R}$.
The obvious solution is that $I^\text{disc}_{\half+i\nu,\ell}$ is different from the full $I_{\half+i\nu,\ell}$. Nevertheless, it would be useful to better understand the emergence of the free theory in dS, described in section \ref{Examples4pt}, as the limit of an interacting QFT in dS.

 Let us go back  to \eref{eq:PWE} and use \eqref{eq:PsiKG} to write,
\begin{align}
 \label{eq:repeatPWE}
\mathcal{G}_\text{disc}(z) &= \sum_{\ell=0,1}\int_{\half }^{\half+i\infty}  \frac{d\D}{2\pi i} I_{\D,\ell}^\text{disc} \Psi_{\D,\ell}(z) 
+  \sum_{\substack{n\in\mathbb{N} \\ \ell=0,1}} \tilde{I}_{n,\ell}^\text{disc} \Psi_{n,\ell}(z)
\\
&=\sum_{\ell=0,1} \int_{\half-i\infty}^{\half+i\infty} \frac{d\D}{2\pi i} I^\text{disc}_{\D,\ell} K_{1-\D,\ell} G_{\D,\ell}(z) 
+ \sum_{\substack{n\in\mathbb{N} \\ \ell=0,1}} \tilde{I}^\text{disc}_{n,\ell} \Psi_{n,\ell}(z)~.
\nonumber
\end{align}
Now, we can deform the $\D$-contour to the right and pick up residues of the poles on the positive real line. The poles at integer $\D$ precisely cancel the contribution from the discrete series because $\tilde{I}^\text{disc}_{n,\ell} = 
{\rm Res}_{\Delta=n} I^\text{disc}_{\D,\ell}$.
We are left with the contribution of the poles at $\D=2\D_{re}+2k+\ell$ for $k\in\mathbb{N}$, 
\begin{align}
\mathcal{G}^\text{disc}(z) =\left|z\right|^{2\D_{re}} &= -\sum_{\ell=0,1} \sum_{k=1}^\infty \operatorname{Res}_{\D= 2\D_{re}+\ell+2k}(I_{\D,\ell}^\text{disc}) K_{1-(2\D_{re}+\ell+2k),\ell} G_{2\D_{re}+\ell+2k,\ell} (z) \label{CBsum} \nonumber
\\
&\rdef \sum_{\ell=0,1}\sum_{k=1}^\infty c^2_{\tiny{\OO \OO^\dagger [\Od\OO]_{k,\ell}} }G_{2\D_{re}+\ell+2k,\ell} (z).
\end{align}
The second line defines OPE coefficients $c^2_{\tiny{\OO \OO^\dagger \OO^\dagger [\Od\OO]_{k,\ell}} }$. The latter  must be positive because the double-trace exchanged operators $ [\Od\OO]_{k,\ell}$ are hermitian.

\subsubsection*{(Non)-Convergence}
Although the sum \eqref{CBsum} converges for any external dimension $\D_\OO=\D_{re}+i\D_{im}$, the integral~\eref{eq:repeatPWE} is not always convergent. 
Let us take a closer look at this issue.
We need to study the asymptotic behavior of partial waves $\Psi$ and the associated coefficients $I$. Using  Stirling's approximation, 
\begin{equation}
I^{\text{disc}}_{\half+i\nu,\ell} 
\limu{\nu \to \infty}
Q \nu^{4\D_{re}-1}\,,\qquad 
\tilde{I}^{\text{disc}}_{n,\ell} 
\limu{n \to \infty}
\label{Ilargenu}  \frac{Q}{\pi} (-1)^{\ell+n} n^{4\D_{re}-1}~,
\end{equation}
where we defined
\be
Q\equiv \frac{\Gamma(\half-\D_{re}-i\D_{im})\Gamma(\half-\D_{re}+i\D_{im})}{2^{4\D_{re}-1}\Gamma(\D_{re}+i\D_{im})\Gamma(\D_{re}-i\D_{im})} ~.
\end{equation}
Using \eqref{eq:PsiKG} and  the 
  known large $\D$ behavior of   conformal blocks~\cite{Rattazzi:2008pe,Hogervorst:2013sma},
 \beq
 G_{\Delta,\ell}(z)\limu{\Delta \to \infty} (-1)^\ell \frac{(4\rho)^{\D}}{\sqrt{1-\rho ^2}}~,
 \eeq
 one can find the asymptotic behavior of the partial waves:
\begin{align}
\Psi_{\half+ i \nu,\ell}(z) &\limu{\nu \to \infty} 2(-1)^\ell \sqrt{\frac{\pi}{\nu}} \frac{(4\rho)^{\half}}{\sqrt{1-\rho ^2}} \cos(x\,\nu - \frac{\pi}{4})~, \\
\Psi_{n,\ell}(z) &\limu{n \to \infty} 2\sqrt{\frac{\pi}{n}} \frac{(-1)^\ell+(-1)^{n}\cosh(2\pi \D_{im})}{\sqrt{1-\rho ^2}} \rho^{n}~,
\end{align}
where we used the $\rho$-coordinate defined as  
\begin{equation}
\rho(z)= \frac{z}{\left(\sqrt{1-z}+1\right)^2}~, \qquad x=\log(\rho(z))~.
\end{equation}
Finally, the large $\nu$ behavior of the integrand in~\eref{eq:repeatPWE} is
\begin{equation}
\sim \nu^{4\D_{re}-\frac{3}{2}} \cos(x\,  \nu -\frac{\pi}{4})~,
\end{equation}
which means the integral is not convergent for $\D_{re}>\frac{1}{8}$.\footnote{One way to make this integral convergent is to introduce a Gaussian regulator $e^{-\epsilon \nu^2}$ with $\epsilon \to 0$. }
On the other hand, the structure is somewhat familiar. This is like the Fourier transform  of a monomial and it corresponds to the behaviour $\sim |x|^{\half-4\D_{re}}$ as  $x\to 0$.
Notice that $x \to 0$ corresponds to $z \to 1$ or equivalently $x_2 \to x_3$, which is the $t$ channel OPE limit. In fact, 
it is instructive to compute the behavior as $z\to 1$ of each term in \eref{eq:repeatPWE}. Using~\cite{Fitzpatrick:2012yx,Komargodski:2012ek} 
 \beq
 G_{\Delta,\ell}\left(z\right)\limu{  \Delta \to \infty} (-1)^\ell 4^\D \sqrt{\frac{\D}{\pi}}K_0(2\D\sqrt{1-z}) ~\qquad
 \text{when}\,\,\,~(1-z)^{-\half}\sim \Delta~,
\eeq
we find
\bsub
\begin{align}
 \int_{\half }^{\half+i\infty}  \frac{d\D}{2\pi i} I_{\D,\ell}^\text{disc} \Psi_{\D,\ell}(z) & 
 \limu{z\to 1} \frac{Q \cos(2\pi \D_{re}) \Gamma^2(2\D_{re})}{2\pi}
  \frac{(-1)^\ell}{(1-z)^{2\D_{re}}}
 \\
  \sum_{\substack{n\in\mathbb{N}  }} \tilde{I}_{n,\ell}^\text{disc} \Psi_{n,\ell}(z)&\limu{z\to 1} \frac{Q \cosh(2\pi \D_{im}) \Gamma^2(2\D_{re})}{2\pi}
  \frac{(-1)^\ell}{(1-z)^{2\D_{re}}}~.
\end{align}
\esub
Although every term diverges as $z\to 1$, the leading singular behavior cancels between the spin 0 and spin 1 contributions. This had to happen because the correlator $\mathcal{G}^\text{disc}(z)=|z|^{2\D_{re}}$ is regular.

Consider now the $u$ channel OPE limit $z \to \infty$. For the case $\D_{im}=0$, 
one can easily obtain the the partial waves for negative $z$ using  the symmetry:
\beq
\Psi_{\D,\ell}(z) = (-1)^\ell \Psi_{\D,\ell}\left( \frac{z}{z-1}\right)\,,\qquad\qquad z<0~.
\eeq
This gives 
\bsub
\begin{align}
 \int_{\half }^{\half+i\infty}  \frac{d\D}{2\pi i} I_{\D,\ell}^\text{disc} \Psi_{\D,\ell}(z) & 
 \limu{z\to -\infty} \frac{Q \cos(2\pi \D_{re}) \Gamma^2(2\D_{re})}{2\pi}
   (-z)^{2\D_{re}}
 \\
  \sum_{\substack{n\in\mathbb{N}  }} \tilde{I}_{n,\ell}^\text{disc} \Psi_{n,\ell}(z)&\limu{z\to -\infty} \frac{Q    \Gamma^2(2\D_{re})}{2\pi}
 (-z)^{2\D_{re}}
\end{align}
\esub
which means that every term in \eref{eq:repeatPWE} contributes to the leading divergence of $\mathcal{G}^\text{disc}(z)=|z|^{2\D_{re}}$ as $z\to \infty$.
In general, we expect   $\mathcal{G}(z) \approx \mathcal{G}^\text{disc}(z)$ as $z\to \infty$ because the identity dominates the $u$ channel OPE. 
Therefore, we expect  the full partial wave coefficients $I_{\half+i\nu,\ell}$ and  $\tilde{I}_{n,\ell}$ to scale as in \eqref{Ilargenu} for large  $\nu$ or $n$.\footnote{Note that the precise asymptotic behavior must be different to be compatible with unitarity. 
Nevertheless, we expect the same asymptotic power law behavior.}

This argument shows that the integral over the principal series in the   partial wave decomposition \eqref{eq:PWE} does not converge absolutely.   This issue poses an important obstacle to any numerical bootstrap approach.
In what follows, we will overcome this obstacle by integrating the crossing equation over $z$ against functions that vanish sufficiently fast at $z=0$ and $z=1$.

\subsection{Regularized crossing equation} \label{sec:Regularized Crossing}
In this section, we want to explore the consequences of the crossing equation~\reef{crossing}. We start with a method to regularize the convergence issue mentioned in the previous section. 
To do so, we shall use the following linear functional:
\begin{equation}\label{eq:omega}
\omega [f]  \ldef \int_0^1 dz\; z^\gamma (1-z)^\sigma f(z)~,
\end{equation}
where $\gamma$ and $\sigma$ should be large enough. One can think of $\gamma$ and $\sigma$ as two free parameters similar to  the number of derivatives in the usual numerical conformal bootstrap  that by varying them one finds a list of crossing equations. The longer the list of pairs  $(\gamma,\sigma)$,  the better the bounds become. 

Since the partial wave coefficient of all the channels of the correlator 
$$\langle \mathcal{O}(x_1)\mathcal{O}(x_2)\mathcal{O}(x_3)\mathcal{O}(x_4) \rangle$$
 are the same, the crossing equation will look like\footnote{Since we have focused on correlator with identical operators, the contribution from $\ell=1$ vanishes and the sum over discrete series only includes even integers.} 
\begin{equation}\label{eq:Crossing}
\int^\infty_0 \frac{d\nu}{2\pi} \;  I_{\half+i\nu,0} F^{s-t}_{\half+i\nu,0}(z) + \sum_{n\in2\mathbb{N}} \tilde{I}_{n,0} F^{s-t}_{n,0}(z) = z^{2\D_\OO} - (1-z)^{2\D_\OO}~,
\end{equation}
where we define
\begin{equation}
 F^{s-t}_{\D,\ell}(z) \equiv (1-z)^{2\D_{\OO}} \Psi _{\D,\ell}(z)- z^{2\D_{\OO}} \Psi _{\D,\ell}(1-z)~,
\end{equation}
and used  $\Psi^t_{\D,\ell}(z)=\Psi^s_{\D,\ell} (1-z)=\Psi_{\D,\ell}(1-z)$. 
Acting with the functional $\omega$ introduced in \eref{eq:omega} on   this equation and using the identity \eqref{eq:3F2 from 2F1}, one finds a new form of the crossing equation
\begin{equation}\label{eq:crossexample}
\int^\infty_0 \frac{d\nu}{2\pi} \;  I_{\half+i\nu,0} \tilde{F}^{s-t}_{\half+i\nu,0}+  \sum_{n\in2\mathbb{N}} \tilde{I}_{n,0} \tilde{F}^{s-t}_{n,0} + D(\gamma,\sigma)=0~,
\end{equation}
where
\be
D(\gamma,\sigma)= \frac{\Gamma (\gamma +1) \Gamma (2 \Delta_\OO +\sigma +1)-\Gamma (\sigma +1) \Gamma (2 \Delta_\OO+ \gamma +1)}{\Gamma (2 \Delta_\OO+\gamma +\sigma +2)}
\ee
and
\begin{align}
\label{eq:Fstdef}
  \tilde{F}^{s-t}_{\D,0} & = \frac{K^{\D_{\OO},\D_{\OO}}_{1-\D,0}}{\Gamma(\D+2\D_{\OO}+\gamma+\sigma+2)} \\
   	& \times \Scale[1.05]{\left[\Gamma(\D+\gamma+1)\Gamma(2\D_{\OO}+\sigma+1) \, \FFF {\D,\D,\D+\gamma+1}{2\D,\D+2\D_{\OO}+\gamma+\sigma+2}{1}-\gamma\leftrightarrow\sigma \right]} \nn \\
 & +  \;\; \D\leftrightarrow 1-\D~. \nn
\end{align}
The formula for $\tilde{F}_{n,\ell}^{s-t}$ instead reads
\begin{align}
\label{eq:Fndef}
  \tilde{F}^{s-t}_{n,0} & =\frac{K^{\D_{\OO},\D_{\OO}}_{1-n,0}}{\Gamma(n+2\D_{\OO}+\gamma+\sigma+2)} \\
   	&  \times \Scale[1.05]{\left[\Gamma(n+\gamma+1)\Gamma(2\D_{\OO}+\sigma+1)\,\FFF{n,n,n+\gamma+1}{2n,n+2\D_{\OO}+\gamma+\sigma+2}{1}-\gamma\leftrightarrow\sigma \right]}~.\nn
\end{align}
The advantage of the functional~\reef{eq:omega} is that we could compute its action on partial waves in terms of the hypergeometric function ${}_3F_2(1)$.

In appendix \ref{eq:tildefbnd}, we show that
\beq
 \tilde{F}^{s-t}_{\half+i\nu,0} \limu{\nu \to \infty} \nu^{-2-4\D_{\OO}-2\text{min}(\sigma,\gamma)}\,,
\eeq
which together with \eqref{Ilargenu} implies that the $\nu$ integral in the regularized crossing equation \eqref{eq:Crossing} is convergent as long as
\begin{equation}\label{eq:gamma sigma condition} 
\text{min}(\sigma,\gamma) > -1 ~.
\end{equation}

\subsection{An invitation to the numerical bootstrap}\label{sec: dS numerical bootstrap}
The crossing symmetry plus positivity (from unitarity) leads to bounds on the space of conformal field theories. 
In this section, following the strategy of the conformal bootstrap, we will show that the same is true for QFT in dS. 
Consider for definiteness the $s-t$ crossing equation in~\eref{eq:Crossing} that is anti-symmetric under exchange of $\gamma \leftrightarrow \sigma$. Therefore, it is sufficient to concentrate on  the case  $\gamma>\sigma$.

At this point, we can rule out putative theories by applying linear functionals to the equation~\reef{eq:crossexample}. As an example of a putative theory, assume that the spectral density obeys $I_{\hd+i\nu,0}=0$ for $|\nu|<\nu^*$.
Now, if one finds a linear functional $\alpha$ satisfying
 \begin{align}
 \alpha\left[\tilde{F}^{s-t}_{\half+i\nu,0}(\gamma,\sigma)\right]& > 0~, \qquad \text{for all $|\nu|>\nu^*$},\nonumber \\
\alpha\left[\tilde{F}^{s-t}_{n,0}(\gamma,\sigma)\right]& > 0~, \qquad \text{for all $n\in \mathbb{N}$},\label{eq:alpha conditions} \\
 \alpha\left[D(\gamma,\sigma)\right]&=1~,\nonumber
 \end{align}
then~\reef{eq:crossexample} cannot be satisfied by a unitary QFT in dS (since in a unitary QFT we must have $I_{\hd+i\nu,0}\geq 0$ and $\tilde{I}_{n,0}\geq 0$). 

One may also  find  bounds on partial wave coefficients. For example, imagine that one can find a linear functional $\alpha$  obeying the first two positivity conditions of~\reef{eq:alpha conditions}, but now $\alpha\left[D(\gamma,\sigma)\right]=-1$. Then there exists  an upper bound on every discrete series partial wave coefficient,
\begin{equation}
 \tilde{I}_{n,0}\leq \frac{1}{ \alpha\left[\tilde{F}^{s-t}_{n,0}(\gamma,\sigma)\right] }~,
\end{equation}
and this bound can be optimized by maximizing $\alpha\left[\tilde{F}^{s-t}_{n,0}(\gamma,\sigma)\right]$.
We leave for the future a systematic implementation using linear programming methods or the semidefinite solver SDPB~\cite{Simmons-Duffin:2015qma}.

We conclude this section with a \textit{proof-of-concept} example of a ruled-out theory.
Consider equation \eqref{eq:crossexample} for an external operator of dimension
$\Dp = \half\pm\frac{1}{8}$ and let $\gamma=2.1$ and $\sigma=2$. 
It turns out that $\tilde{F}^{s-t}_{\half+i\nu,0}(\gamma,\sigma)  $ is positive for all $\nu \geq \nu^*$ where  
\be
\Dp= \half + \frac{1}{8}~:\quad \nu^* = 8.53~, \qquad \Dp= \half - \frac{1}{8}~:\quad \nu^* = 12.80~,
\ee
and $\tilde{F}^{s-t}_{n,0}(\gamma,\sigma)$ is also positive for all even $n\in \mathbb{N}$.\footnote{Note that odd values of $n$ do not contribute for a four point function of identical hermitian operators because $\tilde{F}^{s-t}_{n,0}$ vanishes identically.} Imagine a theory with vanishing $I_{\hd+i\nu,0}$ for $\nu<\nu^*$. Then there is an upper bound on $\tilde{I}_{2,0}$: 
\begin{equation}\label{eq:Itilde bound}
\tilde{I}_{2,0} < \frac{ - D(\gamma=2.1,\sigma=2)}{ \tilde{F}^{s-t}_{2,0}(\gamma=2.1,\sigma=2)} \approx  \left\{
  \begin{array}{@{}ll@{}}
    6.43174, & \text{for}~~\D_\OO = \half + \frac{1}{8} \\
    2.38638, & \text{for}~~\D_\OO = \half - \frac{1}{8}
  \end{array}\right.
\end{equation} 
One can improve this bound using linear programming methods. For example, taking linear combinations with a specific set of eight  different values of $\{\gamma,\sigma\}$, we found stronger bounds:
\begin{equation}\label{eq:Itilde bound LP}
\tilde{I}_{2,0}  < \frac{-\alpha \left[D(\gamma,\sigma)\right]}{\alpha\left[\tilde{F}^{s-t}_{2,0}(\gamma,\sigma)\right]} \approx  \left\{
  \begin{array}{@{}ll@{}}
    5.67049, & \text{for}~~\D_\OO = \half + \frac{1}{8} \\
    2.18236, & \text{for}~~\D_\OO = \half - \frac{1}{8}
  \end{array}\right.
\end{equation}

We hope this simple example convinces the reader that these equations have the potential to put non-trivial bounds on the space of QFTs in dS. 
Optimistically, with a proper systematic treatment, they are sufficient to identify interesting theories at kinks or islands of the allowed theory space.

\section{Conclusion and future directions}\label{sec:Discussion}

The study of QFT in time-dependent background geometries is a formidable challenge. In general, the best one
can do is to study weakly coupled theories using perturbation theory. In fact, even free QFT can be intractable
if the background spacetime is not sufficiently symmetric. A maximally symmetric spacetime like de Sitter opens
the opportunity for a non-perturbative treatment inspired by conformal bootstrap methods. The present work is a
humble first step towards a non-perturbative treatment of QFT in rigid dS$_{d+1}$ utilizing the conformal bootstrap ideology.

The main toolbox of this work is the Hilbert space decomposition derived from first principles in section~\ref{sec:QFT in dS Pre}. In section~\ref{sec: Bulk two-point function}, we make use of the resolution of the identity operator in eq.~\reef{completeKPre} to recover the \toolazy decomposition of the \textit{bulk two-point function} with a positive spectral density. The spectral density carries information about the bulk-boundary expansion spelled out in section~\ref{2ptLateTime}. In sections~\ref{sec:two-point sphere}, we find an equivalent \textit{inversion formula} for spectral density by analytic continuation from the sphere. Through this inversion formula, one can find the spectral density of a theory such as free scalar field or bulk CFT and explore their boundary operator content -- see section~\ref{sec:ExampleInversion}. 

The dS boundary four-point functions admit a partial wave expansion. The coefficients of this expansion satisfy positivity conditions due to the unitarity of the bulk theory. This has been shown in section~\ref{sec:FPF}. We illustrate the concept of partial wave 
coefficients by explicitly calculating  them in Generalized Free Field theory and to the first order of perturbation theory for $\lambda \phi^4$ interaction. Along the way, it turns out that the local terms are unavoidable for a unitary Mean Field Theory.  Finally, we focus on two-dimensional de Sitter spacetime and find non-trivial bounds for partial wave coefficients  in a particular setting as a \textit{proof-of-concept}.

Clearly, there are many open questions left for the future. Let us list some of
them:

\begin{itemize}
\item The {\bf Hilbert space} of a QFT in dS$_{d+1}$ must decompose in unitary irreducible representations of $SO(d+1,1)$. There are two concrete cases where this question can certainly be answered using group theory. The first is CFT in dS where one should be able to decompose conformal multiplets of $SO(d+1,2)$ into irreps of $SO(d+1,1)$, as we illustrated in appendix \ref{Sec:DiscreteSeries} for the case of dS$_2$. 
The second is free QFT in dS where one should be able to decompose the Fock space into irreps of $SO(d+1,1)$. In this case, it would also be interesting to study the effect of perturbative interactions on the structure of the Hilbert space. We hope to return to this question
in the near future.

\item What is the set of {\bf  boundary operators} present in a generic interacting QFT in dS? For CFT in dS, we saw that all boundary operators are hermitian with real scaling dimension $\Delta$. On the other hand, a (sufficiently) massive free scalar in dS gives rise to a pair of hermitian conjugate boundary operators of dimension $\Delta=\frac{d}{2} \pm i\mu$ with $\mu \in \mathbb{R}$. How do these two special cases change under continuous deformations of the QFT? 
In practice, we can study deformations of the CFT by relevant bulk operators and of the free theory by turning on interactions.\footnote{One intriguing feature of the free limit of an interacting QFT is the appearance of local terms in the two-point function of boundary operators $\langle \OO \Od \rangle$ when $\Delta_\OO = \frac{d}{2}+i\mu$. This seems to be a discontinuous effect because conformal symmetry forces  $\langle \OO \Od \rangle=0$ as long as  real part of the scaling dimension $\Re \Delta_\OO \neq \frac{d}{2}$ and we expect $0<\Re \Delta_\OO - \frac{d}{2}\ll1$ for a weakly coupled massive scalar field in dS.}

\item The generalization of the {\bf Källén-Lehmann} decomposition of the bulk two-point functions for local operators with spin would be very helpful to shed light on the two previous questions. 
We are planning to report on this soon.

\item  We introduced {\bf regularised crossing equations} to ameliorate the convergence properties of the integral over the  continuous  label $\nu$ of principal series irreps. It is important to develop a more systematic approach to this issue. In particular, we did not address the case of higher dimensions $d>1$.

\item Pragmatically, the main open task is to set up a {\bf numerical conformal bootstrap} approach to the crossing equations for boundary four-point functions of QFT in dS.
We gave a proof of principle by deriving a bound in a toy example but it is important to develop a systematic algorithm. To use SDPB~\cite{Simmons-Duffin:2015qma} we will need to devise a polynomial approximation to the partial waves (or their regularized version). 

\item There is an alternative approach based on {\bf $\mathbf{6j}$ symbols} that does not use conformal partial waves.
For  simplicity let us focus on the first equation in~\reef{crossing}. Integrating both sides over all points $x_i$ against $\Psi_{\D,\ell}^t(x_i)$ and using orthogonality of partial waves, we find
\be
\begin{aligned}\label{t to s}
I_{\Delta,\ell} &= \frac{1}{n_{\Delta,\ell}} \sum_{\ell^\prime} \int \frac{d\Delta^\prime}{2\pi i} \; I_{\Delta^\prime,\ell^\prime} \; \mathcal{J}_d(\tilde{\Delta}^\prime,\ell^\prime,\tilde{\Delta},\ell|\tilde{\Delta}_\OO,\tilde{\Delta}_\OO,\tilde{\Delta}_\OO,\tilde{\Delta}_\OO) + \mathcal{D}^{st }_{\Delta,\ell}~, \\
\mathcal{D}^{st }_{\Delta,\ell} &\equiv  
\frac{1} {  n_{\Delta,\ell}} 
\int \frac{d^dx_1 \cdots d^dx_4} {\text{vol}(SO(d+1,1))} \; \left(D^{s}(x_i)-D^{t}(x_i)\right) 
\Psi^{t}_{\tilde{\Delta},\ell}(x_i)
\end{aligned}
\ee
where we used the notation of~\cite{Liu:2018jhs} for the $6j$ symbol $\mathcal{J}_d$. 
The disconnected contribution  $\mathcal{D}^{st }_{\Delta,\ell}$ can be computed in a similar fashion to the  MFT partial wave coefficients in~\reef{eq:IMFT}~\cite{Karateev:2018oml}.
The equation \eqref{t to s} says that $I_{\D,\ell}$ is invariant under convolution with the $6j$ symbol. It would be interesting to explore this constraint together with positivity of $I_{\D,\ell}$.

\item What are the {\bf interesting questions} about QFT in dS? In standard CFT, the basic CFT data are scaling dimensions and OPE coefficients and most bootstrap studies derive bounds on these quantities.
For QFT in dS, partial wave coefficients $I_{\D,\ell}$ play a similar role to OPE coefficients in CFT.
However, the former include a set of non-negative functions of the continuous label $\nu$ of principal series irreps. 
What type of bounds should we aim for such functions?
It would be useful to develop  more intuition from perturbative computations.
Ideally, we would like to find questions that can isolate some physical theory inside an island of the allowed space of QFTs.

\item It would be interesting to understand the {\bf flat space limit} of dS correlators
\cite{Maldacena:2011nz,Arkani-Hamed:2018kmz}.
Perhaps there is a limiting procedure that takes dS partial wave coefficients $I_{\D,\ell}$ into flat space partial amplitudes $f_\ell(s)$, where the square of the center of mass energy  $s \sim \nu^2/R^2$.
This is similar to known formulas for AdS \cite{Polchinski:1999ry, Gary:2009ae, Okuda:2010ym, Penedones:2010ue, Raju:2012zr, Paulos:2016fap, Hijano:2019qmi, Komatsu:2020sag}.

\item  The consequences of {\bf perturbative unitarity} are currently being investigated in a program known as the \emph{cosmological bootstrap}~\cite{Arkani-Hamed:2018kmz,Baumann:2020dch,Pajer:2020wnj,Goodhew:2020hob}. Is it possible to make contact between our work and the perturbative cosmological bootstrap? Perhaps recent advances concerning cutting rules in (A)dS~\cite{Meltzer:2019nbs,Meltzer:2020qbr,Meltzer:2021zin} can play a role here.

\item {\bf Massless fields} in dS are known to give rise to infrared divergences in perturbation theory~\cite{Polyakov:2012uc,Krotov:2010ma,Akhmedov:2017ooy,Akhmedov:2013xka}. Recently, the authors of~\cite{Gorbenko:2019rza} claimed to have resolved this issue.
It would be interesting to analyse this problem within our non-perturbative approach. 

\item Can {\bf quantum gravity} in dS be studied with our conformal bootstrap approach?
In the case of AdS, there is a rather systematic way to go from QFT  to quantum gravity. In fact, the conformal bootstrap equations for the boundary correlators are unchanged. The sole effect of quantum gravity in the bulk is the appearance of new boundary operator: the stress tensor. The stress tensor is a special operator because its correlation functions are constrained by Ward identities.
It is tempting to imitate this strategy in dS. As a first step, one should study a bulk massless spin 2 field and analyse the correlators of its associated boundary operators.
It would also be very interesting to compare this approach to previous proposals for a dS/CFT correspondence~\cite{Strominger:2001pn,Witten:2001kn, Anninos:2011ui}.

\section*{Acknowledgements}
We thank Anton de la Fuente and Joao Silva for collaboration in the early stages of this work.
We are grateful to Dionysios Anninos, Alek Bedroya, Victor Gorbenko, Aditya Hebbar, Denis Karateev, Shota Komatsu, Andrea Manenti, Dalimil Mazáč, David Meltzer, Enrico Pajer, David Simmons-Duffin and Amirhossein Tajdini  for useful discussions. We would like to especially thank Manuel Loparco  for the valuable feedback on the draft and for pointing out several typos. The authors are supported by the Simons Foundation grant 488649 (Simons Collaboration on the Nonperturbative Bootstrap) and by the Swiss National Science Foundation through the project 200020\_197160 and through the National Centre of Competence in Research SwissMAP.

\end{itemize}

\appendix

\section{Special functions}\label{sec:SpecialFunctions}

In this appendix, we list a number of identities that are used throughout this paper.
\subsection{Common special functions}
\subsection*{Gamma function}
The large limit of the gamma function is given by Stirling's approximation as 
\begin{equation}\label{eq:streling}
\lim_{\abs{z}\to\infty}\Gamma(z) = \sqrt{\frac{2\pi}{z}} e^{-z} z^z.
\end{equation}
This is true for any x on the complex plane but the negative real line. The convergence is weaker as x approaches the negative real line and to get to the asymptotic regime, one needs to go beyond $\abs{z} \sim 1/\theta$ in which $\theta$ is the angle with the negative real line. In the case in which $z=x+iy$ with $x,y \in \Real$ and $\abs{y}\gg1$, one has:
\be
\Gamma(x+iy) = (1-i)\sqrt{\pi} e^{\frac{i \pi x}{2}-i y -\frac{\pi y}{2}} y^{i y + x - \half}~.
\ee

\subsection*{Barnes's lemma}
The following identity~\cite[Theorem 2.4.3]{aar} is known as Barnes's second lemma:
\begin{multline}
\label{eq:second barnes lemma}
\frac{1}{2\pi i}\int _{-i\infty }^{i\infty} ds \;\Gamma(-s) \frac{\Gamma(a+s)\Gamma (b+s)\Gamma (c+s)\Gamma (1-e-s)}{\Gamma (f+s)}\\
 = \frac{\Gamma (a)\Gamma (b)\Gamma (c)\Gamma (1-e+a)\Gamma (1-e+b)\Gamma (1-e+c)}{\Gamma (f-a)\Gamma (f-b)\Gamma (f-c)}
\end{multline}
which holds when $e=a+b+c-d+1$.

\subsection*{Gegenbauer}
The Gegenbauer function is defined as~\cite[8.932.1]{grad}
\begin{equation}\label{eq:def gegenbauer}
C_{J}^{\alpha}(z)= \frac{\Gamma(J+2\alpha)}{\Gamma(J+1)\Gamma(2\alpha)} \; \FF{-J}{J+2\alpha}{\alpha+\half}{\frac{1-z}{2}}~
\end{equation}
which matches with the Gegenbauer polynomials when $J$ is a non-negative integer with the following orthogonality relations:
\begin{equation}\label{eq:ortho gegenbauer}
\int_{-1}^1 dx  \; (1-x^2)^{\alpha-\half} C_n^\alpha(x) C_m^\alpha(x) = \frac{\pi 2^{1-2\alpha} \Gamma(n+2\alpha)}{\Gamma(n+1)\Gamma(n+\alpha) \Gamma(\alpha)^2} \delta_{mn}~.
\end{equation}
for $n$ and $m$ non-negative integer. 
\subsection*{Hypergeometric function}
In this paper, we may use two equivalent notation of hypergeometric functions:
\be
{}_pF_q(a_1,\cdots,a_p;b_1,\cdots,b_q;z) = \FFG{a_1,\cdots,a_p}{b_1,\cdots,b}{z}~.
\ee
Integrating a hypergeometric function against a monomial yields~\cite[7.511]{grad}:
\begin{equation}\label{eq:power 2f1 integral}
\int^\infty_0 dt \; t^{\alpha-1}  \;_2F_1(a,b;c;-t) = \frac{\Gamma(c)\Gamma(\alpha)\Gamma(a-\alpha)\Gamma(b-\alpha)}{\Gamma(a)\Gamma(b)\Gamma(c-\alpha)}~
\end{equation}
where we assume the $0<\Re(\alpha)<\min\{\Re(a),\Re(b)\}$ for the integral to be convergent. 

It is very useful to change the last argument of hypergeometric to its inverse. This can be done by the identity~\cite[sec 2.9]{HyperRef}
\be\label{eq:2f1inverseZ}
_{2}F_1(a,b;c;z) = \frac{\Gamma(b-a)\Gamma(c)}{\Gamma(b)\Gamma(c-a)} (-z)^{-a} \FF{a}{a-c+1}{a-b+1}{\frac{1}{z}}~+~ a\leftrightarrow b~.
\ee
There is also two other identities that happen to be helpful in some of the calculations in this work~\cite[sec 2.9]{HyperRef}:
\be\label{eq:2f1identityW}
_{2}F_1(a,b;c;z) = (1-z)^{c-a-b} \; _{2}F_1(c-a,c-b;c;z)~,
\ee
and
\be\label{eq:2f1identity1over1-z}
_{2}F_1(a,b;c;z) = \frac{\Gamma(b-a)\Gamma(c)}{\Gamma(b)\Gamma(c-a)} (1-z)^{-a}\, \FF{a}{c-b}{a-b+1}{\frac{1}{1-z}}~+~ a\leftrightarrow b~.
\ee

Let us collect some results that involve the branch cut of the hypergeometric function ${}_2F_1(a,b,c;z)$ across the cut $z \in [1,\infty)$.  In particular, we want to find the discontinuity (Disc) and the average (Ave)
  along the cut, which are defined as
\beq
\text{Disc} \, [f(z)] \ldef f(z+i\epsilon)-f(z-i\epsilon)
\qaq
\text{Ave} \, [f(z)] \ldef \half\left(f(z+i\epsilon) + f(z-i\epsilon) \right)~.
\eeq
Using~\reef{eq:2f1inverseZ} and 
\begin{equation}\label{eq: Disc Poly}
\Disc{z^a} = 2i \sin{\pi a}  \, (-z)^a~,
\end{equation}
 we find that
\label{eq:DiscOf2F1}
\begin{align}
\text{Disc}\, [_2F_1(a,b;c;z)] &= \cos(\pi a) \frac{\Gamma(b-a)\Gamma(c)}{\Gamma(c-a)\Gamma(b)} z^{-a}  \FF{a}{a-c+1}{a-b+1}{\frac{1}{z}} \; + \; a \leftrightarrow b\\
\text{Ave}\, [_2F_1(a,b;c;z)] &= 2i\sin(\pi a) \frac{\Gamma(b-a)\Gamma(c)}{\Gamma(c-a)\Gamma(b)} z^{-a}  \FF{a}{a-c+1}{a-b+1}{\frac{1}{z}} \; + \; a \leftrightarrow b~.\nonumber
\end{align}

Another way to find the discontinuity is to consider the integral representation~\cite[15.6.1]{DLMF}
\begin{equation}
_{2}F_{1}(a,b;c;z)=\frac{\Gamma(c)}{\Gamma(b)\Gamma(c-b)}\int _{0}^{1}x^{b-1}(1-x)^{c-b-1}(1-zx)^{-a}\,dx\qquad \Re (c)>\Re (b)>0.
\end{equation}
together with~\ref{eq: Disc Poly}. This yields
\begin{align}\label{eq:Discof2F1New}
\Disc{_{2}F_{1}(a,b;c;z)} =\frac{2\pi i \Gamma(c)\;  z^{1-c} \, (z-1)^{c-b-a}}{\Gamma(a)\Gamma(b)\Gamma(c-a-b+1)}\; \FF{1-b}{1-a}{c-a-b+1}{1-z}
\end{align}
which is in agreement with \reef{eq:DiscOf2F1} using~\cite[15.8.4]{DLMF}.

Finally, the generalized hypergeometric function $_3F_2$ has the following integral representation:
\begin{equation}\label{eq:3F2 from 2F1}
_3F_2(a_1,a_2,a_3;b_1,b_2;t)=\frac{\Gamma(b_2)}{\Gamma(a_3) \Gamma(b_2-a_3)} \int^1_0 z^{a_3-1} (1-z)^{-a_3+b_2-1}\, _2F_1(a_1,a_2;b_1;z)~.
\end{equation}

\subsection{Estimates for $\tilde{F}$ at large $\Delta$}
\label{eq:tildefbnd}

In this section, we will provide some estimations for the quantity $\tilde{F}$ that is defined in~\reef{eq:Fstdef} and appears in the one-dimensional bootstrap equation~\reef{eq:crossexample}. Since we focus on identical operator four-point functions, only expressions with $\ell = 0$ show up. However, a generic four-point  function and the case of $\ell = 1$ can be studied similarly. The function $\tilde{F}$ consists of four terms:
\bsub
\beq
\tilde{F}^{s-t}_{\Delta,\ell = 0} = \mca{I}(\Delta,\gamma,\sigma) - \mca{I}(\Delta,\sigma,\gamma) + \mca{I}(1-\Delta,\gamma,\sigma) - \mca{I}(1-\Delta,\sigma,\gamma)
\eeq
with
\begin{multline}\label{eq:CurlyIDelta}
  \mca{I}(\Delta,\gamma,\sigma) = K_{1-\Delta,0}^{\Delta_{\OO},\, \Delta_{\OO} } \frac{\Gamma(\Delta+\gamma+1)\Gamma(2\Delta_{\OO}+\sigma+1)}{\Gamma(\Delta+2\Delta_{\OO} + \gamma+\sigma+2)} \\
    \times {}_3F_2\!\left[{{\Delta,\, \Delta,\, \Delta+\gamma+1}~\atop~{2\Delta,\, \Delta+2\Delta_{\OO} + \gamma+\sigma+2}},1 \right]~.
\end{multline}
\esub
Convergence of the hypergeometric functions requires that
\beq
\label{eq:conv}
1 + 2\Delta_{\OO} + \gamma > 0
\qaq
1 + 2\Delta_{\OO} + \sigma > 0~.
\eeq
In order to study the convergence of the bootstrap problem, we need to consider the large-$\nu$ limit for $\Delta = 1/2 + i \nu$ and the large-$n$ limit of $\Delta = n \in 2\mbb{N}$. Let us treat these cases separately.

\subsubsection*{Principal series}

First of all, let us set $\Delta = 1/2 + i \nu$ and analyze the limit $\nu \to \infty$. Notice that the four terms in $\tilde{F}$ are related to $\mca{I}(\Delta,\gamma,\sigma)$ via the permutations $\nu \mapsto -\nu$ and/or $\gamma \lra \sigma$. Hence if we understand the large-$\nu$ asymptotics of $\mca{I}(\Delta,\gamma,\sigma)$, it is straightforward to deduce the large-$\nu$ behavior of the full function $\tilde{F}$. 

For the case at hand, it will prove convenient to rewrite the ${}_3F_2(1)$ using a hypergeometric transformation, which yields
\begin{equation}
\begin{aligned}
  \label{eq:Ire}
  \mca{I}(\Delta,\gamma,\sigma) =&K_{1-\Delta,0}^{\Delta_{\OO},\, \Delta_{\OO} } \frac{\Gamma(\Delta+\gamma+1)\Gamma(2\Delta_{\OO}+\sigma+1)^2}{\Gamma(\Delta+2\Delta_{\OO} + \gamma+\sigma+2)}\\
  \times&  \frac{\Gamma(2\Delta)}{\Gamma(\Delta)\Gamma(1+\Delta + 2\Delta_{\OO} + \sigma)}
    {}_3F_2\!\left[{{\Delta,\, 1+2\Delta_{\OO} + \sigma,\, 2+\gamma + 2\Delta_{\OO} + \sigma}~\atop~{2+\gamma+\Delta+2\Delta_{\OO}+\sigma,\,1+\Delta + 2\Delta_{\OO} + \sigma}},1 \right]~.
\end{aligned}
\end{equation}
The new ${}_3F_2(1)$ converges when $\Re(\Delta) > 0$, which holds in particular on the axis $\Re(\Delta) = 1/2$.
To begin, let us analyze the different factors appearing in $\mca{I}$ from eq.~\reef{eq:Ire}. The $K$-function goes as
\beq
K_{\hd-i\nu,0}^{\Delta_{\OO},\, \Delta_{\OO} } \; \limu{\nu \to \infty} \; e^{-i\pi/4} \sqrt{\pi}\; \frac{4^{- i\nu}}{\sqrt{\nu}}~.
\eeq
Next, the gamma functions go as
\beq
\frac{\Gamma^4}{\Gamma^3} \; \limu{\nu \to \infty} \; \frac{e^{i \pi \kappa}}{\sqrt{\pi}} \Gamma(1+2\Delta_{\OO}+\sigma)^2 \, \frac{4^{i \nu} }{\nu^{3/2 + 4\Delta_{\OO} + 2\sigma}},
\quad
\kappa = \frac{5}{4} - 2\Delta_{\OO} - \sigma~.
\eeq
It remains to find the $\nu \to \infty$ asymptotics of the ${}_3F_2(1)$ hypergeometric function. But it is easy to show that
\beq
    {}_3F_2\!\left[{{\Delta,\, 1+2\Delta_{\OO} + \sigma,\, 2+\gamma + 2\Delta_{\OO} + \sigma}~\atop~{2+\gamma+\Delta+2\Delta_{\OO}+\sigma,\,1+\Delta+ 2\Delta_{\OO} + \sigma}},1 \right]\Big|_{\Delta = \tfrac{1}{2} + i \nu} \; \limu{\nu \to \infty} \; 1~.
    \eeq
    One way to show this is using the series representation of the ${}_3F_2(1)$, which converges for the case in question. Schematically it is of the form
    \beq
        {}_3F_2(1) = 1 + \sum_{n=1}^\infty a_n(\Delta)
        \quad
        \text{with}
        \quad
        a_n(\Delta) \;\limu{\Delta \to \infty} \; \frac{1}{\Delta^n}~,
        \eeq
where the terms with $n \geq 1$ are unimportant in the limit $|\Delta| \to \infty$.
        Bringing everything together, we conclude that
        \beq
        \mca{I}(\th + i \nu,\gamma,\sigma) \; \limu{\nu \to \infty} \;  \frac{\Gamma(1+2\Delta_{\OO} + \sigma)^2}{\nu^{2+4\Delta_{\OO} + 2\sigma}}
        \eeq
        up to some $O(1)$ numerical factor.
        Finally, we conclude that
\beq
\tilde{F}^{s-t}_{\th + i \nu,\ell =0} \; \limu{\nu \to \infty} \; 1/\nu^{2+4\Delta_{\OO} + 2\text{min}(\gamma,\sigma)}~.
\eeq

\subsubsection*{Discrete series}

The analysis for $\Delta = n \in 2\mbb{N}$ is similar. First note that $\tilde{F}_n$ only consists of two terms:
\beq
\tilde{F}^{s-t}_{n,\ell = 0} = \mca{I}(n,\gamma,\sigma) - \mca{I}(n,\sigma,\gamma)
\eeq
where $\mathcal{I}(n,\gamma,\sigma)$ is defined in~\reef{eq:CurlyIDelta}. For large $n$, the $K$-function behaves as:
\beq
K_{1-n,0}^{\Delta_{\OO},\, \Delta_{\OO}} \; \limu{n \to \infty}  \frac{\sqrt{\pi} }{2^{2n-2}\sqrt{n}}~.
\eeq
The large $n$ limit of the rest of the terms in $I(n,\gamma,\sigma)$ are thus very similar to the above expression replacing $\nu\to n$. In the end, one finds:
\beq
\mca{I}(n,\gamma,\sigma) \; \limu{n \to \infty} \; \frac{\Gamma(1+2\Delta_{\OO} + \sigma)^2}{n^{2+4\Delta_{\OO} + 2\sigma}}~.
\eeq
Including the second term with $\gamma \lra \sigma$,  we find that
\beq
\tilde{F}^{s-t}_{n,\ell =0} \; \limu{n \to \infty} \; 1/n^{2+4\Delta_{\OO} + 2\text{min}(\gamma,\sigma)}~.
\eeq

\section{Action of $SO(d+1,1)$ generators}
\label{sec:Action of Generators}
In this appendix, we find the action of $SO(d+1,1)$ generators on scalar sates $\ket{\D,k}$ and $\ket{\D,x}$. For simplicity we focus on scalar states, but the generalization to traceless symmetric spinning states is straightforward. In this appendix, we will use notation $a.b = a_\mu b^\mu$, $k.\partial_k = k^\mu\frac{\partial}{\partial k^\mu}$ and $\partial_\mu = \frac{\partial}{\partial x^\mu}$ except it is mentioned otherwised.

Take the wavefunction defined in~\reef{eq:wavefunction nonpert} or~\reef{eq:wavefdef}:
\begin{align}
\Phi _{x}(x,\eta)= \langle \Omega| {\phi}(x,\eta) | \Delta,k\rangle = e^{-ik.x} (-\eta)^{\frac{d}{2}}  h_{i \nu}(|k| \eta)~
\end{align}
where $|\Omega\rangle$ is the Bunch-Davis vacuum. We again used $\D=\hd+i\nu$. The action of the conformal generators on the field can be expressed as the differential operator
\begin{align}
[Q, \phi(x)] = \hat{Q} \;\phi(x)~,
\end{align}
in which $Q$ is the corresponding Hilbert space charge operator of the Killing vector differential operator $\hat{Q}$. We use notation $\hat{A}$ for the differential operator to distinguish it from the Hilbert space operator $A$. Acting on wavefunctions with $\hat{Q}$ and using the fact that the Bunch-Davis vacuum is invariant under our isometries i.e. $Q|\Omega\rangle =0$, one can find the action of the charges on the states $|\Delta,k\rangle$. For example, in the case of $Q= P_\mu$:
\begin{align}
\hat{P}_\mu \;\Phi_{k}(x,\eta) &=  \partial_\mu \Phi_{k}(x,\eta)  = -ik_\mu  \Phi_{k}(x,\eta) = -ik_\mu \langle \Omega| \phi(x)| \Delta,k\rangle \\
&=  - \langle \Omega| \phi(x) P_\mu| \Delta,k\rangle~.\nonumber
\end{align}
We find the familiar relation 
\be
P_\mu |\Delta,k\rangle = ik_\mu |\Delta,k\rangle~.
\ee

With the same approach, one can find the action of the other generators on the chosen basis. The action of the dilatation operator is
\begin{align}
\hat{D}  \langle \Omega| \phi(x)| \Delta,k\rangle &=
(\eta\partial_\eta + x.\partial_{x}) \left[ (-\eta)^{\frac{d}{2}}h_{i\nu}(|k|\eta) e^{i k. x} \right] \\ 
&= \left[ (-i k. x + \frac{d}{2})h_{i\nu}(|k|\eta) (-\eta)^{\frac{d}{2}} e^{i k. x} + k.\partial_k h_{i\nu}(|k|\eta) (-\eta)^{\frac{d}{2}} e^{i k. x} \right]\nonumber\\
&= (k.\partial_{k} + \frac{d}{2})  \left[(-\eta)^{\frac{d}{2}}h_{i\nu}(|k|\eta) e^{i k. x}\right]~, \nonumber\\
\hat{D}  \langle \Omega| \phi(x)| \Delta,k\rangle  &= - \langle \Omega| \phi(x) D| \Delta,k\rangle~,
\end{align}
in which we use the fact that $h_{i\nu}(|k|\eta)$ is symmetric under exchange of $|k|\leftrightarrow\eta$ and $k.\partial_{k} f(|k|) = |k| \partial_{|k|} f(|k|)$ to change the time derivative to momentum derivative. Hence the action of the Dilatation operator on this basis is
 \begin{equation}
D |\Delta,k\rangle = -(k.\partial_{k} + \frac{d}{2}) |\Delta,k\rangle~.
\end{equation}
Finally, for the case of $Q=K_\mu$ one finds
\begin{align}\label{zakhar}
\hat{K}_\mu  \langle \Omega| \phi(x)| \Delta,k\rangle &=
c_2((\eta^2-x.x) \partial_{x^\mu}  + 2  x_\mu \eta \partial _\eta + 2x_\mu x.\partial_{x}) \left[ (-\eta)^{\frac{d}{2}} H^{(2)}_{i\nu}(-|k|\eta) \; e^{i k. x} \right] \nonumber\\\ 
&= c_2\left[(-ik_\mu(\eta^2-x.x)+ 2x_\mu(-ik. x+\frac{d}{2})) H_{i\nu}+ 2(-|k|\eta) H^\prime_{i\nu} \right](-\eta)^{\frac{d}{2}} e^{i k. x}~,\nonumber\\
\hat{K}_\mu  \langle \Omega| \phi(x)| \Delta,k\rangle  &= - \langle \Omega| \phi(x) K_\mu| \Delta,k\rangle~.
\end{align}
where $c_2= \frac{\sqrt{\pi}}{2}e^{\frac{\pi\nu}{2}}$ and we dropped the Hankel function type index and dependence on $-\eta |k|$ to avoid clutter. We also write $\partial_{-\eta |k|} H_{i\nu}(-\eta |k|) = H^\prime$. In parallel, we have
\begin{align*}
\partial_{k^\mu} \Phi  &= c_2 (-i x_\mu H_{i\nu} + \frac{k_\mu}{k^2} (-\eta |k|) H^\prime_{i\nu})(-\eta)^{\frac{d}{2}} e^{i k. x}~, \\
\vspace{1cm}
\partial_{k^\alpha}\partial_{k^\mu}\Phi &= c_2 [-x_\alpha x_\mu H_{i\nu} - i x_\mu \frac{k_\alpha}{k^2} (-\eta |k|) H^\prime_{i\nu} - i x_\alpha  \frac{k_\mu}{k^2} (-\eta |k|) H^\prime_{i\nu} \\
&+(\frac{\delta_{\mu\alpha}}{k^2}-\frac{k_\mu k_\alpha}{k^4}) (-\eta |k|) H^\prime_{i\nu} + (\frac{k_\mu k_\alpha}{k^4}) (-\eta |k|)^2 H^{\prime \prime}_{i\nu}](-\eta)^{\frac{d}{2}} e^{i k. x},\\
&= c_2 [-x_\alpha x_\mu H_{i\nu} - i x_\mu \frac{k_\alpha}{k^2} (-\eta |k|) H^\prime_{i\nu} - i x_\alpha  \frac{k_\mu}{k^2} (-\eta |k|) H^\prime_{i\nu} \\
&+(\frac{\delta_{\mu\alpha}}{k^2}-2\frac{k_\mu k_\alpha}{k^4}) (-\eta |k|) H^\prime_{i\nu} - (\frac{k_\mu k_\alpha}{k^4}) (-\eta |k|)^2  H_{i\nu} -(\frac{k_\mu k_\alpha}{k^4}) \nu^2  H_{i\nu}](-\eta)^{\frac{d}{2}} e^{i k. x}~,
\end{align*}
in which we exchange second derivative of Hankel function with terms of the first derivative and zero derivatives using the its generating differential equation. This leads to
\begin{align*}
k^\alpha \partial_{k^\alpha}\partial_{k^\mu}\Phi &=  c_2 (-\eta)^{\frac{d}{2}} e^{i k. x} \big[ - \frac{k_\mu}{k^2}(-\eta |k|) H^\prime_{i\nu}  - i (-\eta |k|) H^\prime_{i\nu}  (x_\mu + (k.x)\frac{k_\mu}{k^2})\\
 &\quad- (k.x)x_\mu H_{i\nu} - \eta^2 k_\mu H_{i\nu}  - \nu^2 \frac{k_\mu}{k^2} H_{i\nu}\big]~,\\
 \vspace{1cm}
k_\mu \partial^{k^\alpha}\partial_{k^\alpha}\Phi &= c_2 (-\eta)^{\frac{d}{2}} e^{i k. x} k_\mu \big[ \frac{d-2}{k^2} (-\eta |k|)H^\prime_{i\nu} - i (-\eta |k|)\frac{2k.x}{k^2} H^\prime_{i\nu}\\
  &\quad - x.x \, H_{i\nu}  - \eta^2 H_{i\nu} - \frac{\nu^2}{k^2} H_{i\nu} \big]~.
\end{align*}
Then the particular linear combination of $(-2k^\alpha \partial_{k^\alpha}\partial_{k^\mu} + k_\mu \partial^{k^\alpha}\partial_{k^\alpha} - d \partial_{k^\mu})\Phi$ is equal to
\begin{align*}
c_2 (-\eta)^{\frac{d}{2}} e^{i k. x} \left[2 i (-\eta |k|) H^\prime_{i\nu} x_\mu  - 2 (k.x + i\frac{d}{2}) x_\mu H_{i\nu} +(\eta^2 - x.x) k_\mu H_{i \nu} +\frac{\nu^2}{k^2 } k_\mu H_{i \nu}  \right]~.
\end{align*}
Considering what we found in~\reef{zakhar}, we arrive at the following expression for $K_\mu$ acting on $| \Delta,k\rangle$:
 \begin{equation}x
K_\mu|\Delta,k\rangle = i \left[k_\mu \partial^{k^\alpha}\partial_{k^\alpha} - 2k^\alpha \partial_{k^\alpha}\partial_{k^\mu} - d \partial_{k^\mu} - \frac{\nu^2}{k^2 } k_\mu\right] |\Delta,k\rangle~.
\end{equation}
The action of $M_{\mu\nu}$ is the trivial action of $SO(d)$ rotation group on scalars that we do not spell it out here. One can explicitly check the action of quadratic Casimir $C=D^2-\frac{1}{2}(K^\mu P_\mu + P^\mu K_\mu - M_{\mu\nu}M^{\mu\nu})$ will give the desired relation~\eref{eq:CasimirEigenvalueNew}.

We may now derive the action of conformal generators on the position space states
\begin{equation}\label{eq:primary state pp }
|\Delta,x\rangle = \int d^dk \; e^{i k.x} |k|^{\Delta-\frac{d}{2}} \; \PSK
\end{equation}
mentioned in equations~\reef{eq: PKD position} and~\reef{eq: PKD positionP}.

One may ask why the integral above is not simply the Fourier transformation and it has an extra factor of $|k|^{\Delta-\frac{d}{2}}$. This is due to the fact that we wanted this state to be like a primary state at point x with dimension $\Delta$. In fact, one can put a general function instead and after imposing the right transformations under dilatation or special conformal isometries, will find the suggested factor. One can easily check the action of $P_\mu$ on these states: 
\begin{equation}
\begin{split}
P_\mu\PSX &= \int d^dk \; e^{i k.x} |k|^{\Delta-\frac{d}{2}} \; P_\mu\PSK \\
&= i \int d^dk \; e^{i k.x} |k|^{\Delta-\frac{d}{2}} \; k_\mu\PSK  \\
&= \partial_\mu \int d^dk \; e^{i k.x} |k|^{\Delta-\frac{d}{2}} \; \PSK \\
\Aboxed{&= \partial_\mu\PSX}
\end{split}
\end{equation}

We also may check the action of dilatation operator $D$: 
\begin{equation}
\begin{split}
D\PSX &=  \int d^dk \; e^{i k.x} |k|^{\Delta-\frac{d}{2}} \; D\PSK\\
&= - \int d^dk \; e^{i k.x} |k|^{\Delta-\frac{d}{2}} \; (k.\partial_{k}+\frac{d}{2})\PSK  \\
&= i \partial^\mu  \int d^dk \; e^{i k.x} |k|^{\Delta-\frac{d}{2}} \partial_{k^\mu} \PSK \; - \frac{d}{2}\PSK\\
&= -i \partial^\mu \int d^dk \; \partial_{k^\mu} (e^{i k.x} |k|^{\Delta-\frac{d}{2}}) \PSK \; - \frac{d}{2}\PSK\\
&= -i \partial^\mu \int d^dk \; \left[i x_\mu + (\Delta-\frac{d}{2}) \frac{k_\mu}{k^2}\right] (e^{i k.x} |k|^{\Delta-\frac{d}{2}}) \PSK\; - \frac{d}{2}\PSK \\
&= (x.\partial_{x} + d +(\Delta -\frac{d}{2}) - \frac{d}{2}) \PSX\\
\Aboxed{&= (x.\partial_{x} + \Delta) \PSX}
\end{split}
\end{equation}
where we performed integral by parts and dropped the boundary terms (at $k\to\infty$).
The action of $K_\mu$ on \eref{eq:primary state} is
\begin{equation}
\begin{split}
K_\mu\PSX &=   \int d^dk \; e^{i k.x} |k|^{\Delta-\frac{d}{2}} \; K_\mu\PSK \\
&= i \int d^dk \; e^{i k.x} |k|^{\Delta-\frac{d}{2}} \; (-2k^\alpha \partial_{k^\alpha}\partial_{k^\mu} + k_\mu \partial^{k^\alpha}\partial_{k^\alpha} - d \partial_{k^\mu} -\frac{\nu^2}{k^2 } k_\mu) \PSK
\end{split}
\end{equation}
One might rewrite each of the four terms using integral by parts
\begin{equation}
\begin{split}
\text{1st term} &=-2 \partial^\alpha  \int d^dk \; e^{i k.x} |k|^{\Delta-\frac{d}{2}} \;  \partial_{k^\alpha}\partial_{k^\mu}  \PSK\\
&= 2  \partial^\alpha \int d^dk \; \left[i x_\alpha + (\Delta-\frac{d}{2}) \frac{k_\alpha}{k^2}\right] e^{i k.x} |k|^{\Delta-\frac{d}{2}} \; \partial_{k^\mu}  \PSK\\
&=2i(x.\partial_{x}+\Delta+\frac{d}{2})\int d^dk \; e^{i k.x} |k|^{\Delta-\frac{d}{2}} \; \partial_{k^\mu}  \PSK\\
&=-2i(x.\partial_{x}+\Delta+\frac{d}{2})\int d^dk \; e^{i k.x} |k|^{\Delta-\frac{d}{2}} \;  \left[i x_\mu + (\Delta-\frac{d}{2}) \frac{k_\mu}{k^2}\right]  \PSK\\
&= 2 (x_\mu x.\partial_{x} +(\Delta+\frac{d}{2}+1) x_\mu) \PSX - 2i(x.\partial_{x}+\Delta+\frac{d}{2}) (\Delta-\frac{d}{2}) A_\mu\\
&=2 (x_\mu x.\partial_{x} +(\Delta+\frac{d}{2}+1) x_\mu)  |\PSX - 2i(\Delta+\frac{d}{2}) (\Delta-\frac{d}{2}) A_\mu - 2i  (\Delta-\frac{d}{2}) x^\alpha  \partial_\mu A_\alpha~,
\end{split}
\end{equation}

\begin{equation}
\begin{split}
\text{2nd term} &= \partial_\mu  \int d^dk \; e^{i k.x} |k|^{\Delta-\frac{d}{2}} \;  \partial^{k^\alpha}\partial_{k^\alpha} \PSK\\
&=- \partial_\mu  \int d^dk \; e^{i k.x} |k|^{\Delta-\frac{d}{2}} \;  \left[i x_\alpha + (\Delta-\frac{d}{2})\frac{k_\alpha}{k^2} \right]\partial_{k^\alpha}  \PSK\\
&= \partial_\mu  \int d^dk \; e^{i k.x} |k|^{\Delta-\frac{d}{2}} \; \left[i x^\alpha + (\Delta-\frac{d}{2})\frac{k^\alpha}{k^2} \right] \left[i x_\alpha + (\Delta-\frac{d}{2})\frac{k_\alpha}{k^2} \right] \PSK\\
&+ (d-2)(\Delta-\frac{d}{2})  \partial_\mu  \int d^dk \; e^{i k.x} |k|^{\Delta-\frac{d}{2}} \; \frac{1}{k^2} \PSK\\
&= \partial_\mu  \int d^dk \; e^{i k.x} |k|^{\Delta-\frac{d}{2}} \; [-x^2 +2i(\Delta-\frac{d}{2}) \frac{k.x}{k^2}+\frac{(\Delta-\frac{d}{2})^2}{k^2}] \PSK \\  &\quad+ (d-2)(\Delta-\frac{d}{2})  \partial_\mu  \psi\\
&= (-x^2 \partial_\mu -2x_\mu)  \PSX +2i(\Delta-\frac{d}{2})  \partial_\mu (x^\alpha A_\alpha) \\ &\quad+ i [(\Delta-\frac{d}{2})^2  + (d-2)(\Delta-\frac{d}{2}) ]  A_\mu~,
\end{split}
\end{equation}

\begin{equation}
\begin{split}\text{3rd term} &= id   \int d^dk \; e^{i k.x} |k|^{\Delta-\frac{d}{2}} \;  \left[i x_\mu + (\Delta-\frac{d}{2})\frac{k_\mu}{k^2} \right] \PSK,\\
&= -d x_\mu \PSX+id (\Delta-\frac{d}{2}) A_\mu~,\\
\text{4th term} &= -i \nu^2 A_\mu~.
\end{split}
\end{equation}
in which we defined $A_\mu \equiv \int d^dk \; e^{i k.x} |k|^{\Delta-\frac{d}{2}} \frac{k_\mu}{k^2} \PSK$ and $\psi \equiv  \int d^dk \; e^{i k.x} |k|^{\Delta-\frac{d}{2}} \frac{1}{k^2} \PSK$. We also used the following identities: $\partial_\mu \psi = i A_\mu$ and $x^\alpha \partial_\alpha A_\mu  = x^\alpha \partial_\mu A_\alpha$. Note that we assume the boundary terms coming from integral by parts vanish. 
Putting all these together we find
\begin{equation}
\boxed{K_\mu \PSX = \left(2 x_\mu  x.\partial_{x} -x^2 \partial_\mu +2\Delta \,x_\mu \right) \PSX }
\end{equation}

In conclusion, we showed that under the action of conformal generators, position space state $\PSX$ defined in \eref{eq:primary state} behaves like a primary state in a conformal theory. Hence, any correlation function made of $n$ operators sandwiched between vacuum state $|\Omega\rangle$ and primary state $\PSX$, behaves like  ($n+1$)-point correlation function with an insertion of primary operator $\mathcal{O}(x)$ with dimension $\D$:
\be
\langle \Omega | \OO_1(x_1)\cdots\OO_n(x_n) \ket{\D,x} \sim \langle \Omega |\OO_1(x_1)\cdots\OO_n(x_n) \OO(x) |\Omega\rangle~.
\ee

\section{Spectral density inversion formula}
\label{inver form apx}

In section~\ref{sec:two-point sphere}, the analytic continuation of a two-point function on $S^{d+1}$ to de Sitter was discussed. This appendix explains the proof of the inversion formula~\reef{The inversion formula}, which played an important role in that section. In passing, we discuss its convergence and large $J$ limit.

\subsection{Froissart-Gribov trick}
 The standard Gegenbauer inversion formula on $S^{d+1}$ was shown in Eq.~\reef{eq:EuclIF} in the main text. In what follows we will derive the inversion formula~\reef{The inversion formula} for complex $J$ through what is known as the Froissart-Gribov trick, which is a standard tool in S-matrix theory. We refer~\cite{Caron-Huot:2017vep} and~\cite{Correia:2020xtr} for recent discussions. 

Let us write $\alpha = d/2$ in what follows, and furthermore let
 \[
 \omega(x) \ldef (1-x^2)^{\alpha-1/2}~.
 \]
Suppose that the function $G(x)$ appearing in~\reef{eq:EuclIF} is analytic in a neighborhood of $[-1,1]$. Furthermore, suppose that we are given a function $Q^\alpha_J(z)$ that is analytic in a neighborhood of $[-1,1]$ but has the following discontinuity:
\beq
\label{eq:discQ}
\text{Disc}\, \Big[(z^2-1)^{\alpha-1/2} Q^\alpha_J(z)\Big] = -2\pi i\,  \omega(x) C^\alpha_J(x)
\quad
\text{for}
\quad
z \in [-1,1]~.
\eeq
Given such a function, we have the following identity:
 \begin{equation}
 \label{eq:contourQ}
\int_{-1}^1\!dx\, \omega(x)\, C_J^\alpha(x) G(x) = \frac{1}{2\pi i}\oint_c dz \, (z^2-1)^{\alpha-1/2} Q_J^{\alpha}(z)   G(z) 
 \end{equation}
 in which the contour $c$ is a closed loop around the line segment $[-1,1]$, circled in the counterclockwise direction. It turns out that there exists a unique function satisfying~\reef{eq:discQ}, namely
 \beq
 \label{eq:Qdef}
 Q_J^{\alpha}(z) \ldef    \int_{-1}^{1} dx^\prime \, \left(\frac{1-x'^2}{z^2-1} \right)^{\alpha-1/2}  \frac{C_J^{\alpha}(x^\prime)}{z-x^\prime}
 \eeq
 which by construction obeys~\reef{eq:discQ}; in fact, it can be shown that $Q_J^\alpha$ is the unique function obeying~\reef{eq:discQ}. In order to find an explicit representation of $Q_J^\alpha$ we first of all notice that $Q_J^\alpha$ obeys the same ODE as the Gegenbauer function $C_J^\alpha(x)$, namely
\[
\left[ (1-x^2) \frac{d^2}{dx^2} -(2\alpha+1) x \frac{d}{dx} +J(J+2\alpha) \right]f(x) = 0
\]
which has a two-dimensional solution space. Either by computing the integral~\reef{eq:Qdef} explicitly, or by imposing~\reef{eq:discQ}, one concludes that $Q_J^\alpha(z)$ can be written as
\bsub
\beq
\label{eq:Qdefp}
Q^\alpha_J(z) = \frac{\mca{N}}{(z-1)^{J+2\alpha}}\,\FF{J+\alpha+\half}{J+2\alpha}{2J+2\alpha+1}{ \frac{2}{1-z}}
\ee
where
\be
\mca{N} =  \frac{\pi \Gamma(J+2\alpha)}{2^{J+2\alpha -1}\Gamma(\alpha)\Gamma(J+\alpha+1)}~.
\eeq
An equivalent form is
\beq
Q^\alpha_J(z) = \frac{\mca{N}}{z^{J+2\alpha}}\, \FF{\frac{J}{2}+\alpha}{ \frac{J+1}{2} + \alpha}{ J+\alpha+1}{ \frac{1}{z^2}}
\eeq
\esub
which agrees with~\cite{Correia:2020xtr}, taking into account a different choice of normalization used there. Moreover we see that
\[
Q_J^\alpha(z) \; \limu{z \to \infty} \; 1/z^{J+2\alpha}
\]
so for sufficiently large $J$ the function decreases rapidly at infinity. 

The formula~\reef{eq:contourQ} already provides a formula for $a_J$ that is analytic in $J$:
\beq
\label{eq:rubbersoul}
a_J = \frac{2^{2\alpha-1}J!(J + \alpha)\Gamma(\alpha)^2}{\pi \Gamma(J+2\alpha)}  \frac{1}{2\pi i} \oint_{c} dz\,  (z^2-1)^{\alpha-1/2}\, Q_J^\alpha(z) G(z)~.
\eeq
However, we can further massage the RHS of~\reef{eq:rubbersoul} to obtain a form that is more convenient for computations. We already saw that the function $Q_J^\alpha(z)$ decreases faster than $1/z^J$ at large $z$, so at least for large $J$ we can deform the contour and drop any arcs at infinity. Next, we expect that the function $G(z)$ has a branch cut on the real axis past the point $z=1$, say at $[1,\infty)$. Physically, this cut reflects the kinematics of the $S^{d+1}$ correlator, since $z=1$ amounts to measuring the correlator at coincident points $X = X'$. The function $G(z)$ has to be finite on $(-1,1)$, since these points are physical. Finally $z=-1$ describes the correlator at antipodal points $X = -X'$, where it is completely regular. Consequently, we do not expect $G$ to have a branch cut on the negative real axis $(-\infty,-1]$.
Blowing up the contour $c$, we can therefore write
\beq
\label{eq:monkees}
a_J = \frac{J!\Gamma(\alpha)}{2^J \Gamma(J+\alpha)} \frac{1}{2\pi i} \int_{1}^\infty dx\,  \frac{(x+1)^{\alpha-\half}}{(x-1)^{J+\alpha+\half}} \, \FF{J+2\alpha}{J+\alpha+\half}{2J+2\alpha+1}{\frac{2}{1-x}} \text{Disc}\, [G(x)]~.
\eeq
After setting $\alpha \to d/2$, this is precisely the inversion formula from Eq.~\reef{The inversion formula}. If $G(x)$ has any poles or other branch cuts beyond $[1,\infty)$, additional terms need to be added to formula~\reef{eq:monkees}. 

The derivation presented here suffers from one minor issue. In writing~\reef{eq:contourQ} we had to assume that $G(x)$ extends to an analytic function in a small neighborhood around $[-1,1]$. Yet~\reef{eq:monkees} allows for the possibility that $G(z)$ has a branch cut starting at $z=1$, and indeed typical $S^{d+1}$ correlators have $z=1$ as a branch point. In practice, if $G(z)$ is not too singular near $z=1$ then the inversion formula still holds. 

\subsection{Example: $a_J$ of the massive boson}\label{sec:free massive aJ appendix}
We now check the proposed inversion formula in the case of the free field of mass $m^2 R^2 = \Delta_\phi(d-\Delta_\phi)$. In the $x$-coordinate, the propagator reads
\begin{align}\label{eq:free propagator}
  G_{\text{f}}(x)   =  \frac{1}{R^{d-1}}\frac{1}{4\pi^{d/2+1}} \frac{\Gamma(\hd)\Gamma(\Delta_\phi)\Gamma(d-\Delta_\phi)}{\Gamma(d)}
  \, \FF{\D_\phi}{d-\D_\phi}{\frac{d+1}{2}}{\frac{1+x}{2}}~.
\end{align}
The coefficients $a_J$ are computed in~\cite{Marolf:2010zp}, and the result is printed in~\reef{eq:marolfaj}. Here we will reproduce their result using the inversion formula.
The discontinuity of the $G_\mrm{f}(x)$ can be computed in various ways, for instance using~\reef{eq:DiscOf2F1}.
Finding discontinuity of two-point function reduces to calculating discontinuity of hyeprgeometric function in \reef{eq:free propagator}. Using \reef{eq:Discof2F1New}, one finds
\begin{equation}\label{Disc aj free}
\Disc{G_\mrm{f}(x)}=\frac{2^d \pi i\, R^{1-d}}{4 \pi^{1+\frac{d}{2}}}\frac{\Gamma(\frac{d}{2}) \Gamma(\frac{d+1}{2})}{\Gamma(d)\Gamma(\frac{3-d}{2})}\,  (x^2-1)^{\frac{1}{2}-\frac{d}{2}} \, \FF{1+\D_\phi-d}{1-\D_\phi}{\frac{3-d}{2}}{\frac{1-x}{2}}~.
\end{equation}
Before calculating the inversion formula integral, let us comment on its convergence. By examining the limits $x \to 1^{+}$ and $x \to \infty$, we conclude that~\reef{eq:monkees} converges iff
\[
x \to 1^{+}:
\quad
\Re(J+\Delta_\phi) > 0,
\quad
\Re(J + d-\Delta_\phi) > 0
\quad
\text{as well as}
\quad
x \to \infty:
\quad
d < 3~.
\]

Let us now calculate the integral~\reef{eq:monkees}. Inside the integrand, we replace the ${}_2F_1$ appearing in $\text{Disc}\, G(x)$ with the help of the Barnes hypergeometric integral representation
\begin{equation}
_{2}F_{1}(a,b,c,z)= \frac{\Gamma(c)}{2\pi i\Gamma(a)\Gamma(b)} \int^{\gamma+i\infty}_{\gamma-i\infty} ds\; \frac{\Gamma(s)\Gamma(a-s)\Gamma(b-s)}{\Gamma(c-s)} (-z)^{-s}~,
\end{equation}
where $\gamma$ is chosen in such a way that the three families of poles in the $s$-plane that move to the left and right are separated.  After the change of variable $x \rightarrow t=\frac{2}{x-1}$ and using the identity~\reef{eq:power 2f1 integral}, we can compute the $t$-integral exactly. This yields
\begin{multline}
a_J = \frac{\Gamma(\frac{3-d}{2})}{2^{J+d}\pi i \Gamma(1-\D_\phi)\Gamma(1+\D_\phi-d)} \frac{\Gamma(2J+d+1)}{\Gamma(J+d)\Gamma(J+\frac{d}{2}+\frac{1}{2})} \\
\times \int^{\gamma+i\infty}_{\gamma-i\infty}  ds \, 
\frac{- \Gamma(1-s) \Gamma(s)\Gamma(1+\D_\phi-d-s)\Gamma(1-\D_\phi-s) \Gamma(J+s+d-1)}{\Gamma(J-s+2)}~.
\end{multline}
The remaining Mellin-Barnes integral can be done using~\reef{eq:second barnes lemma}, which yields
\beq
a_J =\frac{R^{1-d}}{4 \pi^{1+\frac{d}{2}} 2^{2J}} \frac{\Gamma^2(\frac{d}{2}) \Gamma(\frac{d+1}{2})}{\Gamma(d)}  \frac{\Gamma(2J+d+1)}{\Gamma(J+\frac{d}{2})\Gamma(J+\frac{d}{2}+\frac{1}{2})} \frac{1}{(J+d-\D_\phi)(J+\D_\phi)}~.
\eeq
Using some simplifications, we indeed recover the result~\reef{eq:marolfaj}.

\subsection{Large $J$ behavior}\label{LargeJ}

As discussed in section~\ref{sec:two-point sphere}, we studied the analytic continuation of $a_J$ using the inversion formula~\reef{The inversion formula} to find the spectral density of the theory. As we change the contour in~\reef{eq:int over gtilde}, we need to know the large $J$ behavior of $a_J$ and to be precise, we want to find the upper bound of $a_{J}$ as we approach the limit $|J|\rightarrow \infty$. We will argue that the $J \to \infty$ behavior is related to the $x \to 1$ (or $\xi \to \infty$) limit of the correlator. We have already encountered this in one example: for the bulk CFT correlator~\reef{eq:CRJ}, we computed that
\bsub
\beq
G_\delta(x) = \frac{1}{(1-x)^\delta}
\quad
\Rightarrow
\quad
\rho_\delta(\thd + i \nu) \; \limu{\nu \to \infty} \; \frac{2^{d+2}\pi^{(d+3)/2}}{\Gamma(\delta)\Gamma(\delta-\hd+\th)}\, \nu^{2\delta -d}
\eeq
or using~\reef{eq:jac0} and setting $\nu \to J$, at least formally we obtain
\beq
\label{eq:hypo}
a_J \; \limu{J \to \infty} \; 1/J^{d-2\delta}~.
\eeq
\esub
We want to put this relation~\reef{eq:hypo} on a more solid footing by means of Eq.~\reef{The inversion formula}.

Let us spell out the assumptions going in the derivation below. We assume that the discontinuity of $G(x)$ behaves as
\beq
\label{eq:moutarde}
x \geq 1:
\quad
\mrm{Disc}\, G(x) = \left(\frac{x+1}{x-1}\right)^\delta \, \wh{G}(x)
\quad
\text{for some}
\quad
\delta <1\,.
\eeq
Here
$\wh{G}(x)$ is a bounded and slowly varying function on $[1,\infty)$, having a finite limit as $x \to 1$.  It turns out that the large-$x$ behavior of $\widehat{G}(x)$ is not really important, provided that $\wh{G}(x)$ does not grow faster than any power law.
The restriction $\delta < 1$ is necessary to guarantuee convergence of the inversion formula at finite $J$, and the second assumption (which is stronger in $d<2$ but weaker for $d \geq 2$) is needed to have a uniform $J \to \infty$ limit, as we will see. For values $\delta \geq 1$ the integrand needs to be regulated, and we will not discuss this case at present.

Given the above, we write the inversion formula for this case as
\be
\label{eq:starttt}
a_J \approx  \frac{1}{4^J J^{d/2-1}} \int_{1}^\infty\!\frac{dx}{(x-1)^\delta} \left(\frac{2}{1+x}\right)^{J+1-\delta}\, \mca{F}_J(x) \, \wh{G}(x)
\ee
where we defined
\be
\mca{F}_J(x) \ldef \FF{J+1}{J+\hd+\half}{2J+d+1}{\frac{2}{1+x}}~.
\ee
We have dropped some $J$-independent factors in the prefactor, as they will not play a role later. Eq.~\reef{eq:starttt} can be obtained from the inversion formula by a hypergeometric transformation. The function $\mca{F}_J(x)$ is a manifestly decreasing function of $x$ that has a finite limit as $x \to 1$ (unless $d=1$, in which case $\mca{F}_J(x)$ diverges logarithmically) and obeys $\mca{F}_J(x) \to 1$  as $x \to \infty$.

We now claim that in the $J \to \infty$ limit, $a_J$ is dominated by the part of the integral near $x = 1$. To wit, fix some $c > 1$ and split the integral into two parts:
\[
a_J = a_J^{(1)} + a_J^{(2)},
\quad
a_J^{(1)} = \int_{1}^c \left[ \ldots \right]
\qaq
a_J^{(2)} = \int_{c}^\infty \left[ \ldots \right]~.
\]
Using the above assumptions, it is easy to show that
\beq
J \gg 1:
\quad
\big| a_J^{(2)} \big| \leq \frac{C}{2^J J^{d/2}}
\eeq
for some constant $C > 0$. This contribution is exponentially small, whereas $a_J^{(1)}$ will scale as a power law. In order to estimate $a_J^{(1)}$, we first estimate $\mca{F}_J(x)$ using steepest descent. In order to do so we employ the integral representation
\beq
\label{intrepFJ}
\mca{F}_J(1+y) = \frac{\Gamma(d+2J+1)}{\Gamma\!\left(J+\frac{d}{2}+\frac{1}{2}\right)^2} \int_0^1\!dt\; \frac{(y+2) (t(1-t))^{\frac{d-1}{2}} }{2+y-2t} \left(\frac{t(1-t) (y+2)}{2+y-2t}\right)^J~.
\eeq
At large $J$, the integral is dominated by the contribution near
\[
t = t_*(y) = \frac{2+y-\sqrt{y(2+y)}}{2}.
\]
After evaluating the integral using steepest descent, at large $J$ and fixed $y$ we then obtain
\beq
\mca{F}_J(1+y) \; \limu{J \to \infty} 4^J\, \wh{F}(y)\, e^{-Jq(y)}
\ee
where $\widehat{F}(y)$ is a rather complicated function of $y$ that does not depend on $J$ and 
\be
q(y) = \ln 2-\ln\left[2-(2+y)\sqrt{y(2+y)}+y(3+y) \right] \approx \sqrt{2y} + O(y)~.
\eeq
Because of the exponential, values of $x = 1+y$ for which $q(y) \gtrsim 1/J$ are suppressed in the integral~\reef{eq:starttt} (which is cut off at $x=c$). In terms of the variable
\[
v \ldef \sqrt{2y}J
\]
this condition reads $v \lesssim 1$. 
The relevant limit is then
\beq
\mca{F}_J\left (1+\frac{v^2}{2J^2}\right) \limu{J \to \infty} 
J^{1-\hd}2^{d+2J}\frac{1}{\sqrt{\pi}} \int_0^\infty dr \, r^\frac{d-3}{2} e^{-r-\frac{v^2}{4r}} =
J^{1-\hd}2^{\frac{3+d+4J}{2}}\frac{1}{\sqrt{\pi}}  v^\frac{d-1}{2} K_{\frac{d-1}{2}}(v)
\eeq
where we used the integral representation \eqref{intrepFJ} with $t=1-r/J$ because the integral is dominated by $1-t \sim 1/J$.
We can therefore remove the cutoff $c$, perform the indicated change of variable and take the limit $J \gg 1$. Keeping track of powers of $J$, this results in the following estimate:
\beq
a_J^{(1)} \; \limu{J \to \infty} \; \frac{1}{{J^{d-2\delta}} } \frac{2^{\frac{3 + d}{2} + \delta} \wh{G}(0)}{\sqrt{\pi}}\int_0^\infty \frac{dv}{v^{2\delta-\frac{d+1}{2}}}K_{\frac{d-1}{2}}(v) 
= \frac{ \wh{G}(0)}{{J^{d-2\delta}} } \frac{2^{1+d- \delta} \Gamma(1-\delta)
\Gamma(\frac{1+d-2\delta}{2})}{\sqrt{\pi}} ~.
\eeq
This is the desired result, provided that the integral on the RHS converges. It does so precisely because of the assumption made in~\reef{eq:moutarde}.  This concludes the proof. 

\section{From $SO(2,2)$ to $SO(2,1)$}
\label{Sec:DiscreteSeries}

A generic quantum field theory on dS have the symmetries dicatated by background metric of dS i.e.\@ $SO(d+1,1)$. A conformal theory, on the other hand, has more symmetries. The fact that its energy-momentum tensor is traceless enhances it symmetry group to $SO(d+1,2)$. In this appendix, we study how the unitary irreducible representations of $SO(d+1,2)$ decompose into irreps of the subgroup $SO(d+1,1)$ in the case $d=1$. 

Take the generators of $SO(d+1,2)$ to be the Lorentz generators $J_{AB}$   in embedding space $\mathbb{R}^{d+1,2}$, with the metric $\eta = \text{diag}(-1,-1,+1,\dots, +1)$ in which $A,B \in \{-1,0,1,\dots,d+1\}$. These satisfy these commutation relations \eqref{Jcom} and are anti-hermitian $J_{AB}^\dagger = -J_{AB}$.

The generators of the $SO(d+1,2)$ conformal group can be written as
\begin{align}
\tilde{D} &= -i J_{-10} \\
\tilde{P}_a &= -i J_{-1 a } + J_{0 a }  \\
\tilde{K}_a &= -i J_{-1 a } -  J_{0 a }  \\
\tilde{M}_{ab} &=  -i J_{ab}\,.
\end{align}
where $a,b \in {1,2,\dots,d+1}$ and we used tildes to distinguish from the $SO(d+1,1)$ generators defined by~\eqref{eq:relabelGenerators}. The hermiticity properties are then
\begin{align}
\tilde{D}^\dagger = \tilde{D} \,,\qquad
(\tilde{P}_a)^\dagger  = \tilde{K}_a   \,,\qquad
(\tilde{M}_{ab})^\dagger   =  M_{ab}\,.
\end{align}
Notice that the conventions here differ from those in the main text, namely~\reef{eq:relabelGenerators}, which led to anti-hermitian generators. 

Let us now focus in the case $d=1$ which corresponds to $SO(2,1) \cong SL(2,\mathbb{R})$ (at the level of the algebra).
In this case, it is convenient to use the following basis for the algebra  
\be
\begin{aligned}
S^z&=-iJ_{12}=\tilde{M}_{12}~,\\
S^+&=-iJ_{01}- J_{02} = \frac{ \tilde{K}_{2} - \tilde{P}_{2} }{2} + i \frac{ \tilde{K}_{1} - \tilde{P}_{1} }{2}~,\\
S^-&=-iJ_{01}+ J_{02} =- \frac{ \tilde{K}_{2} - \tilde{P}_{2} }{2} + i \frac{ \tilde{K}_{1} - \tilde{P}_{1} }{2}~.
\end{aligned}
\ee
This leads to the usual $SL(2,\mathbb{R})$ commutation relations
\beq
[S^z,S^\pm]=\pm S^\pm\,,\qquad\qquad
[S^+,S^-]=-2S^z\,,
\eeq
and Casimir
\beq
C=(S^z)^2-\frac{1}{2}(S^+S^-+S^-S^+)\,.
\eeq
The hermiticity properties are
\beq
(S^z )^\dagger = S^z \,,\qquad \qquad 
(S^+ )^\dagger = S^- \,.
\eeq
Principal series representations have Casimir eigenvalue $C=-\frac{1}{4} -\nu^2\le -\frac{1}{4}$.
Complementary series have $-\frac{1}{4}\le C\le0$. Discrete series have $C=k(k-1)$ with $k =1,2,\dots$.

A highest weight representation of $SO(2,2)$ is the vector space generated by the states
\beq
|n,\bar{n}\rangle = (\tilde{P}_1-i\tilde{P}_2)^n (\tilde{P}_1+i\tilde{P}_2)^{\bar{n}} |\Delta,\ell \rangle\,,\qquad
n,\bar{n} \in \{0,1,2,\dots \}\,,
\eeq
with $|\Delta,\ell \rangle$ a primary state,\footnote{Notice that here we use $\Delta$ to denote the eigenvalue of the $SO(2,2)$ dilatation generator $\tilde{D}$.
The notation $\tilde{\Delta}$, used in the main text, would be appropriate but we shall use simply $\Delta$ to avoid cluttering the equations in this appendix.}
\beq
 \tilde{K}_{1} |\Delta,\ell \rangle=\tilde{K}_{2} |\Delta,\ell \rangle=0\,,\qquad
  \tilde{M}_{12} |\Delta,\ell \rangle=\ell |\Delta,\ell \rangle\,,\qquad
    \tilde{D} |\Delta,\ell \rangle=\Delta |\Delta,\ell \rangle\,.
\eeq

We would like to diagonalize the Casimir $C$ in this vector space.
First notice that $S^z$ is already diagonal
\beq
S^z |n,\bar{n}\rangle =(n-\bar{n} +\ell) |n,\bar{n}\rangle\equiv s |n,\bar{n}\rangle~.
\eeq
The action of the Casimir takes the form
\beq
C |n,\bar{n}\rangle =q(n) |n,\bar{n}\rangle +w(n) |n-1,\bar{n}-1\rangle +\frac{1}{4} |n+1,\bar{n}+1\rangle~,
\eeq
where
\be
\begin{aligned}
q(n)&= - n  (\Delta-\ell+2\bar{n} ) -  \bar{n}   ( \Delta+\ell ) +\left(\ell^2 - \Delta \right)\\
&=\Delta  (-\ell-2 n+s-1)+\ell (s-2 n)+2 n (s-n)~.
\end{aligned} 
\ee
and
\be
\begin{aligned}
w(n)&=4 \sum_{k=1}^n ( \Delta+\ell +2k-2) \sum_{q=1}^{\bar{n}}  (\Delta -\ell+2q-2)\\
&=4 n (\Delta +\ell+n-1) (\ell+n-s) (\Delta +n-s-1)~.
\end{aligned} 
\ee
These functions were computed using the commutators
\be
\begin{aligned}
[C,P]&=K( \tilde{D}+S^z) -P( \tilde{D} - S^z)~,\\
[C,\bar{P}]&=\bar{K}( \tilde{D}-S^z) -\bar{P}( \tilde{D} + S^z)~, \\
[K,\bar{P}]&=4( \tilde{D} - S^z)~,\\
[\bar{K},P]&=4( \tilde{D} + S^z)~,\\
[K,P]&=0~,\\
[\bar{K},\bar{P}]&=0~.
\end{aligned}
\ee
where $P\equiv \tilde{P}_1-i\tilde{P}_2$,  $\bar{P}\equiv \tilde{P}_1+i\tilde{P}_2$, 
$K\equiv \tilde{K}_1-i\tilde{K}_2$ and  $\bar{K}\equiv \tilde{K}_1+i\tilde{K}_2$.
In practice, we used
\begin{equation*}
C |n,\bar{n}\rangle = \sum_{k=1}^n P^{n-k} [C,P] P^{k-1} \bar{P}^{\bar{n}} |\Delta,\ell \rangle + \sum_{k=1}^{\bar{n}} P^{n} \bar{P}^{\bar{n}-k}[C,\bar{P}] \bar{P}^{k-1}  |\Delta,\ell \rangle  +  P^n \bar{P}^{\bar{n}} C|\Delta,\ell \rangle  
\end{equation*}
together with
\be\label{KthroughPs}
\begin{aligned}
K
\bar{P}^{\bar{n}} |\Delta,\ell \rangle &=
\sum_{q=1}^{\bar{n}} \bar{P}^{\bar{n}-q}[K , \bar{P}] 
\bar{P}^{q-1} |\Delta,\ell \rangle =
4 \sum_{q=1}^{\bar{n}}  (\Delta +q-1-\ell+q-1)
\bar{P}^{\bar{n}-1} |\Delta,\ell \rangle\\
C|\Delta,\ell \rangle &= \left(\ell^2 - \Delta \right) |\Delta,\ell \rangle + \frac{1}{4} 
P\bar{P}|\Delta,\ell\rangle~.
\end{aligned}
\ee

Simultaneous eigenstates of $S^z$ (with eigenvalue $s\le \ell$) and the Casimir $C$ can be written as
\beq
|\psi \rangle= \sum_{n=0}^\infty a_n  |n,\ell-s+n\rangle~.
\eeq
Then, $C|\psi \rangle=\lambda |\psi \rangle$ leads to the recursion equation
\beq
\lambda a_n = q(n) a_n +w(n+1) a_{n+1} +\frac{1}{4}a_{n-1}~.
\label{recan}
\eeq
The eigenvalues $\lambda$ will be fixed by requiring that the solution to this equation has finite norm
\beq
\begin{aligned}
\langle \psi |\psi \rangle =& \sum_{n=0}^\infty |a_n|^2 \langle n,\ell-s+n  |n,\ell-s+n\rangle\\
=& \sum_{n=0}^\infty |a_n|^2 4^{\ell-s+2n} n! (\ell-s+n)! (\Delta+\ell)_n(\Delta-\ell)_{n+\ell-s}
\end{aligned}
\eeq
where we used 
\beq
\langle n,\bar{n} |n,\bar{n}\rangle = 4^{n+\bar{n}} n! \bar{n}! 
(\Delta+\ell)_{n} (\Delta-\ell)_{\bar{n}}~.
\eeq
This expression for the norm follows from (using \eqref{KthroughPs})
\be
\begin{aligned}
\langle n,\bar{n} |n,\bar{n}\rangle &= 
\langle \Delta,\ell | K^{\bar{n}}\bar{K}^n
P^n \bar{P}^{\bar{n}} |\Delta,\ell \rangle \\ 
&= 
4 \bar{n} (\Delta -\ell +\bar{n}-1)
\langle \Delta,\ell | K^{\bar{n}-1}\bar{K}^n
P^n \bar{P}^{\bar{n}-1} |\Delta,\ell \rangle
\\ 
&= 
4 \bar{n} (\Delta -\ell +\bar{n}-1)\langle n,\bar{n}-1 |n,\bar{n}-1\rangle~. 
\end{aligned}
\ee
It is convenient to define
\beq
c_n = a_n \sqrt{4^{\ell-s+2n} n! (\ell-s+n)! (\Delta+\ell)_n(\Delta-\ell)_{n+\ell-s}}
\eeq
so that the inner product becomes
\beq
\langle \psi |\psi' \rangle = \sum_{n=0}^\infty c_n^* c_n'\,.
\eeq
The recursion relation then becomes
\begin{multline}
\frac{ \left(\Delta +\lambda +\ell (\Delta +2 n-s)+2 n^2+2 n (\Delta
   -s)-\Delta  s\right)}{\sqrt{n (\Delta +\ell+n-1) (\ell+n-s) (\Delta
   +n-s-1)}}c_n\\
   - \sqrt{\frac{(n+1) (\Delta +\ell+n) (\ell+n-s+1) (\Delta
   +n-s)}{n (\Delta +\ell+n-1) (\ell+n-s) (\Delta +n-s-1)}} c_{n+1}=c_{n-1}~.
   \nonumber
\end{multline}
This implies the following asymptotic behavior
\beq
c_n = \frac{R}{n^{\frac{1-i\nu}{2}}} \left[ 1+ O(1/n) \right] + \text{c.c}~.
\eeq
where $4\lambda+1=-\nu^2$.  The complex parameter $R$ cannot be determined from an asymptotic analysis of the recursion relation. Here we assumed that the parameter $\nu$ is real as required for principal series representations.  
In this case, the state $|\psi \rangle$ is delta-function normalizable. Let us see how this works
\be
\begin{aligned}
\langle \psi |\psi' \rangle  &\sim 2|R R'| \sum_{n}^\infty \frac{1}{n} 
\left[ \cos\left(\frac{\nu-\nu'}{2} \log n +\phi-\phi'\right)+
\cos\left(\frac{\nu+\nu'}{2} \log n +\phi+\phi'\right)
\right]\\
&\sim  2|R R'| \int^\infty dy 
\left[ \cos\left(\frac{\nu-\nu'}{2} y +\phi-\phi'\right)+
\cos\left(\frac{\nu+\nu'}{2} y +\phi+\phi'\right)
\right]\\
&\sim 4\pi |R R'| 
\left[ \delta(\nu-\nu')+\delta(\nu+\nu')
\right]
\end{aligned}
\ee
where we used $R=|R|e^{i\phi}$ and $R'=|R'|e^{i\phi'}$. Notice that the appearance of the $\delta-$functions follows solely from the asymptotic behavior of the coefficients $c_n$. On the other hand, orthogonality between eigenstates of different Casimir eigenvalue is guaranteed.
We conclude that the $SO(2,2)$ highest weight unitary irreducible representations contains $SO(2,1)$ principal series representations for all values of $\nu \in \mathbb{R}$ (with $\nu$ and $-\nu$ identified).

For complementary and discrete series representations, we have $4\lambda+1=v^2$ with $v>0$.
This leads to 
\beq
c_n = \frac{R_+}{n^{\frac{1+v}{2}}} \left[ 1+ O(1/n) \right] + \frac{R_-}{n^{\frac{1-v}{2}}} \left[ 1+ O(1/n) \right]~.
\label{normasymp}
\eeq
Generically, this leads to non-normalizable states
\begin{align}
\langle \psi |\psi' \rangle  \sim \sum_{n}^\infty \
n^{-1+\frac{v+v'}{2}} \to \infty\,.
\end{align}
Of course, if $R_-=0$ then we obtain a normalizable state. 
In fact, we will now construct some exact solutions with  $R_-=0$. We suspect these exhaust the solutions with  $R_-=0$ but have no proof of this fact.

Discrete series irreps are highest/lowest weight for $S^z$ and therefore, they must contain a state that is annihilated by $S^+$/$S^-$.
This condition leads to a first order recursion relation. Firstly, notice that 
\begin{align}
-2iS^+|n,\bar{n}\rangle&=\left(K -  P\right) P^n \bar{P}^{\bar{n}} |\Delta,\ell \rangle =4 \bar{n} (\Delta -\ell +\bar{n}-1)|n,\bar{n}-1 \rangle - |n+1,\bar{n}\rangle
\end{align}
where we used \eqref{KthroughPs}. Therefore, $S^+ |\psi \rangle =0$ leads to 
\beq
4 (\ell-s+n) (\Delta -s+n-1) a_n  - a_{n-1}=0~.
\eeq
In particular, the equation with $n=0$ can only be satisfied if $s=\ell$ or $a_0=0$.\footnote{There is another formal solution with $\ell>s=\Delta-1$. The unitarity bound $\Delta \ge  |\ell|$ then implies that the second possibility only works for $\Delta=\ell$ and $s=\ell-1$. But then the states with non-zero $\bar{n}$ have zero norm (they are descendants of the state associated to the divergence of the conserved current).}
If $a_0 \neq 0$ then we find
\beq
a_n = \frac{a_0}{4^n n! (\Delta-\ell)_n}
\eeq
with associated norm
\beq
\langle \psi |\psi \rangle =|a_0|^2\sum_{n=0}^\infty  \frac{(\Delta+\ell)_n}{ (\Delta-\ell)_n}
\sim \sum_{n}^\infty  n^{2\ell}
\eeq
which converges for (half-integer) $\ell \le -1$.
Indeed, this solves the recursion relation \eqref{recan} with $\lambda=\ell(\ell+1)$.
This matches exactly the expectation from the discrete series.
In fact there are more solutions of the form $a_n=0$ for $n < s- \ell$ and 
\beq
a_{s-\ell +\bar{n} } = \frac{a_{s-\ell} }{4^{\bar{n}} \bar{n}! (\Delta-\ell)_{\bar{n}}}\,,\qquad \qquad
\bar{n} \ge 0.
\eeq
The norm of this state is  (here $n=\bar{n}+s-\ell$)
\beq
\langle \psi |\psi \rangle =|a_{s-\ell}|^2\sum_{\bar{n}=0}^\infty  4^{n-\bar{n}} \frac{n!}{\bar{n}!} \frac{(\Delta+\ell)_n}{ (\Delta-\ell)_{\bar{n}}}
\sim \sum_{\bar{n}}^\infty  \bar{n}^{2s}\,.
\eeq
We conclude that there is a normalizable highest weight state for every $s=-1,-2,\dots,-|\ell|$.

Similarly, looking for lowest weight states obeying  $S^- |\psi\rangle=0$ we find states with  $s=  1,2 , \dots, |\ell|$.
We conclude that for each $SO(2,2)$ conformal family based on a primary of non-zero spin $\ell$, there are $|\ell|$ discrete series irreps of  $SO(2,1)$ with Casimir eigenvalue $\lambda = s(s-1)$ with $s \in \{1,2,\dots,|\ell|\}$. This seems to be confirmed by numerical experiments where we diagonalize  matrix truncations of the Casimir operator.

There is also a complementary series irrep with Casimir eigenvalue  $ \lambda  = \Delta(\Delta-1)$ for conformal families with $\Delta <\frac{1}{2}$. In this case, there is an exact solution 
\beq
a_n = \frac{(\ell-s)!}{4^n\, n! (n+\ell-s)!}a_0~.
\eeq
which matches the expansion \eqref{normasymp} with $R_-=0$. This gives a normalizable state in the complementary series.
Notice that this state is really normalizable as opposed to delta-function normalizable like the principal series states.
Finally, notice that the unitarity bound $\Delta > |\ell|$ implies that this complementary series irrep only exists for $\ell=0$ conformal families.
The presence of this state matches the comments after equation \eqref{eq:Spectral CFT2} about the Källén-Lehmann decomposition  of the two-point function of a CFT primary operator with scaling dimension smaller than $\frac{d}{2}$.


\phantomsection
{
\footnotesize{
\bibliographystyle{head/utphys}
\bibliography{head/biblio}
 }
}

\end{document}